\documentclass[12pt]{article}
\usepackage[dvips]{graphicx}
\usepackage{epsfig}
\usepackage{latexsym,amsmath,amsfonts,amssymb}
\usepackage[latin1]{inputenc}
\usepackage[american]{babel}
\usepackage[dvips]{graphicx}
\usepackage{bbm}
\def\prl{Phys. Rev. Lett.}
\def\pr{Phys. Rev.}

\def\im{Invent. Math.}

\newcommand{\be}{\begin{equation}}
\newcommand{\ee}{\end{equation}}
\newcommand{\beq}{\begin{equation}}
\newcommand{\eeq}{\end{equation}}
\newcommand{\bea}{\begin{eqnarray}}
\newcommand{\eea}{\end{eqnarray}}

\newcommand{\ba}{\begin{eqnarray}}
\newcommand{\ea}{\end{eqnarray}}
\begin{document}
\baselineskip=15.5pt
\pagestyle{plain}
\setcounter{page}{1}
%--------+---------+---------+---------+---------+---------+---------+
%Body

% Ofer's definitions

\def\del{{\partial}}
\def\vev#1{\left\langle #1 \right\rangle}
\def\cn{{\cal N}}
\def\co{{\cal O}}
%\newfont{\Bbb}{msbm10 scaled 1200}     %instead of eusb10
%\newcommand{\mathbb}[1]{\mbox{\Bbb #1}}
\def\IC{{\mathbb C}}
\def\IR{{\mathbb R}}
\def\IZ{{\mathbb Z}}
\def\RP{{\bf RP}}
\def\CP{{\bf CP}}
\def\Poincare{{Poincar\'e }}
\def\tr{{\rm tr}}
\def\tp{{\tilde \Phi}}

\def\TL{\hfil$\displaystyle{##}$}
\def\TR{$\displaystyle{{}##}$\hfil}
\def\TC{\hfil$\displaystyle{##}$\hfil}
\def\TT{\hbox{##}}
\def\HLINE{\noalign{\vskip1\jot}\hline\noalign{\vskip1\jot}}
\def\seqalign#1#2{\vcenter{\openup1\jot
   \halign{\strut #1\cr #2 \cr}}}
\def\lbldef#1#2{\expandafter\gdef\csname #1\endcsname {#2}}
\def\eqn#1#2{\lbldef{#1}{(\ref{#1})}%
\begin{equation} #2 \label{#1} \end{equation}}
\def\eqalign#1{\vcenter{\openup1\jot
     \halign{\strut\span\TL & \span\TR\cr #1 \cr
    }}}
\def\eno#1{(\ref{#1})}
\def\href#1#2{#2}
\def\half{{1 \over 2}}

%--------+---------+---------+---------+---------+---------+---------+
%Hirosi's macros:
\def\ads{{\it AdS}}
\def\adsp{{\it AdS}$_{p+2}$}
\def\cft{{\it CFT}}

\newcommand{\ber}{\begin{eqnarray}}
\newcommand{\eer}{\end{eqnarray}}

\newcommand{\beqar}{\begin{eqnarray}}
\newcommand{\cN}{{\cal N}}
\newcommand{\cO}{{\cal O}}
\newcommand{\cA}{{\cal A}}
\newcommand{\cT}{{\cal T}}
\newcommand{\cF}{{\cal F}}
\newcommand{\cC}{{\cal C}}
\newcommand{\cR}{{\cal R}}
\newcommand{\cW}{{\cal W}}
\newcommand{\eeqar}{\end{eqnarray}}
\newcommand{\tht}{\thteta}
\newcommand{\lm}{\lambda}\newcommand{\Lm}{\Lambda}
\newcommand{\eps}{\epsilon}

%--------+---------+---------+---------+---------+---------+---------+

\newcommand{\nonu}{\nonumber}
\newcommand{\oh}{\displaystyle{\frac{1}{2}}}
\newcommand{\dsl}
   {\kern.06em\hbox{\raise.15ex\hbox{$/$}\kern-.56em\hbox{$\partial$}}}
\newcommand{\id}{i\!\!\not\!\partial}
\newcommand{\as}{\not\!\! A}
\newcommand{\ps}{\not\! p}
\newcommand{\ks}{\not\! k}
\newcommand{\D}{{\cal{D}}}
\newcommand{\dv}{d^2x}
\newcommand{\Z}{{\cal Z}}
\newcommand{\N}{{\cal N}}
\newcommand{\Dsl}{\not\!\! D}
\newcommand{\Bsl}{\not\!\! B}
\newcommand{\Psl}{\not\!\! P}
\newcommand{\eeqarr}{\end{eqnarray}}
\newcommand{\ZZ}{{\rm \kern 0.275em Z \kern -0.92em Z}\;}
%--------------------------------Alfonso's definitions%%%%%%%%%%%%%

% DEFINITIONS

\def\del{{\delta^{\hbox{\sevenrm B}}}} \def\ex{{\hbox{\rm e}}}
\def\azb{A_{\bar z}} \def\az{A_z} \def\bzb{B_{\bar z}} \def\bz{B_z}
\def\czb{C_{\bar z}} \def\cz{C_z} \def\dzb{D_{\bar z}} \def\dz{D_z}
\def\im{{\hbox{\rm Im}}} \def\mod{{\hbox{\rm mod}}} \def\tr{{\hbox{\rm Tr}}}
\def\ch{{\hbox{\rm ch}}} \def\imp{{\hbox{\sevenrm Im}}}
\def\trp{{\hbox{\sevenrm Tr}}} \def\vol{{\hbox{\rm Vol}}}
\def\rl{\Lambda_{\hbox{\sevenrm R}}} \def\wl{\Lambda_{\hbox{\sevenrm W}}}
\def\fc{{\cal F}_{k+\cox}} \def\vev{vacuum expectation value}
\def\nodiv{\mid{\hbox{\hskip-7.8pt/}}}
\def\ie{{\em i.e.}}
\def\ie{\hbox{\it i.e.}}

\def\CC{{\mathchoice
{\rm C\mkern-8mu\vrule height1.45ex depth-.05ex
width.05em\mkern9mu\kern-.05em}
{\rm C\mkern-8mu\vrule height1.45ex depth-.05ex
width.05em\mkern9mu\kern-.05em}
{\rm C\mkern-8mu\vrule height1ex depth-.07ex
width.035em\mkern9mu\kern-.035em}
{\rm C\mkern-8mu\vrule height.65ex depth-.1ex
width.025em\mkern8mu\kern-.025em}}}

\def\RR{{\rm I\kern-1.6pt {\rm R}}}
\def\NN{{\rm I\!N}}
\def\ZZ{{\rm Z}\kern-3.8pt {\rm Z} \kern2pt}
\def\IB{\relax{\rm I\kern-.18em B}}
\def\ID{\relax{\rm I\kern-.18em D}}
\def\II{\relax{\rm I\kern-.18em I}}
\def\IP{\relax{\rm I\kern-.18em P}}
\newcommand{\CS}{{\scriptstyle {\rm CS}}}
\newcommand{\CSs}{{\scriptscriptstyle {\rm CS}}}
\newcommand{\rc}{\nonumber\\}
\newcommand{\bear}{\begin{eqnarray}}
\newcommand{\eear}{\end{eqnarray}}
\newcommand{\W}{{\cal W}}
\newcommand{\F}{{\cal F}}
\newcommand{\x}{{\cal O}}
\newcommand{\LL}{{\cal L}}

\def\mani{{\cal M}}
\def\calo{{\cal O}}
\def\calb{{\cal B}}
\def\calw{{\cal W}}
\def\calz{{\cal Z}}
\def\cald{{\cal D}}
\def\calc{{\cal C}}
\def\to{\rightarrow}
\def\ele{{\hbox{\sevenrm L}}}
\def\ere{{\hbox{\sevenrm R}}}
\def\zb{{\bar z}}
\def\wb{{\bar w}}
\def\nodiv{\mid{\hbox{\hskip-7.8pt/}}}
\def\menos{\hbox{\hskip-2.9pt}}
\def\dr{\dot R_}
\def\drr{\dot r_}
\def\ds{\dot s_}
\def\da{\dot A_}
\def\dga{\dot \gamma_}
\def\ga{\gamma_}
\def\dal{\dot\alpha_}
\def\al{\alpha_}
\def\cl{{closed}}
\def\cls{{closing}}
\def\vev{vacuum expectation value}
\def\tr{{\rm Tr}}
\def\to{\rightarrow}
\def\too{\longrightarrow}

% Umut likes:

\def\a{\alpha}
\def\b{\beta}
\def\c{\gamma}
\def\d{\delta}
\def\e{\epsilon}           % Also, \varepsilon
\def\f{\phi}               %      \varphi
\def\vf{\varphi}  \def\tvf{\tilde{\varphi}}
\def\vp{\varphi}
\def\g{\gamma}
\def\h{\eta}
\def\j{\psi}
\def\k{\kappa}                    % Also, \varkappa (see below)
\def\l{\lambda}
\def\m{\mu}
\def\n{\nu}
\def\o{\omega}  \def\w{\omega}
\def\q{\theta}  \def\th{\theta}                  %     \vartheta
\def\r{\rho}                                     %     \varrho
\def\s{\sigma}                                   %     \varsigma
\def\t{\tau}
\def\u{\upsilon}
\def\x{\xi}
\def\z{\zeta}
\def\pt{\tilde{\varphi}}
\def\tt{\tilde{\theta}}
\def\lab{\label}
\def\6{\partial}
\def\wg{\wedge}
\def\atanh{{\rm arctanh}}
\def\bpsi{\bar{\psi}}
\def\bt{\bar{\theta}}
\def\bvf{\bar{\varphi}}

%
% FONTS

%\newfont{\headfont}{cmbx10 scaled 1440}
\newfont{\namefont}{cmr10}
%\newfont{\initialfont}{cmr10 scaled 1200}
\newfont{\addfont}{cmti7 scaled 1440}
\newfont{\boldmathfont}{cmbx10}
%\newfont{\figfont}{cmr7 scaled 1200}
\newfont{\headfontb}{cmbx10 scaled 1728}
%%%%%%%%%%%%%%%%%%%%%%%%%%%%%%%%%%%%%%%%%%%%%%%%%%%%%%%%%%%%%%%%%%%%%%%%%%%%
%%%%%%%%%%%%%%%Stefano and Francesco fonts%%%%%%%%%%%%%%%%%%%%%%%%%%%%%%%%
%%%%%%%%%%%%%%%%%%%%%%%%%%%%%%%%%%%%%%%%%%%%%%%%%%%%%%%%%%%%%%%%%%%%%%%%%%%%
\newcommand{\re}{\,\mathbb{R}\mbox{e}\,}
\newcommand{\hyph}[1]{$#1$\nobreakdash-\hspace{0pt}}
\providecommand{\abs}[1]{\lvert#1\rvert}
\newcommand{\Nugual}[1]{$\mathcal{N}= #1 $}
\newcommand{\sub}[2]{#1_\text{#2}}
\newcommand{\partfrac}[2]{\frac{\partial #1}{\partial #2}}
\newcommand{\bsp}[1]{\begin{equation} \begin{split} #1 \end{split} \end{equation}}
\newcommand{\calF}{\mathcal{F}}
\newcommand{\calO}{\mathcal{O}}
\newcommand{\calM}{\mathcal{M}}
\newcommand{\calV}{\mathcal{V}}
\newcommand{\bbZ}{\mathbb{Z}}
\newcommand{\bbC}{\mathbb{C}}

\numberwithin{equation}{section}

\newcommand{\Tr}{\mbox{Tr}}    % trace over gauge indices

%%%%%%%%%%%%%%%%%%%%%%%%%%%%%%%%%%%%%%%%%%%%%%%%%%%%%%%%%%%%%%%%%%%%%%%%%%%%
%%%%%%%%%%%%%%%%%%%%%%%%%%%%%%%%%%%%%%%%%%%%%%%%%%%%%%%%%%%%%%%%%%%%%%%%%%%%

%
\renewcommand{\theequation}{{\rm\thesection.\arabic{equation}}}
\begin{titlepage}
\rightline{SISSA 54/2009/EP}  
%\rightline{US-FT-3/07}
\vspace{0.1in}

\begin{center}
\Large \bf Gravity duals of  2d supersymmetric  gauge theories
\end{center}
\vskip 0.2truein
\begin{center}
Daniel Are\'an ${}^{\dagger}$\footnote{arean@sissa.it}, 
Eduardo Conde  ${}^{*}$\footnote{eduardo@fpaxp1.usc.es}, 
 and
Alfonso V. Ramallo${}^{*}$\footnote{alfonso@fpaxp1.usc.es}\\
\vspace{0.2in}
${}^{*}$\it{
Departamento de  F\'\i sica de Part\'\i  culas, Universidade
de Santiago de
Compostela\\and\\Instituto Galego de F\'\i sica de Altas
Enerx\'\i as (IGFAE)\\E-15782, Santiago de Compostela, Spain
}
\\
\vspace{0.2in}
${}^{\dagger}$  \it{ SISSA and INFN-Sezione di Trieste\\ Via Beirut 2; 
I-34014\\Trieste, Italy}

\vspace{0.2in}
\end{center}
\vspace{0.2in}
%\begin{center}
\centerline{{\bf Abstract}}
We find new supergravity solutions generated by D5-branes wrapping a four-cycle and preserving four and two supersymmetries. We first consider the configuration in which the fivebranes wrap a four-cycle in a Calabi-Yau threefold, which preserves four supersymmetries and is a gravity dual to the Coulomb branch of two-dimensional gauge theories with ${\cal N}=(2,2)$  supersymmetry. We also study the case of  fivebranes wrapping a co-associative four-cycle in a manifold of $G_2$-holonomy, which provides a gravity dual of ${\cal N}=(1,1)$ supersymmetric Yang-Mills theory in two dimensions.  We also discuss the addition of unquenched fundamental matter fields to these backgrounds and find the corresponding gravity solutions with  flavor brane sources. 

\smallskip
\end{titlepage}
\setcounter{footnote}{0}

\tableofcontents

\section{Introduction}

The construction of supergravity solutions with a reduced number of supersymmetries has been  one of the main research directions in the continuous effort to make the gauge/gravity correspondence \cite{jm, MAGOO} closer to more realistic theories of Nature. 

An approach that has been very fruitful in recent years is the analysis of supergravity solutions that correspond to branes wrapping supersymmetric cycles inside a non-compact manifold of special holonomy \cite{Maldacena:2000mw}. These solutions have various background fluxes turned on and provide us with gravity duals of supersymmetric Yang-Mills (SYM) theories living on the unwrapped part of the brane. As notable examples of these backgrounds, let us mention the one obtained in \cite{CV}, and interpreted in \cite{MN} as the gravity dual of ${\cal N}=1$ SYM theory in four dimensions. This background corresponds to  fivebranes wrapping a two-cycle. Similarly, in refs. \cite{Gauntlett:2001ps, Bigazzi:2001aj} the supergravity dual of ${\cal N}=2$ SYM in $d=4$ was found, also from fivebranes wrapping a two-cycle. Moreover, by wrapping a fivebrane in a three-cycle we can generate the supergravity dual of SYM in three space-time dimensions with different amounts  of supersymmetry. This program was carried out in refs. \cite{Chamseddine:2001hk, Maldacena:2001pb, Schvellinger:2001ib} for ${\cal N}=1$ 3d SYM theory, whereas the background dual to ${\cal N}=2$ SYM in $d=3$ was found in refs. \cite{Gomis:2001aa, Gauntlett:2001ur}. 

In this paper we will continue with this line of research by considering D5-branes wrapping four-cycles and preserving different amounts of supersymmetry, whose corresponding dual field theories are two-dimensional. We will consider first the case in which the branes wrap a four-cycle of a Calabi-Yau threefold and four supersymmetries are preserved. The supersymmetry of the corresponding two-dimensional dual field theory is ${\cal N}=(2,2)$. We will then analyze the configuration in which the special holonomy manifold has $G_2$ holonomy and the number of supersymmetries preserved is two, which leads to ${\cal N}=(1,1)$ supersymmetry in the dual field theory.  We will argue that these backgrounds are the gravitational duals of a slice of the Coulomb branch of the corresponding gauge theories. Moreover, we will also analyze the addition of unquenched flavor to both setups. 

When dealing with backgrounds generated by wrapped branes a useful tool is the use of 
an appropriate lower-dimensional gauged supergravity \cite{Maldacena:2000mw}, in which the brane worldvolume is a domain wall object of codimension one in the lower dimensional space-time. Thus, to obtain fivebrane solutions the appropriate gauged supergravity must be seven-dimensional. Moreover, in order to find supersymmetric solutions of wrapped branes in this approach one has to identify the spin connection along the wrapped cycle with some particular gauge fields of the gauged supergravity. This is an implementation of  the so-called ``topological twist", needed to realize supersymmetry in the D-brane worldvolume \cite{Bershadsky:1995qy}. In this paper we will use $SO(4)$ gauged supergravity \cite{salam, it1}, which turns out to be the one needed to  accommodate  the twistings required for our solutions.

Once the metric and gauge fields in seven dimensions are known, one can obtain the metric and RR three-form of the ten-dimensional background by using the corresponding uplifting formulae \cite{uplift}. In  general, the results obtained by this procedure for our systems are rather complicated and the corresponding expressions for the metric and three-form that we will get  involve coordinates that are  non-trivially fibered.  However, there is a change of variables, generalizing the one in \cite{DiVecchia:2002ks} for the gravity dual of ${\cal N}=2$, $d=4$ SYM,  which makes the results more transparent and neat. In this new system of coordinates the directions parallel and transverse to the special holonomy manifold are clearly distinguished. The price that one has to pay for this extra clarity is that all functions of the ansatz depend on two non-compact variables. Despite this fact, we will formulate our setup in terms of the ten-dimensional variables and we will impose the preservation of supersymmetry directly in ten-dimensions. This condition will lead us to a system of first-order BPS equations in partial derivatives for the functions entering our ansatz. After performing the corresponding uplifting and change of variables, the gauged supergravity approach provides a particular non-trivial analytic solution for this system of BPS equations. Other analytic solutions of the BPS system, not derived from gauged supergravity, will also  be  obtained. 

The addition of matter degrees of freedom in the fundamental representation of the gauge group is another generalization of the gauge/gravity correspondence of obvious interest. The (by now) standard method to add this new matter sector in the correspondence consists of the inclusion of flavor branes, which should extend along all  the gauge theory directions and wrap a non-compact cycle in the special holonomy manifold in order to make its worldvolume symmetry a global symmetry from the gauge theory point of view \cite{KK}. If the number of flavors, $N_f$, is small compared with the number of colors $N_c$, the flavor branes can be treated as probes in the background created by the color branes. On the contrary, when the number of flavors is of the order of the number of colors ($N_f\sim N_c$) one necessarily has to include the backreaction of the flavor branes on the geometry. In this case the flavor branes should be considered as dynamical sources of the different supergravity fields. 

In this paper we will try to add flavor to the two types of wrapped fivebrane backgrounds studied. We will include the backreaction following the approach of ref. \cite{Casero:2006pt}, in which the localized brane sources are substituted by a continuous distribution  (see also \cite{noncritical}).  This approach has been successfully  applied in several brane setups (\cite{Casero:2007pz}-\cite{Caceres:2009bk}). Here, we will be able to find a satisfactory implementation of this flavoring procedure for our background with four supersymmetries. However, in the case of the background dual to ${\cal N}=(1,1)$ supersymmetric field theory we will face new difficulties to determine the appropriate deformation introduced by the flavor branes. Nevertheless, we will be able to find the general structure of this deformation and we will find the corresponding backreacted background in terms of an unknown function satisfying a set of conditions.

The rest of this paper is organized as follows. In section \ref{22section} we will present our brane setup for the case in which four supersymmetries  are preserved and we will specify our ansatz for the ten-dimensional metric and RR three-form. We will obtain a system of BPS differential equations and we will study the solution obtained from seven-dimensional  gauged supergravity. The steps followed to find this solution will be detailed in appendix \ref{gaugedsugra7d}. We will end section \ref{22section}  by  presenting the flavored version of the dual to the ${\cal N}= (2,2)$ theory. In section \ref{11-10d} we analyze the background preserving two supersymmetries. After an initial motivation, we will formulate our ansatz and we will  find the corresponding system of BPS equations. Again, gauged supergravity provides a non-trivial solution of these equations, which will be explored from the ten-dimensional point of view in section \ref{11-10d}, leaving the details of the gauged supergravity analysis to appendix \ref{gaugedsugra7d}. We will finish section \ref{11-10d} by presenting  our approach to add flavor to the ${\cal N}=(1,1)$ gravity dual. Section \ref{conclusions} contains our conclusions and summarizes our main results.

The paper  ends with three appendices containing technical details which can be skipped in a first reading. In appendix \ref{gaugedsugra7d} the gauged supergravity approach is presented, while in appendix \ref{additional} we study additional solutions of the ten-dimensional unflavored BPS systems. Finally, in appendix \ref{eoms} we check that the second order equations of motion for the gravity plus brane sources systems are satisfied by any solution of the  first-order BPS equations.

\section{The  dual of the ${\cal N}= (2,2)$ theory}
\label{22section}

The first case we will study corresponds to a background of type IIB supergravity  generated by a stack of $N_c$ D5-branes wrapping a four-cycle ${\cal C}_4$ of a  Calabi-Yau (CY) cone of complex dimension three. To simplify matters we will restrict to the case in which  ${\cal C}_4$ is the product ${\cal C}_2 \times {\cal C}_2 $ of two two-cycles  ${\cal C}_2$. The corresponding brane array is:
\begin{center}
\begin{tabular}{|c|c|c|c|c|c|c|c|c|c|c|}
\multicolumn{3}{c}{ }&
\multicolumn{6}{c}
{$\overbrace{\phantom{\qquad\qquad\qquad\qquad\qquad}}^{\text{CY}_3}$}\\
\hline
&\multicolumn{2}{|c|}{$\mathbb{R}^{1,1}$}
&\multicolumn{2}{|c|}{$S^2$}
&\multicolumn{2}{|c|}{$S^2$}
&\multicolumn{2}{|c|}{$N_2$}
&\multicolumn{2}{|c|}{$\mathbb{R}^{2}$}\\
\hline
$N_c$ D$5$ &$-$&$-$&$\bigcirc$&$\bigcirc$&$\bigcirc$&$\bigcirc$
&$\cdot$&$\cdot$&$\cdot$&$\cdot$\\
\hline
\end{tabular}
\end{center}
where the $S^2$'s represent the directions of the two-cycles ${\cal C}_2$ and $N_2$ are the directions of the normal bundle to ${\cal C}_4$. Notice that the symbols ``$-$" and 
``$\cdot$" represent unwrapped worldvolume directions and directions transverse to the brane respectively, whereas a circle denotes a wrapped worldvolume direction.

 In order to write a concrete ansatz for the metric of this array, let us parametrize the two two-cycles by two angles $(\theta_i, \phi_i)$, with $0\le \theta_i\le \pi$ and $0\le \phi_i<2\pi$ ($i=1,2$), and let $\sigma$ represent the radial coordinate of the CY cone. We will also parametrize the transverse $\mathbb{R}^{2}$ space by another radial coordinate $\rho$, as well as by another angle $\chi$ ($0\le \chi <2\pi$). Given these coordinates, the ansatz we will adopt for the string frame metric is:
\bear
&&ds^2\,=\,e^{\Phi}\,\,\Big[\,dx_{1,1}^2\,+\,{z\over m^2}\,\,
\Big(\,d\theta_1^2\,+\,\sin^2\theta_1\,d\phi_1^2\,+\,
d\theta_2^2\,+\,\sin^2\theta_2\,d\phi_2^2\,\Big)\,\Big]\,+\,
\rc\rc
&&\,\,\,\,\,+\,
{e^{-\Phi}\over m^2 z^2}\,\Big[\,d\sigma^2\,+\,\sigma^2\,
\Big(\,d\psi+\cos\theta_1d\phi_1\,+\,\cos\theta_2d\phi_2\,\Big)^2\,\Big]+
{e^{-\Phi}\over m^2}\,\Big (\,d\rho^2\,+\,\rho^2\,d\chi^2\,\Big)\,\,,
\qquad\qquad
\label{D5-newmetric}
\eear
where $m$ is a constant with units of mass which, for convenience, we will take as:
\beq
m^2\,=\,{1\over g_s \alpha' N_c}\,\,,
\label{m}
\eeq
with $g_s$ and $\alpha'$ being respectively the string coupling constant and the Regge slope of superstring theory.  Notice that $\psi$  ($0\le \psi< 2\pi$) is an angular coordinate along $N_2$, which is fibered over ${\cal C}_2\times {\cal C}_2$ as in the metric of the conifold. In (\ref{D5-newmetric}) $dx_{1,1}^2$ is the Minkowski metric in 1+1 dimensions, $\Phi$ is the dilaton of the type IIB theory and $z$ is a function that controls the size of the cycle. The
dilaton $\Phi$ and the function $z$ should be considered as functions of the two radial variables $(\rho, \sigma)$:
\beq
\Phi\,=\,\Phi(\rho,\sigma)\,\,,\qquad\qquad
z\,=\,z(\rho,\sigma)\,\,.
\eeq
We shall adopt an  ansatz for the RR three-form $F_3$ in which it is
represented in terms of a two-form potential $C_2$ as:
\beq
F_3=dC_2\,\,,
\eeq
where $C_2$  will  be taken as:
\beq
C_2\,=\,g\,d\chi\wedge (\,d\psi+\cos\theta_1 d\phi_1+\cos\theta_2d\phi_2\,)\,\,,
\label{C2-unflavored22}
\eeq
with $g$ being also a function of the variables $(\rho, \sigma)$. The RR field strength 
corresponding to the potential (\ref{C2-unflavored22}) is given by:
\beq
F_3\,=\,g\,d\chi\wedge(\,
\sin\theta_1 d\theta_1\wedge d\phi_1\,+\,\sin\theta_2d\theta_2\wedge d\phi_2\,)\,+\,
dg\wedge d\chi\wedge(\,d\psi+\cos\theta_1 d\phi_1+\cos\theta_2d\phi_2\,)\,\,.
\label{F3ansatz}
\eeq
We will determine the functions $z(\rho, \sigma)$, $\Phi(\rho,\sigma)$ and $g(\rho, \sigma)$ of our ansatz by requiring that the background preserves four supersymmetries. 
This condition can be fulfilled by imposing the vanishing of the supersymmetric variations of the dilatino $\lambda$ and gravitino $ \psi_{M}$ of type IIB supergravity which, for the type of background we intend to study here, are given by:
\bear
&&\delta \lambda\,=\,{1\over 2}\,
\Big[\,\Gamma^{M}\,\partial_{M}\,\Phi\,-\,
{e^{\Phi}\over 12}\,\,F_{M_1M_2M_3}\,\Gamma^{M_1M_2M_3}\,
\tau_1\,\Big]\,\epsilon\,\,,\rc\rc
&&\delta\psi_{M}\,=\,\Big[\,D_{M}\,+\,{e^{\Phi}\over 48}\,\,
F_{M_1M_2M_3}\,\Gamma^{M_1M_2M_3}\,
\Gamma_{M}\,\tau_1\,\Big]\,\epsilon\,\,,
\label{Susy-variations}
\eear
where $\epsilon$ is a doublet of Majorana-Weyl spinors of fixed ten-dimensional chirality and $\tau_1$ is the first Pauli matrix (which acts on the doublet $\epsilon$). It turns out that the supersymmetry preserving conditions $\delta\lambda=\delta\psi_M=0$ can be solved for the Killing spinors $\epsilon$ if we impose on them a certain set of projections.  In order to specify these projections, let us choose the following vielbein basis for the metric (\ref{D5-newmetric}):
\bear
&&e^{0,1}\,=\,e^{{\Phi\over 2}}\,dx^{0,1}\,\,,\qquad
e^{2}\,=\,{e^{{\Phi\over 2}}\over m}\,\sqrt{z}\,d\theta_1\,\,,\qquad\qquad
e^{3}\,=\,{e^{{\Phi\over 2}}\over m}\,\sqrt{z}\,\sin\theta_1\,d\phi_1\,\,,\rc\rc
&&e^{4}\,=\,{e^{{\Phi\over 2}}\over m}\,\sqrt{z}\,d\theta_2\,\,,\qquad
e^{5}\,=\,{e^{{\Phi\over 2}}\over m}\,\sqrt{z}\,\sin\theta_2\,d\phi_2\,\,,\qquad
e^{6}\,=\,{e^{-{\Phi\over 2}}\over m z}\,d\sigma\,\,,\rc\rc
&&e^{7}\,=\,{e^{-{\Phi\over 2}}\sigma\over m z}\,\Big(\,
d\psi+\cos\theta_1 d\phi_1+\cos\theta_2d\phi_2\,\Big)\,\,,\rc\rc
&&e^{8}\,=\,{e^{-{\Phi\over 2}}\over m }\,d\rho\,\,,\qquad\qquad
e^{9}\,=\,{e^{-{\Phi\over 2}}\rho\over m }\,d\chi\,\,.
\label{frame}
\eear
Let us now impose to $\epsilon$ that:
\beq
\Gamma_{2345}\,\epsilon\,=\,-\epsilon\,\,,\qquad\qquad
\Gamma_{4567}\,\epsilon\,=\,\epsilon\,\,,\qquad\qquad
\Gamma_{6789}\,\tau_1\,\epsilon\,=\,\epsilon\,\,,
\label{10d-projections}
\eeq
where $\Gamma_{ a_1 a_2\cdots}$ are antisymmetrized products of constant Dirac matrices in the frame (\ref{frame}). Then, one can show that the Killing spinors of the system are of the form:
\beq
\epsilon\,=\,e^{{\Phi\over 4}}\,e^{-{\psi\over 2}\,\Gamma_{23}\,+\,
{\chi\over 2}\,\Gamma_{89}}\,\,\eta\,\,,
\label{10d-spinor}
\eeq
where $\eta$ are constant spinors satisfying the same set of  projections as in (\ref{10d-projections}). Moreover, the three unknown functions $z$, $\Phi$ and $g$ must satisfy a set of first-order differential equations. If the prime (dot)  denotes the partial derivative with respect to $\rho$ ($\sigma$), this system of equations is:
\bear
&&m^2\,g\,=\,\rho\,z'\,\,,\rc
&&e^{2\Phi}\,=\,{\sigma\over z^2\dot z}\,\,,\rc
&&m^2\,g'\,=\,2\,e^{-2\Phi}\,\rho\,\sigma\,\dot\Phi\,\,,\rc
&&m^2\,\dot g\,=\,-2m^2\,{\sigma\over z^3}\,e^{-2\Phi}\,g\,-\,
2\,{\rho\sigma\over z^2}\,e^{-2\Phi}\,\Phi'
\label{BPS}\,\,.
\eear
Notice the similarity to the equations found in refs. \cite{Angel, Arean:2008az, Ramallo:2008ew}. As in these other cases, the four equations in (\ref{BPS}) are not independent. Indeed, one can easily verify that the last equation can be derived from the other equations of the system. One can also check that, if the system (\ref{BPS}) holds, the second order equations of motion of type IIB supergravity for our ansatz are also satisfied (see appendix \ref{eoms}).  Moreover, 
the system  (\ref{BPS}) can be reduced to the following PDE for the function $z(\rho, \sigma)$:
\beq
\rho\,z^2\,(\dot z\,-\,\sigma\,\ddot z\,)\,=\,\sigma\,(\,
2\rho\,z\,\dot z^2\,+\,z'\,+\,\rho\,z''\,)\,\,.
\label{PDE-22}
\eeq
Notice that, if $z(\rho, \sigma)$ is  known, the other function $g$ of the ansatz, as well as the dilaton $\Phi$, can be obtained from the  first two equations of the BPS system (\ref{BPS}).  Moreover, since the Killing spinors $\epsilon$ satisfy the three conditions (\ref{10d-projections}), our background preserves four supersymmetries and, in fact,  one can easily verify\footnote{The simplest way of deriving this result is by changing to a spinor basis in which the Pauli matrix $\tau^1$ of the last projection in (\ref{10d-projections}) acts diagonally and by using the first projection in (\ref{10d-projections}) and  the fact that the total ten-dimensional chirality is fixed.}
 that there are two supercharges of each two-dimensional chirality, 
as it should for the case of an ${\cal N}=(2,2)$ gauge theory in two dimensions. Actually, one can recognize the first two projections in (\ref{10d-projections}) as the ones required to preserve the K\"ahler structure of the underlying $CY_3$ manifold, while the projection involving the Pauli matrix $\tau_1$  is the one associated to the color D5-branes.

There is an alternative way to obtain the BPS system that makes manifest its geometric nature. This method uses the so-called (generalized) calibration form which, in our case, is a six-form 
${\cal K}$ obtained from fermionic bilinears. Let us represent it in terms of the  frame basis as: 
\beq
{\cal K}\,=\,{1\over 6!}\,{\cal K}_{a_1\cdots a_6}\,e^{a_1\,\cdots a_6}\,\,,
\label{K-def}
\eeq
where  $e^{a_1\,\cdots a_6}=e^{a_1}\wedge e^{a_2}\wedge\cdots \wedge e^{a_6}$. 
The different components of ${\cal K}$  are given by:
\beq
{\cal K}_{a_1\cdots a_6}\,\equiv\,-e^{-{\Phi\over 2}}\,\,\epsilon^{\dagger}\,\tau_1\,\Gamma_{a_1\,\cdots a_6}\,\epsilon\,\,,
\label{K-bilinear}
\eeq
where $\epsilon$ is a Killing spinor of the background, normalized as 
$e^{-{\Phi\over 2}}\,\,\epsilon^{\dagger}\,\epsilon\,=\,1$, and the minus sign in the definition (\ref{K-bilinear}) has been introduced for convenience.  By using the SUSY projections satisfied by our solutions (eq. (\ref{10d-projections})), we get the actual components of 
${\cal K}$  in the frame basis (\ref{frame}), namely:
\beq
{\cal K}\,=\,e^{01}\,\wedge\,\big(\,e^{2345}\,-\,e^{2367}\,-\,e^{4567}\,\big)\,\,.
\label{calK-explicit}
\eeq
The K\"ahler form $J$ of the internal manifold  can be simply written as:
\beq
J\,=\,e^{23}\,+\,e^{45}\,-\,e^{67}\,\,.
\eeq
In terms of $J$, the calibration form ${\cal K}$  in (\ref{calK-explicit}) can be written as:
\beq
{\cal K}\,=\,{\rm Vol}({\rm Min}_{1,1})\,\wedge\, J\wedge J\,\,,
\eeq
where ${\rm Vol}({\rm Min}_{1,1})=e^{\Phi}\,dx^0\wedge dx^1$ is the volume form of the Minkowski part of the metric.  The calibration conditions\footnote{Actually, only the condition involving ${}^*\,F_3$ in (\ref{calibration-conditions}) is a (generalized) calibration condition. In spite of this, in an abuse of language, we will continue referring to the two equations in (\ref{calibration-conditions}) in this way.} are:
\beq
{}^*\,F_3\,=\,-d\big(\,e^{-\Phi}\,{\cal K}\,\big)\,\,,\qquad\qquad
d\big({}^*\,{\cal K})\,=\,0\,\,.
\label{calibration-conditions}
\eeq
By computing the exterior derivatives  in  (\ref{calibration-conditions}) one can check, component by component, that the resulting 
equations coincide with those of the BPS system  (\ref{BPS}). Thus, (\ref{calibration-conditions}) is equivalent to the supersymmetry preserving conditions $\delta\lambda=\delta\psi_M=0$.

\subsection{Integration of the BPS system}
\label{22sol}

Let us now obtain a solution of the BPS system (\ref{BPS}). First of all we notice that, when $\sigma=0$ and $\rho$ varies, $g'$ is zero (see the third equation in (\ref{BPS})). Thus, we can take $g(\rho, \sigma=0)=g_0$, where $g_0$ is a constant. Actually, by using a flux quantization condition we can  fix this constant to be:
\beq
m^2\, g(\rho,\sigma=0)\,=\,1\,\,.
\label{g-sigma0}
\eeq
By using this value of $g$ at $\sigma=0$ one can easily integrate $z(\rho,\sigma=0)$ from the first equation in (\ref{BPS}), namely: 
\beq
z(\rho,\sigma=0)\,=\,\log \rho\,\,,
\label{z-sigma0}
\eeq
where we have fixed the integration constant by requiring that $z(\rho,\sigma=0)$ vanishes when the (dimensionless) variable $\rho$ is equal to one.  To extend this solution to other values of $\sigma$, and to get the other functions of our ansatz, it is useful to look at the realization of our brane setup in seven dimensional gauged supergravity \cite{Cvetic:1999xp}. In this setup, the particular fibering of the $\psi$ coordinate in the metric (\ref{D5-newmetric}) comes up very naturally when the solution is uplifted to ten dimensions, and a solution of type IIB supergravity with metric and three-form as in our ansatz is generated (see appendix \ref{gaugedsugra7d} for a detailed account).  Actually, the functions of the gauge supergravity ansatz depend only on one radial variable and the corresponding BPS equations can be integrated in analytic form. After a suitable change of variables this solution provides a solution of the PDE equations (\ref{BPS}) and (\ref{PDE-22}).  Let us present this solution here, leaving the details for appendix \ref{gaugedsugra7d}. First of all, we define the function $x(z)$ as:
\beq
e^{2x}\,=\,{1-2z+2z^2+ce^{-2z}\over 2 z^2}\,\,,
\label{e2x}
\eeq
where $c$ is a constant. Then, $z(\rho,\sigma)$ is given in implicit form as:
\beq
\sigma^2\,-\,z^2\,e^{2x}\,(\,e^{2z}\,-\,\rho^2\,)\,=\,0\,\,,
\label{implict-22}
\eeq
while $g(\rho,\sigma)$ and the dilaton $\Phi(\rho,\sigma)$  are:
\beq
m^2\,g\,=\,{ \rho^2 \over  \rho^2 +e^{-2x}(e^{2z}-\rho^2)
}\,\,,
\label{g-gaugedsugra}
\qquad\qquad
e^{2\Phi}\,=\,\rho^2\,(\,e^{2x}\,-\,1)\,+e^{2z}\,\,.
\eeq
From these expressions it is easy to prove that $z$, $g$ and $\Phi$ do indeed satisfy the differential equations in  (\ref{BPS}) and (\ref{PDE-22}). 

 The detailed analysis of this solution depends on the value of the integration constant $c$ in (\ref{e2x}). In general, the metric is singular. However, it is argued in appendix \ref{gaugedsugra7d} that only for $c<-1$ the singularity is ``good" in the sense of ref. \cite{Maldacena:2000mw}. For this reason we will restrict ourselves to analyzing this case. One can straightforwardly verify that, when $c<-1$, the function $e^{2x(z)}$ defined in (\ref{e2x}) has a zero for some positive value of the variable $z$, \ie\ that there exists a $z_0$ such that:
\beq
e^{2x(z_0)}\,=\,0\,\,.
\eeq
Let us denote by $\rho_m$ the value of $\rho$ obtained by taking $z=z_0$ in (\ref{z-sigma0}), namely:
\beq
\rho_m\equiv e^{z_0}\,\,.
\eeq
It is now clear that the implicit relation (\ref{implict-22}) can be solved for $\sigma=0$  as:
\beq
z(\rho,\sigma=0)\,=\,
\begin{cases}
\log\rho\,\,, &{\rm if} \,\,\,\,\rho\ge \rho_m\,\,,\cr\cr
z_0\,\,,&{\rm if} \,\,\,\,0\le \rho\le \rho_m\,\,.
\end{cases}
\label{z-total}
\eeq
Thus, the expression (\ref{z-sigma0}) for $z(\rho,\sigma=0)$ is only valid for $\rho\ge \rho_m$.  A clue to understand this result is obtained by looking at the behavior of the function $g(\rho, \sigma=0)$ near $\rho=\rho_m$ in (\ref{g-gaugedsugra}). For $\rho>\rho_m$ the value (\ref{g-sigma0}) is reproduced, while $g(\rho, \sigma=0)$  vanishes for 
$\rho<\rho_m$. This discontinuous change of $g$ at $\rho=\rho_m$, $\sigma=0$ seems to indicate that this is the location of the D5-branes. A confirmation of this fact can be obtained by studying the form of the dilaton. From (\ref{g-gaugedsugra}) and (\ref{z-total}) one gets:
\begin{figure}
\centering
\includegraphics[width=0.35\textwidth]{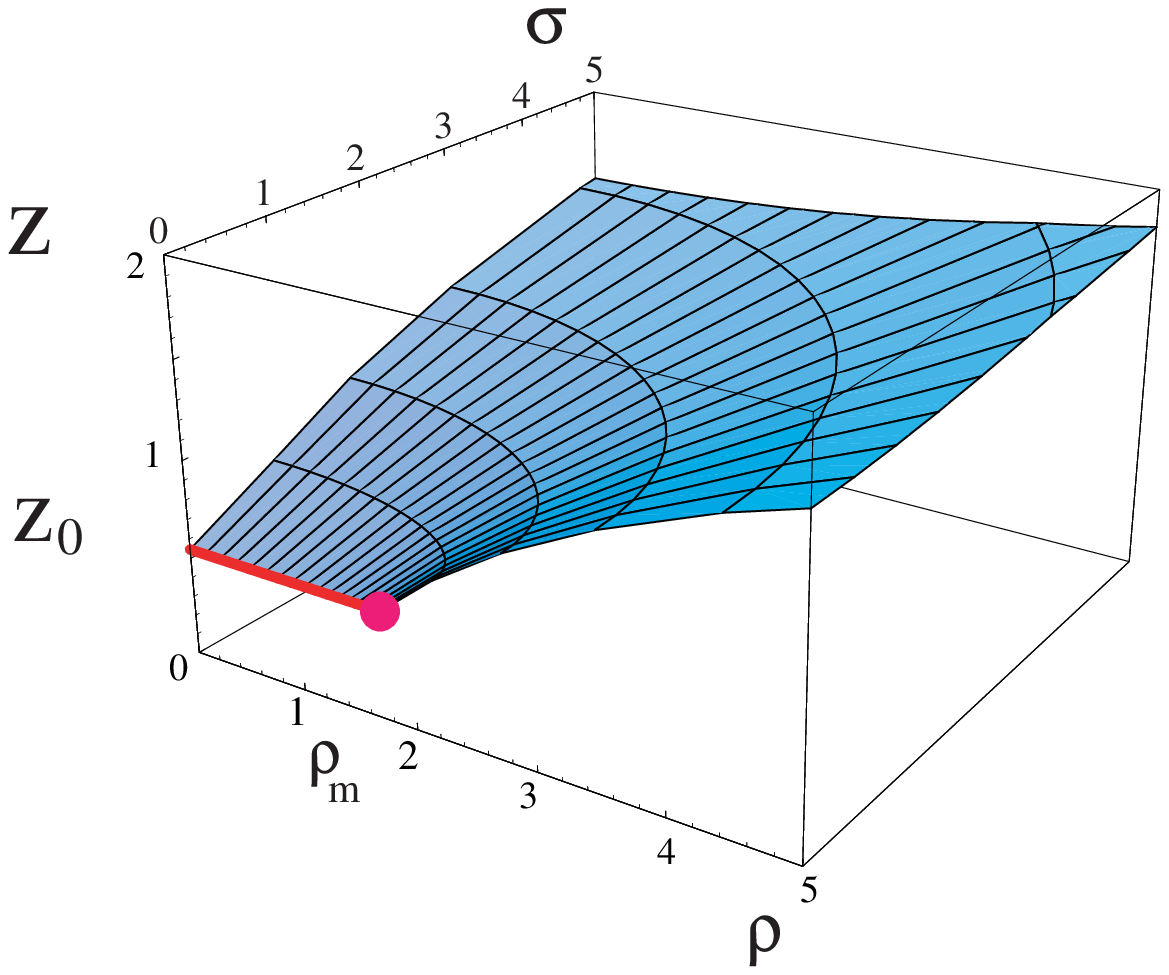} \hfill
\includegraphics[width=0.4\textwidth]{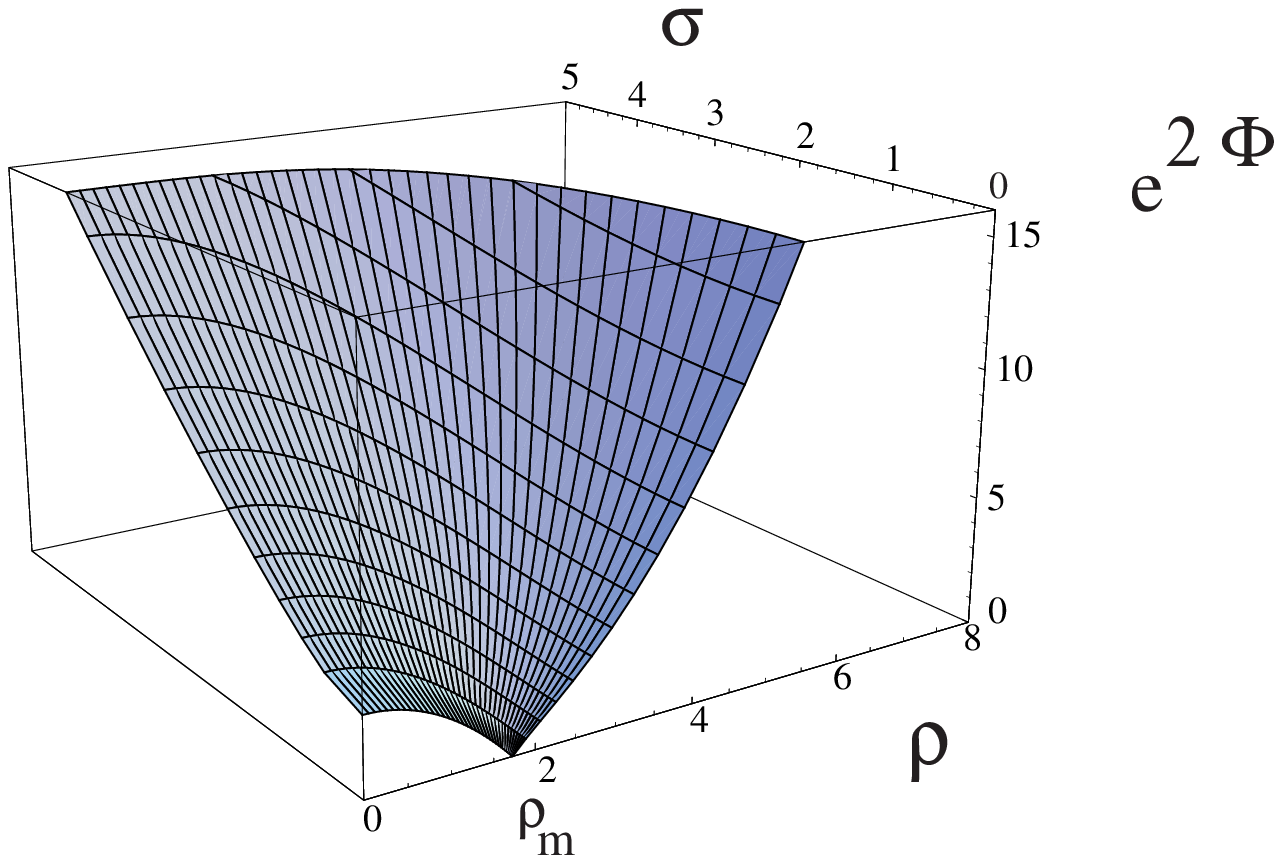}
\caption{On the left we plot $z(\rho,\sigma)$ as given by the implicit equation (\ref{implict-22}). The line in this plot is the segment $z=z_0$ at $\sigma=0$ and $0\le \rho\le\rho_m$. On the right we plot $e^{2\Phi}$ from (\ref{g-gaugedsugra}). In both cases we are taking $c=-1.5$. }
\label{z-phi-22}
\end{figure}

\beq
e^{2\Phi(\rho,\sigma=0)}\,=\,
\begin{cases}
\rho^2\,
\Big[\,1\,-\,{1\over \log\rho}\,+\,{1\over 2(\log\rho)^2}\,\Big]\,+\,{c\over 2 (\log\rho)^2}\,\,,
 &{\rm if} \,\,\,\,\rho\ge \rho_m\,\,,
\cr\cr
e^{2z_0}-\rho^2\,\,,
&{\rm if} \,\,\,\,0\le \rho\le \rho_m\,\,,
\end{cases}
\label{dilaton-total}
\eeq
which, in particular means that $e^{\Phi}$ vanishes for $\sigma=0, \rho=\rho_m$, making the dilaton singular at that point. All these features of our solution are displayed in figure \ref{z-phi-22}, which in particular shows the constant $z=z_0$ segment along the $\sigma=0$ axis.  Taken together, all these results mean that our ${\cal N}=(2,2)$ solution can be interpreted as generated by a distribution of D5-branes smeared along the ring $\rho=\rho_m$, $0\le\chi<2\pi$ and located at the $\sigma=0$ point of the $CY_3$.  Accordingly, our solution can be regarded as the supergravity dual of a slice of the Coulomb branch of ${\cal N}=(2,2)$ two-dimensional SYM.  We will confirm this interpretation in appendix \ref{gaugedsugra7d} by looking at the UV and the IR behavior of our solution. Moreover, in the next subsection we will verify that $\sigma=0$ defines the SUSY locus of a D5-brane probe in our geometry. 

\subsection{Probe analysis}
\label{probe-22}
Let us now study the dynamics of a color D5-brane probe in our background. The action of such a probe is determined by the DBI+WZ action, given by:
\beq
 S\,=\,-T_5\,\int d^6\zeta\,e^{-\Phi}\,\,\sqrt{-\det (\,\hat G_6\,+\,2\pi\alpha'\,F)}\,+\,
 T_5\,\int \hat C_6\,\,,
 \label{actionD5}
\eeq
where $\zeta^{a}$ ($a=0,\cdots, 5$) is a set of worldvolume coordinates, $F$ is the field strength of the worldvolume gauge field,  $\hat G_6$ is the induced metric and a hat over $C_6$ denotes its pullback to the worldvolume. The D5-brane probe we will analyze is extended along the Minkowski directions, wraps the $S^2\times S^2$ four-cycle and is located at a fixed value of the remaining coordinates. Accordingly, we shall choose our
worldvolume coordinates as $\zeta^{a}\,=\,(x^0,x^1, \theta_1, \phi_1, \theta_2, \phi_2)$  and embed our brane at a constant value of $(\sigma, \psi, \rho,\chi)$. We will consider first the case in which the worldvolume gauge field vanishes. The induced metric for such a configuration is:
\bear
&&\hat G_{ab}^{(6)}\,d\zeta^a\,d\zeta^b\,=\,
e^{\Phi}\,\,\Big[\,dx_{1,1}^2\,+\,{z\over m^2}\,\big(\,
d\theta_1^2\,+\,\sin^2\theta_1\,d\phi_1^2\,+\,d\theta_2^2\,+\,\sin^2\theta_2\,d\phi_2^2
\,\big)\,\Big]\,+\,\rc\rc
&&\qquad\qquad\qquad\qquad\qquad
+\,
{e^{-\Phi}\,\sigma^2\over m^2 z^2}\,\big[\,
\cos\theta_1\,d\phi_1\,+\,\cos\theta_2\,d\phi_2\,\big]^2\,\,.
\eear
The RR six-form potential is related to the Hodge dual of $F_3$ as $F_7=-{}^*F_3\,=\,dC_6$. From the first calibration condition in (\ref{calibration-conditions}) one concludes that $C_6$ can be chosen as:
\beq
C_6\,=\,e^{-\Phi}\,{\cal K}\,\,.
\label{C6-K}
\eeq
By using the explicit expression of the calibration form ${\cal K}$ given  in (\ref{calK-explicit}) one immediately obtains the pullback of $C_6$, namely:
\beq
\hat C_6\,=\,{e^{2\Phi} z^2\over m^4}\,\sin\theta_1\,\sin\theta_2\,d^6\zeta\,\,.
\eeq
Plugging these results into the DBI+WZ action (\ref{actionD5}) one gets the value of the action for the static configuration we are considering:
\beq
S\,=\,-T_5\,\int\,d^6\zeta\,{z^2\,e^{2\Phi}\over m^4}\,\,
\sin\theta_1\,\sin\theta_2\,\Big[\,
\sqrt{1+\,{\sigma^2 e^{-2\Phi}\over z^3}\,\big(\,\cot^2\theta_1\,+\,\cot^2\theta_2)}\,-\,1
\,\Big]\,\,,
\label{potential-22}
\eeq
which is nothing but minus the static potential between the stack of $N_c$ color branes that created the background and the additional probe. It is clear from (\ref{potential-22}) that this potential is non-vanishing except when $\sigma=0$, which confirms that this point is a zero-force supersymmetric locus inside the Calabi-Yau space.  This fact can also be verified directly by means of kappa symmetry. 

Let us now assume that our probe is located at $\sigma=0$ and that we switch on a worldvolume gauge field $F_{\mu\nu}$ whose only non-vanishing components are those along the Minkowski directions $x^{\mu}$. Recall that our branes can be at any point of the transverse $\mathbb{R}^2$ space. We will parametrize these flat directions in terms of a complex coordinate ${\cal Z}$, related to $(\rho,\chi)$ as follows:
\beq
{\cal Z}\,=\,{1\over m}\,\rho\,e^{i\chi}\,\,,
\eeq
where we have included the $1/m$ factor to absorb the one we introduced in (\ref{D5-newmetric}) with our notation (notice that  ${\cal Z}$ has dimensions of length). We will assume that ${\cal Z}$ only depends on $(x^0, x^1)$. By expanding the DBI action up to quadratic terms, we get:
\beq
\int_{S^2\times S^2}{\cal L}_{DBI}\Big|_{quadratic}=
-{2(2\pi)^4\,(\alpha')^2\,T_5\over  m^4}\,z^2
\Bigg[\,{1\over 2}\,{\rm Tr}\,\Big[F_{\mu\nu}\,F^{\mu\nu}\Big]+
{1\over (2\pi\alpha')^2}\,{\rm Tr}\,\big[\partial_{\mu}\,{\cal Z}^{\dagger}\,
\partial^{\mu}\,{\cal Z}\,\big]\,\Bigg]\,\,,\qquad\qquad\qquad
\label{quadratic-action}
\eeq
where $z$ is evaluated at $\sigma=0$ and, in order to pass from the abelian to the non-abelian theory, we have substituted $F_{\mu\nu}\,F^{\mu\nu}\,\to\,{\rm Tr}\,\big[F_{\mu\nu}\,F^{\mu\nu}\big]$ (and similarly for ${\cal Z}$). Let us now define the complex scalar field $\Psi$ as follows:
\beq
\Psi\,=\,{{\cal Z}\over 2\pi\alpha'}\,\,.
\eeq
Then, the action (\ref{quadratic-action}) can be written as:
\beq
-{1\over 2 g^{2}_{YM}}\,{\rm Tr}\,\Big[F_{\mu\nu}\,F^{\mu\nu}\Big]\,-\,
{1\over g^{2}_{YM}}\,{\rm Tr}\,\big[\partial_{\mu}\,{\Psi}^{\dagger}\,
\partial^{\mu}\,{\Psi}\,\big]\,\,.
\eeq
where the gauge fields have the canonical normalization and the Yang-Mills coupling takes the form:
\beq
{1\over g^2_{YM}}\,=\,{g_s\alpha' N_c^2\over \pi}\,\,z^2(\rho, \sigma=0)\,\,.
\label{YM-22}
\eeq
From the explicit value of $z(\rho, \sigma=0)$ in (\ref{z-total}) we conclude that $g_{YM}$  decreases as we move towards the UV region (large $\rho$), as expected in a non-abelian Yang-Mills theory. However, we cannot extract more specific information from (\ref{YM-22}) due to the fact that we do not know the  precise holographic radius-energy relation. Moreover, one  also  has  to take  into account the growing of the dilaton as $\rho\to\infty$ (see (\ref{dilaton-total})), which invalidates the supergravity approximation. This is actually a generic  problem of all backgrounds generated by D5-branes (see section \ref{conclusions} for an alternative brane setup).

\subsection{The  dual of the (2,2) theory with flavor}

Let us now try to extend the results of the previous subsection to the construction of a supergravity background which could encode the effects of adding flavor degrees of freedom, \ie\ of fields in the fundamental representation of the gauge group. In order to perform this task we should add new  (flavor) branes that introduce a new open string sector to the theory on the gravity side. These flavor branes are D5-branes that fill the Minkowski space-time and in addition are extended along a non-compact four-cycle of the Calabi-Yau threefold and sit at a fixed point in the transverse $\mathbb{R}^{2}$. Actually, the value $\rho_Q$ of the radial $\mathbb{R}^{2}$ coordinate $\rho$ at which the flavor branes sit  in  $\mathbb{R}^{2}$ parametrizes the mass $m_Q$ of the matter fields we are adding ($m_Q\sim \rho_Q/\sqrt{\alpha'}$).

In order to characterize the four-dimensional submanifolds of the Calabi-Yau threefold that could be suitable to wrap a flavor brane, one should determine the embeddings of the D5-brane that preserve the four supersymmetries of the unflavored background. This analysis can be done systematically by using kappa symmetry and it is very similar to the one performed in \cite{Arean:2004mm} for the embeddings of D7-branes in the conifold Klebanov-Witten  model (see also refs. \cite{Canoura:2005uz,Canoura:2006es}). Let us skip the details of this analysis and just give the final result. Let us suppose that we have a D5-brane extended in $x^0, x^1$ as well as  along a four-dimensional submanifold ${\cal M}_4$  of the $CY_3$. It is quite convenient to define three complex variables $\zeta_i$ ($i=1,2,3$) as:
\beq
\zeta_1\,=\,\tan\Big({\theta_1\over 2}\Big)\,e^{i\phi_1}\,\,,
\qquad
\zeta_2\,=\,\tan\Big({\theta_2\over 2}\Big)\,e^{i\phi_2}\,\,,
\qquad
\zeta_3\,=\,\sigma\,\sin\theta_1\,\sin\theta_2\,e^{-i\psi}\,\,.
\label{zetas-D5}
\eeq
Then, the ${\cal M}_4$'s that correspond to a supersymmetric embedding of a D5-brane are those in which $\zeta_1$, $\zeta_2$ and $\zeta_3$ are constrained by a holomorphic relation. Actually, we will restrict ourselves to the case in which this relation is polynomial and  ${\cal M}_4$ is the geometric locus of the equation:
\beq
\zeta_1^{m_1}\,\zeta_2^{m_2}\,\zeta_3^{m_3}\,=\,{\rm constant}\,\,,\qquad\qquad
\label{polynomial}
\eeq
with $m_1$, $m_2$ and $m_3$ being some constant exponents. When the number of flavor branes is much lower than the number $N_c$ of color branes, one can neglect the backreaction of the former in the geometry and treat them as probes. On the field theory side this approximation corresponds to neglecting quark loops in the 't Hooft large $N_c$ expansion, \ie\ to the so-called quenched approximation. In this paper we will concentrate on studying the opposite (unquenched) limit, in which the number $N_f$ of flavor branes is of the same order as the number $N_c$ of color branes. This limit is the gravity counterpart of the one considered by Veneziano \cite{Veneziano:1976wm} in gauge theories.

When $N_f\sim N_c$ the backreaction cannot be ignored and one has to consider the full coupled gravity plus branes system. Finding the general solution of the equations of motion of this coupled system  for any embedding of the family (\ref{polynomial}) is, of course, a formidable task. For this reason we will try to find a particular class of embeddings for which  the problem becomes simpler and tractable.   Notice that the radial coordinate $\sigma$ and the fibered angle $\psi$ only enter in (\ref{polynomial}) through $\zeta_3$. Thus, if $m_3=0$, then $\sigma$ and $\psi$ are not constrained by (\ref{polynomial}) and they can take all possible values which, in particular, means that the cycle is non-compact as desired.  If, in addition, $m_1$ or $m_2$ also vanish, the description of ${\cal M}_4$ in terms of the other coordinates $(\theta_i, \phi_i)$ greatly simplifies. Indeed, if $m_1=m_3=0$,  eq. (\ref{polynomial}) tells us that 
$(\theta_2, \phi_2)$   are constant and $(\theta_1, \phi_1)$  are unconstrained.   Thus, in this case, if $\xi^{\alpha}$ are a set of worldvolume coordinates, the embedding is characterized by:
\beq
\xi^{\alpha}\,=\,(x^0, x^1, \theta_1, \phi_1, \sigma, \psi)\,\,,
\qquad\qquad
(\theta_2, \phi_2, \rho,\chi\,\,{\rm constant})\,\,.
\label{branch-one}
\eeq
Similarly, when $m_2=m_3=0$ the embedding is:
\beq
\xi^{\alpha}\,=\,(x^0, x^1, \theta_2,\phi_2, \sigma, \psi)\,\,,
\qquad\qquad
(\theta_1, \phi_1, \rho,\chi\,\,{\rm constant})\,\,.
\label{branch-two}
\eeq

It is easy to check directly that superposing any number of flavor branes extended as in (\ref{branch-one}) and (\ref{branch-two}) does not break any of the supersymmetries preserved  by the unflavored system. Indeed, using the fact that the total ten-dimensional chirality of the type IIB theory is fixed:
\beq
\Gamma^{01\cdots 9}\,\epsilon\,=\,-\Gamma_{01\cdots 9}\,\epsilon\,=\,\epsilon\,\,,
\label{10d-chirality}
\eeq
one can straightforwardly demonstrate that the projections (\ref{10d-projections}) imply:
\beq
\Gamma_{012367}\tau_1\,\epsilon\,=\,\epsilon\,\,,
\qquad\qquad\qquad
\Gamma_{014567}\tau_1\,\epsilon\,=\,\epsilon\,\,,
\label{projection-flavor}
\eeq
which are just the conditions required by kappa symmetry to ensure that the embeddings 
(\ref{branch-one}) and (\ref{branch-two}) are supersymmetric\footnote{
Similarly, after using (\ref{10d-chirality}), the last projection in (\ref{10d-projections}) can be written as:
$$
\Gamma_{012345}\tau_1\,\epsilon\,=\,-\epsilon\,\,,
$$
which is the condition expected for color D5-branes extended along the directions 
$(x^0, x^1, \theta_1,\phi_1, \theta_2,\phi_2)$. Notice, however, the different sign with respect to the one in the projections (\ref{projection-flavor}) for the flavor branes, which is reflecting the different orientation of the worldvolume of the latter. 
}. 

We want to find the backreacted geometry in which the metric is still given by the ansatz (\ref{D5-newmetric}). Notice that this ansatz is symmetric under the exchange of the two-spheres parametrized by $(\theta_1, \phi_1)$ and $(\theta_2, \phi_2)$. Accordingly, 
we will  consider a configuration in which an equal number $N_f$  of D5-branes are extended along the two ``branches" (\ref{branch-one}) and (\ref{branch-two}). Thus, the brane setup that we will  consider preserves this symmetry and can be represented by means of the array:
\begin{center}
\begin{tabular}{|c|c|c|c|c|c|c|c|c|c|c|}
\multicolumn{3}{c}{ }&
\multicolumn{6}{c}
{$\overbrace{\phantom{\qquad\qquad\qquad\qquad\qquad}}^{\text{CY}_3}$}\\
\hline
&\multicolumn{2}{|c|}{$\mathbb{R}^{1,1}$}
&\multicolumn{2}{|c|}{$S^2$}
&\multicolumn{2}{|c|}{$S^2$}
&\multicolumn{2}{|c|}{$N_2$}
&\multicolumn{2}{|c|}{$\mathbb{R}^{2}$}\\
\hline
$N_c$ D$5$ (color) &$-$&$-$&$\bigcirc$&$\bigcirc$&$\bigcirc$&$\bigcirc$
&$\cdot$&$\cdot$&$\cdot$&$\cdot$\\
\hline
$N_f$ D$5$ (flavor) &$-$&$-$&$\bigcirc$&$\bigcirc$&$\cdot$&$\cdot$
&$-$&$-$&$\cdot$&$\cdot$\\
\hline
$N_f$ D$5$(flavor)  &$-$&$-$&$\cdot$&$\cdot$&$\bigcirc$&$\bigcirc$
&$-$&$-$&$\cdot$&$\cdot$\\
\hline
\end{tabular}
\end{center}

Contrary to what happens with the metric, in the presence of flavor branes one necessarily has to modify the ansatz of the RR three-form $F_3$.  This is due to the fact that the branes couple to the RR fields by means of their Wess-Zumino term in the action, which leads to a modification of the Bianchi identity for $F_3$. To get the precise form of this modification, let us look at the WZ term of one of each flavor branes:
\beq
S_{WZ}^{flavor, i}\,=\,T_{5}\,\sum^{N_f}\,\int_{{\cal M}_6^{(i)}}\,\,
\hat C_6\,\,,
\eeq
where the index $i=1,2$ labels the two branches (\ref{branch-one}) and (\ref{branch-two}). As the embeddings are mutually supersymmetric for any value of the coordinates transverse to the flavor brane, we can homogeneously distribute the branes in some of their transverse directions.  Actually, we shall locate the branes at a fixed value $\rho=\rho_Q$ of the radial coordinate $\rho$ (which corresponds to a fixed value of the quark mass) and we will distribute them homogeneously along their transverse angular coordinates. Moreover, when $N_f\to\infty$ we can substitute the discrete distribution of  branes by a continuous distribution with the appropriate normalization. This approach of smearing the branes was pioneered in \cite{Casero:2006pt} (see also \cite{noncritical}) and has been applied successfully to construct several  flavored backgrounds \cite{Casero:2007pz}-\cite{Caceres:2009bk}. In our case, 
the smearing in the $i^{th}$ branch amounts to performing the substitution:
\beq
T_{5}\,\sum^{N_f}\,\int_{{\cal M}_6^{(i)}}\,\,
\hat C_6\,\,\Rightarrow\,\,
-T_{5}\,\int_{{\cal M}_{10}}\, {\rm Vol} \big(\,{\cal Y}_{4}^{(i)}\,\big)
\,\wedge C_6\,\,,
\label{smeared-WZ}
\eeq
where $ {\rm Vol} \big({\cal Y}_{4}^{(i)}\,\big)$ is the volume form of the space transverse to the $i^{th}$ branch, normalized as:
\beq
\int_{{\cal M}_4^{(i)}}\,\,{\rm Vol} \big({\cal Y}_{4}^{(i)}\,\big)\,=\,N_f\,\,,
\eeq
with ${\cal M}_4^{(i)}$ being the four-volume orthogonal to the six-dimensional worldvolume ${\cal M}_6^{(i)}$ of the $i^{th}$ branch. The minus sign on the right-hand side of (\ref{smeared-WZ}) is due to  the orientation of the worldvolume of the  flavor branes.  As we place the flavor branes 
at a fixed value $\rho_Q$ of the $\rho$ variable and we smear them along the other compact dimensions, one can check as in \cite{Angel} that:
\beq
{\rm Vol} \big({\cal Y}_{4}^{(1)}\,\big)\,=\,{N_f\over 8\pi^2}\,\delta(\rho-\rho_Q)\,
d\rho\wedge d\chi\wedge \omega_2\,\,,\qquad
{\rm Vol} \big({\cal Y}_{4}^{(2)}\,\big)\,=\,{N_f\over 8\pi^2}\,\delta(\rho-\rho_Q)\,
d\rho\wedge d\chi\wedge \omega_1\,\,,
\eeq
where $\omega_1$ and $\omega_2$ are the volume  forms of the $(\theta_1,\phi_1)$ and $(\theta_2,\phi_2)$ two-spheres, namely:
\beq
\omega_1\,=\,\sin\theta_1\,d\theta_1\wedge d\phi_1
\,\,,\qquad\qquad
\omega_2\,=\,\sin\theta_2\,d\theta_2\wedge d\phi_2\,\,.
\label{omega12}
\eeq
Given the
coupling of the WZ term of the flavor branes to the $C_6$ potential (see (\ref{smeared-WZ})), one has the following modification of the Bianchi identity for $F_3$:
\beq
dF_3\,=\,2\kappa_{10}^2\,T_5\, \Omega\,\,,
\eeq
where the four-form $\Omega$ is the so-called smearing form, which encodes the distribution  of  D5-brane charge of our setup and is given by:
\beq
\Omega\,=\,-{\rm Vol} \big({\cal Y}_{4}^{(1)}\,\big)\,-\,
{\rm Vol} \big({\cal Y}_{4}^{(2)}\,\big)\,=\,-{N_f\over 8\pi^2}\,\delta(\rho-\rho_Q)\,
d\rho\wedge d\chi\wedge\,\big(\, \omega_1\,+\, \omega_2\,)\,\,.
\label{Omega-22}
\eeq
Moreover, taking into account that $1/T_{5}\,=\, (2\pi)^5\,g_s\,(\alpha')^3$ and 
$2\kappa_{(10)}^2\,=\,(2\pi)^7\,g_s^2\,(\alpha')^4$, 
one gets:
\beq
dF_3\,=\,-{g_s\alpha' N_f\over 2}\,\delta(\rho-\rho_Q)\,d\rho\wedge d\chi\wedge\,
\big(\,\omega_1\,+\,\omega_2\,\big)\,\,.
\label{Bianchi-id}
\eeq
It is now easy to modify the ansatz for $F_3$ in (\ref{F3ansatz}) in such a way that (\ref{Bianchi-id}) is satisfied. Indeed, one can readily verify that $F_3$ can be taken to be:
\beq
F_3\,=\,\Big[\,g\,-\,{g_s\alpha' N_f\over 2}\,\Theta(\rho-\rho_Q)\,\Big]\,d\chi\wedge(\,
\omega_1\,+\,\omega_2)\,+\,
dg\wedge d\chi\wedge(\,d\psi+\cos\theta_1 d\phi_1+\cos\theta_2d\phi_2\,)\,\,,
\eeq
with $\Theta$ being the Heaviside function. As stated above, we will continue to assume that the backreacted metric has the form written in (\ref{D5-newmetric}).  Using for this modified ansatz  the same projections as in (\ref{10d-projections}) and the same form of the Killing spinor as in (\ref{10d-spinor}), we now get the following system of first-order BPS equations:
\bear
&&m^2\,\Big[\,g\,-\,{g_s\alpha' N_f\over 2}\,\Theta(\rho-\rho_Q\,)\Big]\,=\,\rho\,z'\,\,,\rc
&&e^{2\Phi}\,=\,{\sigma\over z^2\dot z}\,\,,\rc
&&m^2\,g'\,=\,2\,e^{-2\Phi}\,\rho\,\sigma\,\dot\Phi\,\,,\rc
&&m^2\,\dot g\,=\,-2m^2\,{\sigma\over z^3}\,e^{-2\Phi}\,
\Big[\,g\,-\,{g_s\alpha' N_f\over 2}\,\,\Theta(\rho-\rho_Q\,)\Big]\,\,-\,
2\,{\rho\sigma\over z^2}\,e^{-2\Phi}\,\Phi'\,\,.
\label{BPS-flavored}
\eear
As the set of projections imposed on the Killing spinors in this flavored case is just the same as in (\ref{10d-projections}), any solution of the system (\ref{BPS-flavored}) gives rise to a background that preserves the same four supersymmetries of the unflavored theory.  Clearly, when $N_f=0$ or $\rho<\rho_Q$ the system (\ref{BPS-flavored}) reduces to the one written in (\ref{BPS}) for the unflavored system. Moreover, as explained in detail in appendix \ref{eoms}, any solution of (\ref{BPS-flavored}) also solves  the equations of motion with source terms from the smeared flavor branes. Furthermore, 
as it happens for the unflavored case, we can reduce this system to the following PDE for $z(\rho,\sigma)$:
\beq
{N_f\over 2N_c}\,\sigma \delta(\rho-\rho_Q)\,+\,
\rho\,z^2\,(\dot z\,-\,\sigma\,\ddot z\,)\,=\,\sigma\,(\,
2\rho\,z\,\dot z^2\,+\,z'\,+\,\rho\,z''\,)\,\,.
\label{flavored-PDE}
\eeq
 Notice that the effect of the flavor in (\ref{flavored-PDE}) is a delta-function source located at $\rho=\rho_Q$.

 Notice that  the third equation in  (\ref{BPS-flavored}) implies that $g(\rho, \sigma=0)$ is also constant in this flavored case. Actually, as argued in \cite{Angel} the solution should be the same as in the unflavored case for $\rho<\rho_Q$. Taking this into account we conclude that (\ref{g-sigma0}) also holds when $N_f\not=0$. Using this information it is straightforward to integrate the first equation in  (\ref{BPS-flavored}) and get $z(\rho, \sigma=0)$. Imposing that the solution matches the unflavored one for $\rho<\rho_Q$ and that is continuous at $\rho=\rho_Q$ one gets:
\beq
z(\rho, \sigma=0)\,=\,\Big[\,1\,-\,{N_f\over 2N_c}\,\Theta(\rho-\rho_Q)\,\Big]\,
\log\rho\,+\,{N_f\over 2N_c}\,\Theta(\rho-\rho_Q)\,\log\rho_Q\,\,,
\qquad (\rho\ge \rho_m)
\label{z-sigma0-flavor}\,\,.
\eeq
When $N_f<2N_c$ we see from (\ref{z-sigma0-flavor}) that  the addition of flavor makes $z(\rho, \sigma=0)$ grow slower with $\rho$  which, in view of (\ref{YM-22}) makes the UV decreasing of $g_{YM}$ smaller. Actually, this is what is expected to happen to the Yang-Mills coupling when matter degrees of freedom are added to a gauge theory. When  $N_f>2N_c$ the function $z$ starts to decrease at $\rho=\rho_Q$ and
becomes negative at some large value of $\rho$, where  the solution ceases to be valid. 

\begin{figure}
\centering
\includegraphics[width=1\textwidth]{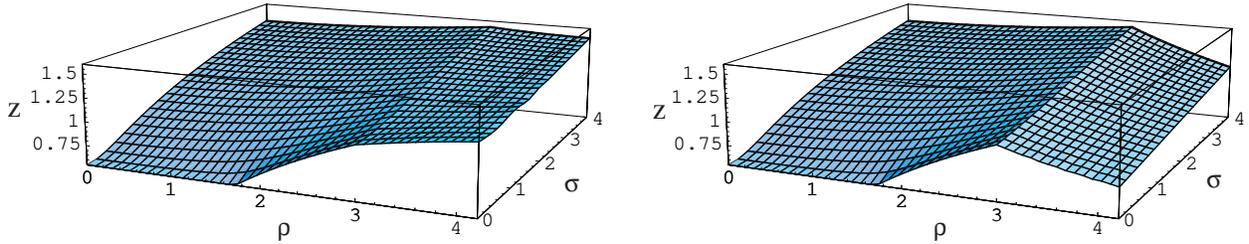} \hfill
\caption{Plot of $z(\rho,\sigma)$ obtained by the numerical integration of (\ref{flavored-PDE}) for $\rho_Q=3$ and $N_f=N_c$ (left) and  $N_f=3N_c$ (right). }
\label{z-phi-22-flv}
\end{figure}

In order to evaluate $z$ for arbitrary values of $\rho$ and $\sigma$ we must integrate numerically the BPS system (\ref{BPS-flavored}). In this integration we assume that the solution reduces to the unflavored one one for $\rho\le \rho_Q$ and that the function $g$ is continuous at $\rho=\rho_Q$. Then, it follows from the first equation in (\ref{BPS-flavored}) that  $z'(\rho, \sigma)$ has a discontinuity at $\rho=\rho_Q$  which is independent of $\sigma$ and given by:
\beq
z'(\rho_Q+\epsilon, \sigma)\,-\,z'(\rho_Q-\epsilon, \sigma)\,=\,
-{N_f\over 2 N_c}\,\,{1\over \rho_Q}\,\,.
\eeq
The result of the numerical  integration is shown in figure \ref{z-phi-22-flv}. Notice the characteristic  wedge shape of $z(\rho,\sigma)$ near $\rho=\rho_Q$.

\section{The dual of ${\cal N}=(1,1)$ theories}
\label{11-10d}

We shall now try to find a background representing D5-branes wrapped on a four-cycle with reduced supersymmetry. In particular we shall concentrate on the case in which the number of supersymmetries is two.  By a simple counting argument one readily concludes that the four-cycle that the D5-branes wrap must belong to a manifold of $G_2$ holonomy. We are thus led to consider the following brane setup:
\begin{center}
\begin{tabular}{|c|c|c|c|c|c|c|c|c|c|c|}
\multicolumn{2}{c}{ }&
\multicolumn{9}{c}
{$\overbrace{\phantom{\qquad\qquad\qquad\qquad\qquad}}^{\text{G}_2}$}\\
\hline
&\multicolumn{2}{|c|}{$\mathbb{R}^{1,1}$}
&\multicolumn{4}{|c|}{$S^4$}
&\multicolumn{3}{|c|}{$N_3$}
&\multicolumn{1}{|c|}{$\mathbb{R}$}\\
\hline
D$5$ &$-$&$-$&$\bigcirc$&$\bigcirc$&$\bigcirc$&$\bigcirc$
&$\cdot$&$\cdot$&$\cdot$&$\cdot$\\
\hline
\end{tabular}
\end{center}
In order to formulate a specific ansatz for the metric and the  three-form of this brane arrangement, let us recall that a geometry of a non-compact manifold with $G_2$ holonomy with a co-associative four-cycle was found in refs. \cite{bs, GPP}. Let us explicitly   write this metric. We start by representing the line element of a four-sphere as:
\beq
d\Omega_4^2\,=\,
{4\over(1+\xi^2)^2}
\left[d\xi^2+{\xi^2\over4}\left((\omega^1)^2+(\omega^2)^2+(\omega^3)^2
\right)\right]\,\,,
\label{dOmega4}
\eeq
where $-\infty<\xi<+\infty$  is a non-compact coordinate and $\omega^i$ ($i=1,2,3$) is a set of $SU(2)$ left-invariant one-forms satisfying $d\omega^i={1\over 2}\epsilon_{ijk} \,\omega^j\wedge\omega^k$. Let us in addition introduce two angular coordinates $\theta$ and $\phi$ ($0\le \theta\le \pi$, $0\le \phi <2\pi$) parametrizing a two-sphere and let $E^1$ and $E^2$ be the following one forms:
\bear
&&E^1=d\theta+{\xi^2\over1+\xi^2}\left(\sin\phi\,\omega^1-\cos\phi\,\omega^2\right)
\,\,,\rc\rc
&&E^2=\sin\theta\left(d\phi-{\xi^2\over1+\xi^2}\,\omega^3\right)+{\xi^2\over1+\xi^2}\,
\cos\theta\left(\cos\phi\,\omega^1+\sin\phi\,\omega^2\right)\,\,.
\label{E1E2}
\eear
Then, the metric of $G_2$ holonomy  of refs. \cite{bs,GPP} can be written as:
\beq
ds^2_{7}\,=\,{(d\sigma)^2\over 1\,-\,{a^4\over \sigma^4}}+
{\sigma^2\over 2}\,d\Omega_4^2
+\, {\sigma^2\over 4}\,\,\Big(\,1\,-\,{a^4\over \sigma^4}\,\Big)\Big[\,(E^1)^2\,+\,(E^2)^2\Big]\,\,,
\label{G2cone}
\eeq
where $a$ is a real constant and the variable $\sigma$ is defined in the range
$a\le \sigma<\infty$. The geometry (\ref{G2cone}) is a resolved  $G_2$ cone with a blown-up four-cycle at the tip of size $a$.  Notice that the $(\theta, \phi)$ two-sphere is fibered over the four-cycle.

We will take the metric (\ref{G2cone}) as the starting point to formulate  our ansatz for the ten-dimensional metric. First of all we add two Minkowski coordinates $x^{0,1}$ and a new non-compact coordinate $\rho$ transverse to the special holonomy manifold. Moreover, we will parametrize the size of the cycle by a function $z$, which depends on both $\sigma$ and $\rho$.  The deformation of the manifold induced by the D5-brane produces a squashing between the four-cycle and the fibered $S^2$. We will adopt a particular ansatz for this squashing, which is just the one that is obtained when the metric  is generated by uplifting from seven-dimensional gauged supergravity (see appendix \ref{gaugedsugra7d}). Accordingly, let us consider a string frame metric of the form:
\beq
ds^2\,=\,e^{ \Phi}\,\Big[\,dx^2_{1,1}\,+\,{z\over m^2}\,d\Omega_4^2\,\Big]\,+\,
{e^{- \Phi}\over m^2 z^{{4\over 3}}}\,\Big[\,d\sigma^2\,+\,\sigma^2\,
\big(\,(E^1)^2+(E^2)^2\,\big)\,\Big]\,+\,{e^{-\Phi}\over m^2}\,(d\rho)^2\,\,,
\label{11-10d-ansatz}
\eeq
where $\Phi(\rho,\sigma)$ is the dilaton and $m$ is the constant with dimension of mass written in (\ref{m}). As in any other supergravity solution representing D5-branes, the background is endowed with a non-trivial RR three-form $F_3$. In order to write its expression in a compact form,  let us define the one-forms ${\cal S}^{\xi}$ and ${\cal S}^i$ $(i=1,2,3)$ as:
\bear
&&{\cal S}^{\xi}\,=\,{2\over 1+\xi^2}\,d\xi\,\,,\rc\rc
&&{\cal S}^{1}\,=\,{\xi\over 1+\xi^2}\,\Big(
\sin\phi\,\omega^1-\cos\phi\,\omega^2\Big)
\,,\rc\rc
&&
{\cal S}^{2}\,=\,{\xi\over 1+\xi^2}\,\Big(
\sin\theta\,\omega^3-\cos\theta\left(\cos\phi\,\omega^1+
\sin\phi\,\omega^2\right)\Big)
\,,\rc\rc
&&
{\cal S}^{3}\,=\,{\xi\over 1+\xi^2}\,\Big(-\cos\theta\,\omega^3-\sin\theta\left(\cos\phi\,\omega^1+
\sin\phi\,\omega^2\right)\Big)
\,.
\label{rotomega}
\eear 
Notice that the metric of the four-sphere can be  simply written as:
\beq
d\Omega^2_4\,=\,\big({\cal S}^{\xi}\big)^2\,+\,\big({\cal S}^{1}\big)^2\,+\,
\big({\cal S}^{2}\big)^2\,+\,\big({\cal S}^{3}\big)^2\,\,.
\eeq
Let us represent $F_3$ in terms of a two-form potential $C_2$ as:
\beq
F_3\,=\,dC_2\,\,.
\label{F3-C2}
\eeq
Then, we shall adopt the following ansatz for $C_2$:
\beq
C_2\,=\,g_1\,\,E^1\wedge E^2\,+\,g_2\,
\big({\cal S}^{\xi}\wedge {\cal S}^{3}\,+\,{\cal S}^1\wedge {\cal S}^{2}\big)\,\,,
\label{C2-ansatz}
\eeq
where $g_1$ and $g_2$ are functions of the variables $\rho$ and $\sigma$.  By using:
\bear
&&d\big(E^1\wedge E^2\big)\,=\,d\big(
{\cal S}^{\xi}\wedge {\cal S}^{3}\,+\,{\cal S}^1\wedge {\cal S}^{2}\big)\,=\,\rc\rc
&&=\,E^1\wedge ({\cal S}^{\xi}\wedge {\cal S}^{2}\,-\,{\cal S}^1\wedge {\cal S}^{3}\big)\,+\,
E^2\wedge ({\cal S}^{\xi}\wedge {\cal S}^{1}\,+\,{\cal S}^2\wedge {\cal S}^{3}\big)\,\,,
\label{dE1E2}
\eear
one finds that the field strength $F_3$ is given by:
\bear
&&F_3\,=\,\big(g_1'\,d\rho+\,\dot g_1\,d\sigma\big)\wedge E^1\wedge E^2\,+\,
\big(g_2'\,d\rho+\,\dot g_2\, d\sigma\big)\wedge\big({\cal S}^{\xi}\wedge {\cal S}^{3}\,+\,{\cal S}^1\wedge {\cal S}^{2}\big)\,+\,\qquad\qquad\rc\rc
&&\qquad\qquad\,+\,\big(g_1+g_2\big)\,\Big[\,
E^1\wedge ({\cal S}^{\xi}\wedge {\cal S}^{2}\,-\,{\cal S}^1\wedge {\cal S}^{3}\big)\,+\,
E^2\wedge ({\cal S}^{\xi}\wedge {\cal S}^{1}\,+\,{\cal S}^2\wedge {\cal S}^{3}\big)\,\Big]\,\,.
\qquad\qquad
\label{F3-10d-unflavored}
\eear

Let us now impose that our ansatz preserves two supersymmetries. As in section \ref{22section} we have to require the vanishing of the supersymmetric variations (\ref{Susy-variations}) of the dilatino and gravitino,  once a certain set of projections are imposed to the Killing spinors. Specifically, let us choose the following vielbein frame basis for the metric (\ref{11-10d-ansatz}):
\bear
&&e^{i}\,=\,e^{{ \Phi\over 2}}\,dx^i\,\,,\qquad (i=0,1)\,\,,\qquad\qquad\qquad\qquad
e^2\,=\,{e^{{ \Phi\over 2}}\over m}\,\,\sqrt{z}\,\,{\cal S}^{\xi}\,\,,\rc\rc
&&e^{j}\,=\,{e^{{ \Phi\over 2}}\over m}\,\,\sqrt{z}\,\,{\cal S}^{j-2}\,\,,
\qquad (j=3,4,5)\,\,,\qquad\qquad
e^6\,=\,{e^{-{ \Phi\over 2}}\over m\,z^{{2\over 3}}}\,\,d\sigma\,\,,\rc\rc
&&e^{k}\,=\,{e^{-{ \Phi\over 2}}\over m\,z^{{2\over 3}}}\,\,\sigma\,E^{k-6}
\,\,,\qquad (k=7,8)\,\,,\qquad\qquad\qquad
e^9\,=\,{e^{-{ \Phi\over 2}}\over m\,}\,\,d\rho\,\,.
\label{10d-frame}
\eear
Then, we shall impose on the Killing spinors $\epsilon$ the following projections along the internal directions:
\bear
&&\Gamma_{25}\,\epsilon\,=\,\Gamma_{34}\,\epsilon\,=\,\Gamma_{78}\,\epsilon\,\,,\rc\rc
&&\Gamma_{45}\,\epsilon\,=\, \Gamma_{67}\,\epsilon\,\,.
\label{internal-projections}
\eear
In addition, we shall also require the condition that corresponds to color D5-branes extended  along the Minkowski directions $x^0$, $x^1$ and along the four-cycle, namely:
\beq
\Gamma_{6789}\tau^1\,\epsilon\,=\,\epsilon\,\,.
\label{D5-proj}
\eeq
Using the fact that the total ten-dimensional chirality is fixed (see (\ref{10d-chirality})) this condition is equivalent to:
\beq
\Gamma_{012345}\tau^1\,\epsilon\,=\,-\epsilon\,\,.
\label{color-brane}
\eeq
We will assume that the Killing spinor $\epsilon$ does not depend on the Minkowski and angular coordinates.  The set of BPS equations obtained in this way is:
\bear
&&\dot{ \Phi}\,=\,{m^2\over 2}\,{z^{{2\over 3}}\,e^{2 \Phi}\,\over \sigma^2}\,g_1'\,-\,
{\sigma e^{-2 \Phi}\over z^{{7\over 3}}}\,\,,\rc\rc
&& \Phi'\,=\,-{m^2\over 2}\,{z^{2}\,e^{2 \Phi}\,\over \sigma^2}\,\dot g_1\,-\,
{3 m^2\over \sigma  z^{{1\over 3}}}(g_1+g_2)\,\,,\rc\rc
&&\dot z\,=\,{3\sigma\over z^{{4\over 3}}}\, e^{-2 \Phi}\,\,,\rc\rc
&&z'\,=\,{3m^2\over \sigma}\, z^{{2\over 3}}\,(g_1+g_2)\,\,,\rc\rc
&&\dot g_2\,=\,{1\over \sigma}\,(g_1+g_2)\,\,,\rc\rc
&&g_2'\,=\,-{\sigma\over m^2 z^{{2\over 3}}}\,e^{-2 \Phi}\,\,.
\label{BPS-10d-unflavored}
\eear
In addition, the vanishing of the  supersymmetry variation of  $\psi_{M}$ implies that the derivatives with respect to $\sigma$ and $\rho$ of the Killing spinor $\epsilon$ are given by:
\bear
&&\dot \epsilon\,=\,{m^2\over 8 z^{{5\over 3}}}\,\Big[\,
2g_2'\,+\,{e^{2 \Phi}\, z^{{7\over 3}}\over \sigma^2}\,g_1'\,\Big]
\,\epsilon\,\,,\rc\rc
&&\epsilon'\,=\, -{m^2\over 8 z^{{1\over 3}}}\,\Big[\,
2\dot g_2\,+\,{e^{2 \Phi}\, z^{{7\over 3}}\over \sigma^2}\,\dot g_1\,+\,
{4(g_1+g_2)\over\sigma} \,\Big]\,\epsilon\,\,.
\label{epsilon-derivatives}
\eear
By comparing the right-hand sides of the two equations in (\ref{epsilon-derivatives}) with the system (\ref{BPS-10d-unflavored}), one discovers that they can be simply written as derivatives of $ \Phi$, namely:
\beq
\dot \epsilon\,=\,{\dot{ \Phi}\over 4}\,\epsilon\,\,,\qquad\qquad
\epsilon'\,=\,{ \Phi\,'\over 4}\,\epsilon\,\,.
\eeq
It follows that $\epsilon$ must be of the form:
\beq
\epsilon\,=\,e^{{ \Phi\over 4}}\,\epsilon_0\,\,,
\eeq
where $\epsilon_0$ is a constant spinor satisfying the same projections as $\epsilon$.  This means that our background is 1/16-supersymmetric, \ie\ it preserves two supersymmetries which, as in the case studied in section \ref{22section},  have different two-dimensional chirality, as it corresponds to  a gravity dual of an ${\cal N}=(1,1)$ supersymmetric gauge theory in two dimensions. 

One can verify that any solution of (\ref{BPS-10d-unflavored}) solves the second-order equations of motion (see appendix \ref{eoms}). It can be also checked 
that the first two equations  in the system (\ref{BPS-10d-unflavored}) are a consequence of the last four.  Moreover, we can write a single PDE for $z(\rho,\sigma)$:
\beq
z^{{4\over 3}}\,\big(\,z \ddot z\,+\,{2\over 3}\,(\dot z)^2\,\big)\,=\,{2\over 3}\,(z')^2-z\,z''\,\,.
\label{PDE-unflavored}
\eeq
If a solution of (\ref{PDE-unflavored}) is known, one can use it in the system (\ref{BPS-10d-unflavored}) to get  the other functions of the ansatz, namely $\Phi$, $g_1$ and $g_2$. 

As in the case studied in section \ref{22section}, the BPS system can be recast in terms of a calibration form ${\cal K}$. Indeed, let ${\cal K}$ be defined as in (\ref{K-def})-(\ref{K-bilinear}) in terms of spinor bilinears.  By using the projections (\ref{internal-projections})-(\ref{D5-proj}) satisfied by the Killing spinor in this case, one gets that ${\cal K}$ can be written:
\beq
{\cal K}\,=\,e^{01}\,\wedge\,\big(\,e^{2345}\,+\,e^{2367}\,+\,e^{3478}\,-\,e^{2468}\,+\,
e^{3568}\,+\,e^{2578}\,+\,e^{4567}\,\big)\,\,,
\label{Phi-expression}
\eeq
where the $e^a$'s are the one-forms of the basis (\ref{10d-frame}). Moreover, one can verify that the system (\ref{BPS-10d-unflavored}) is equivalent to the calibration conditions written in (\ref{calibration-conditions}) for the six-form written in (\ref{Phi-expression}).

\subsection{Integration of the BPS equations}
\label{11-solution}

The gauged supergravity approach provides a particular solution of the PDE equation (\ref{PDE-unflavored}) in implicit form.   The details of this solution are explained in appendix \ref{gaugedsugra7d}. Let us write here the final result. The function $z(\rho,\sigma)$ is the solution of the implicit equation:
\beq
{\rho^2\over 
z^{{1\over 3}}\,\Big[\,
I_{{1\over6}}\,({z\over3})+c\,I_{-{1\over6}}\,({z\over3})\,\Big]^2}\,+\,
{\sigma^2\over 
z^{{5\over 3}}\,\Big[\,
I_{-{5\over6}}\,({z\over3})+c\,I_{{5\over6}}\,({z\over3})\,\Big]^2}\,=\,1\,\,,
\label{implicit-sol}
\eeq
where the $I_n$ are modified Bessel functions of the first kind.  One can check that, indeed, this implicit function solves the PDE (\ref{PDE-unflavored}). In order to write the expression of the dilaton $\Phi$, let us define the function $x(z)$ as:
\beq
e^{x(z)}\,=\, \Bigg[{I_{-{5\over6}}\,({z\over3})+c\,I_{{5\over6}}\,({z\over3})\over
I_{{1\over6}}\,({z\over3})+c\,I_{-{1\over6}}\,({z\over3})}
\Bigg]^{{1\over 2}}\,\,.
\label{exz}
\eeq
Then, the dilaton $\Phi$ is given by:
\beq
e^{\Phi}\,=\,\sqrt{e^{6x}\,\rho^2\,+\,z^{-{4\over 3}}\,e^{-2x}\,\sigma^2}\,\,.
\label{dilaton-11-10d}
\eeq
Let us next  define the angle $\psi$ ($0\le\psi\le\pi)$ as:
\beq
\cot\psi\,=\,z^{{2\over 3}}\,e^{2x}\,\,{\rho\over \sigma}\,\,.
\eeq
Then, the functions $g_1$ and $g_2$ can be written as:
\beq
g_1\,=\,{1\over m^2}\,\,\Big(\,
z^{-{2\over 3}}\,e^{4x-2\Phi}\,\rho\sigma\,-\,\psi\,\Big)\,\,,
\qquad\qquad
g_2\,=\,{\psi\over m^2}\,\,.
\label{g12-11-10d}
\eeq
As in the case with four supersymmetries, the interpretation of the ${\cal N}=(1,1)$ solution (\ref{implicit-sol})-(\ref{g12-11-10d}) depends crucially  on the value of the integration constant $c$. We will argue in detail in appendix \ref{gaugedsugra7d} that the physically sensible solution is obtained when $c<-1$. In this case the function $e^{2x(z)}$ defined in (\ref{exz}) vanishes for some value $z_0$ of its variable $z$, which is the solution of the following transcendental equation:
\beq
I_{-{5\over6}}\Big({ z_0\over3}\Big)+c\,I_{{5\over6}}\Big({
z_0\over3}\Big)=0\,\,.
\label{z0}
\eeq

\begin{figure}
\centering
\includegraphics[width=0.4\textwidth]{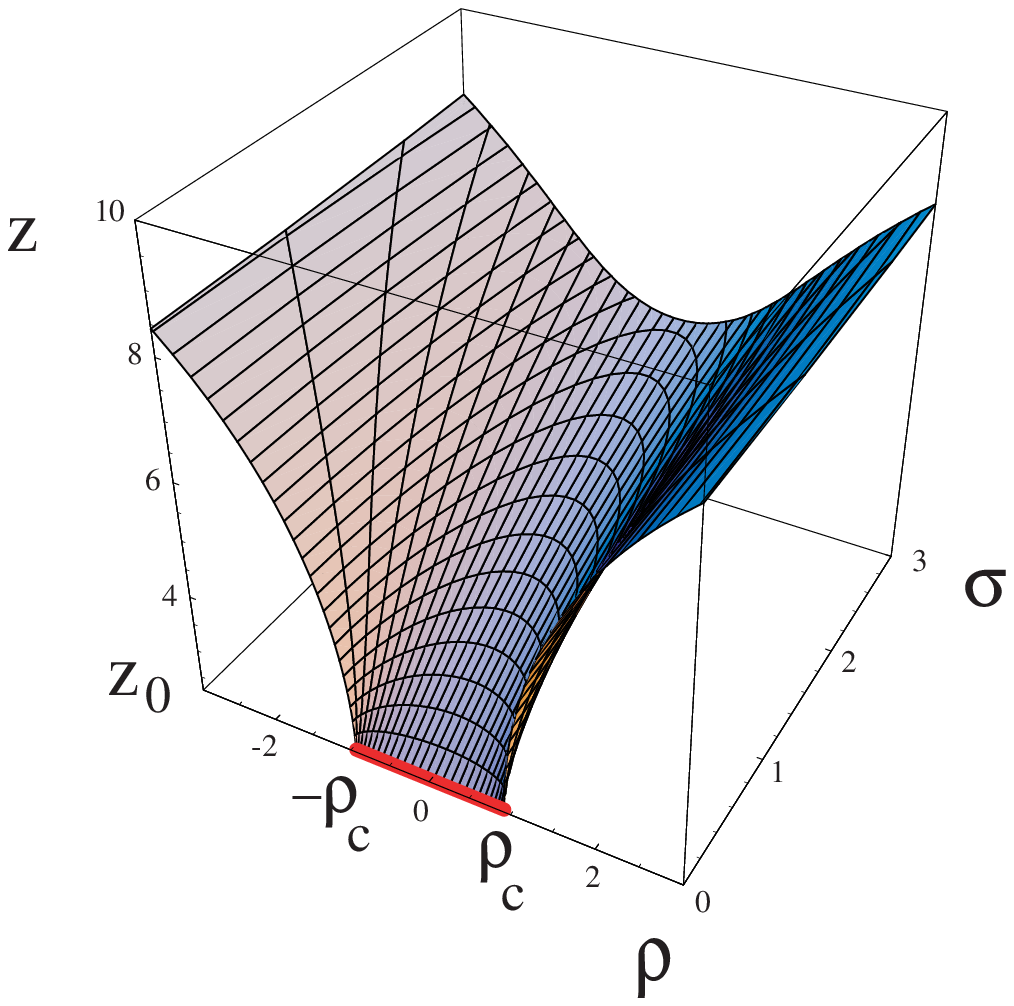} \hfill
\includegraphics[width=0.4\textwidth]{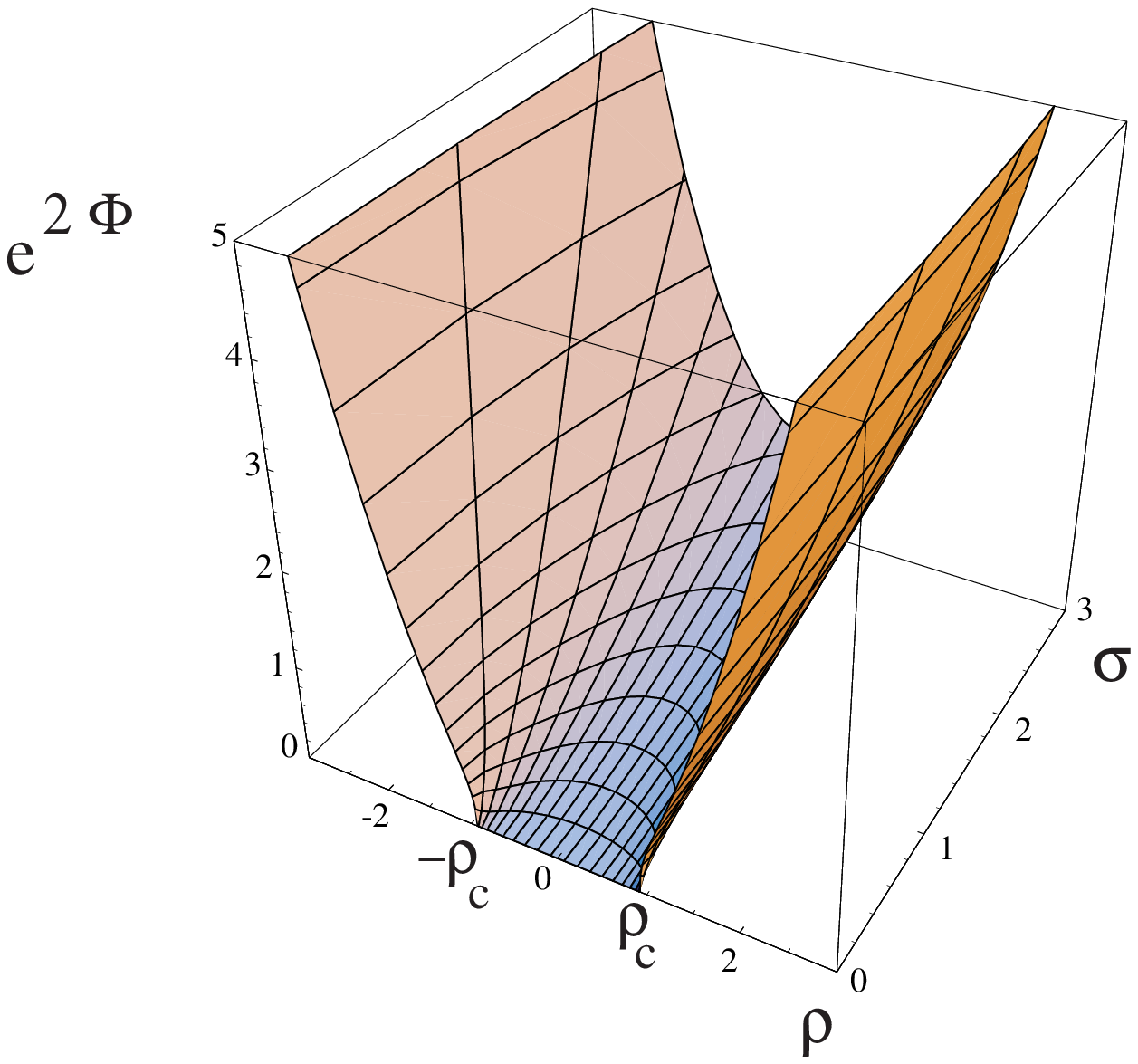}
\caption{Plots of $z(\rho,\sigma)$ (left) and $e^{2\Phi}$ (right) obtained from (\ref{implicit-sol})  and (\ref{dilaton-11-10d}). In both cases we have taken $c=-1.5$. On the left plot we have represented  by a line the segment $z=z_0$, $-\rho_c\le\rho\le \rho_c$ in the $\sigma=0$ plane. }
\label{z-phi-11-flv}
\end{figure}

By inspecting (\ref{implicit-sol}) we deduce that $\sigma=0$ when $z=z_0$. Actually, from the explicit plot of the function $z(\rho, \sigma)$ (see figure (\ref{z-phi-11-flv})) we conclude that $z=z_0$ defines a segment in the $\sigma=0$ axis. This segment is characterized by the conditions $\sigma=0,\,\, -\rho_c\le \rho\le \rho_c$, with:
\beq
\rho_c=z_0^{{1\over 6}}\,\Big|\,
I_{{1\over6}}\,({z_0\over3})+c\,I_{-{1\over6}}\,({z_0\over3})\,\Big|\,\,.
\label{rho_c}
\eeq
One can also verify that $e^{\Phi}$ vanishes along this constant $z=z_0$ segment. Thus our solution has a linear distribution of singularities. One can check that these singularities are good in the sense of \cite{Maldacena:2000mw} (see appendix \ref{gaugedsugra7d}). Therefore, our solution can naturally be  interpreted as being generated by a linear distribution of D5-branes, smeared along a  finite segment on the direction of the coordinate $\rho$ orthogonal to the $G_2$ holonomy manifold (see appendix \ref{gaugedsugra7d} for a more detailed study). Accordingly, the solution  (\ref{implicit-sol})-(\ref{g12-11-10d})  can be regarded as giving the gravity dual of a slice of the Coulomb branch of ${\cal N}=(1,1)$ SYM. We will further check this statement in the next subsection by means of a probe computation.

\subsection{Probe analysis}
\label{probe-11}

Let us now consider a D5-brane probe extended along the Minkowski directions and wrapping the four-sphere. The dynamics of such a probe is governed by the action (\ref{actionD5}). In order to select an appropriate set of worldvolume coordinates, let us parametrize the left-invariant one-forms $\omega^i$ appearing in $d\Omega_4^2$ by means of three angles $\tilde\theta$, $\tilde\varphi$ and $\tilde\psi$ as follows:
\bear
&&\omega^1\,=\,\cos\tilde\psi\,d\tilde\theta\,-\,\sin\tilde\psi\,\sin\tilde\theta\,d\tilde\varphi
\,\,,\rc
&&\omega^2\,=\,\sin\tilde\psi\,d\tilde\theta\,+\,\cos\tilde\psi\,\sin\tilde\theta\,d\tilde\varphi
\,\,,\rc
&&\omega^3\,=\,d\tilde\psi\,-\,\cos\tilde\theta \,d\tilde\varphi\,\,,
\label{left-invariant}
\eear
with $0\le\tilde\theta\le \pi$, $0\le \tilde\varphi<2\pi$, $0\le\tilde\psi<4\pi$. Given this parametrization, we will choose 
$\zeta^{a}\,=\,(x^0, x^1, \xi,\tilde\theta, \tilde\varphi, \tilde \psi)$
as our set of worldvolume coordinates
and we will assume that the remaining ten-dimensional coordinates ($\sigma$, $\theta$, $\phi$ and $\rho$) are constant.  In order to write the induced metric for this embedding, let us point out that, as follows from their definition, the one-forms $E^1$ and $E^2$ can be written as:
\beq
E^1\,=\,d\theta\,+\,\xi\,{\cal S}^1\,\,,\qquad\qquad
E^2\,=\,\sin\theta\,d\phi\,-\,\xi\,{\cal S}^2\,\,,
\eeq
where ${\cal S}^1$ and ${\cal S}^2$ are the one-forms defined in (\ref{rotomega}). 
Then, the pullback of $E^{1,2}$ to the worldvolume for our particular embedding is related to those of ${\cal S}^1$ and ${\cal S}^2$, namely $\hat E^1=\xi\,\hat{\cal S}^1$, 
$\hat E^2=-\xi\,\hat{\cal S}^2$. Hence, it follows that the induced metric, in string frame,  becomes:
\beq
\hat G^{(6)}_{ab}\,d\zeta^{a}\,d\zeta^{b}\,=\,
e^{\Phi}\,dx^2_{1,1}\,+\,{e^{\Phi} \,z\over m^2}\,
\Big(\,1\,+\,{\sigma^2 \over z^{{7\over 3}}\,e^{2\Phi}}\,\,\xi^2\,\Big)\,
\Big(\,\big(\hat {\cal S}^1\big)^2\,+\,\big(\hat {\cal S}^2\big)^2\,\Big)\,+\,
{e^{\Phi}\, z\over m^2}\,
\Big(\,\big(\hat {\cal S}^3\big)^2\,+\,\big(\hat {\cal S}^{\xi}\big)^2\,\Big)\,\,.
\eeq
We will first consider the configuration in which the worldvolume gauge field $F$ vanishes. The determinant of the induced metric entering the DBI action in this case is just:
\beq
\sqrt{-\det \hat G_6}\,=\,{2z^2\,e^{3\Phi}\over m^4}\,\,
\Big(\,1\,+\,{\sigma^2 \over z^{{7\over 3}}\,e^{2\Phi}}\,\,\xi^2\,\Big)\,
{\xi^3\sin\tilde\theta\over (1+\xi^2)^4}\,\,.
\eeq
Therefore, the DBI term in the action is:
\beq
S_{DBI}\,=\,-{2T_5\over m^4}\,
\int d^6\zeta\,\, z^2\,e^{2\Phi}\,
\Big(\,1\,+\,{\sigma^2 \over z^{{7\over 3}}\,e^{2\Phi}}\,\,\xi^2\,\Big)\,
{\xi^3\sin\tilde\theta\over (1+\xi^2)^4}\,\,.
\eeq
In the Wess-Zumino term of the action 
$\hat C_6$ is the pullback of the RR six-form potential which can be chosen in terms of the calibration form as in (\ref{C6-K}). From the explicit expression of ${\cal K}$ in (\ref{Phi-expression}), we get the pullback of $C_6$, namely:
\beq
\hat C_6\,=\,{2 z^2\,e^{2\Phi}\,\xi^3\sin\tilde\theta\over
 m^4\, (1+\xi^2)^4}\,\Big(\,1\,-\,{\sigma^2\xi^2\over z^{{7\over 3}}\,e^{2\Phi}}\,
 \Big)\,d^6\zeta\,\,.
 \eeq
By adding the DBI and WZ contributions we get:
\beq
S\,=\,-{4T_5\over m^4}\,\,\int d^6\zeta\,\,\sigma^2\,
{\xi^5\sin\tilde \theta\over (1+\xi^2)^4\,z^{{1\over 3}}}\,\,,
\eeq
which is just minus the static potential between the stack of color branes and the additional probe. This potential is vanishing for $\sigma=0$, which is the no-force locus of the branes in the internal manifold. It can easily  be  checked by using kappa symmetry that this $\sigma=0$ configuration is also supersymmetric.

 Let us next assume that we are at the $\sigma=0$ point and that  we switch on a worldvolume gauge field $F$ such that its only non-vanishing components are directed along the unwrapped Minkowski directions $x^{0,1}$.  Following the same steps as in 
subsection \ref{probe-22} one can expand the DBI action up to quadratic order and find the expression of the Yang-Mills coupling. One gets:
 \beq
{1\over g_{YM}^2}\,=\,{g_s\,\alpha'\,N_c^2\over 6 \pi}\,z^2(\rho, \sigma=0)\,\,.
\label{gYM-11}
\eeq
The function $z(\rho, \sigma=0)$ can be obtained from the implicit relation (\ref{implicit-sol}). As is clear from the plot in  figure \ref{z-phi-11-flv}, $z(\rho, \sigma=0)$ grows when $\rho$ is large which, according to (\ref{gYM-11}), makes $g_{YM}$  decrease in the UV, as expected. However, as in the case with four supersymmetries, the dilaton blows up at $\rho\to\infty$ and the behavior of the solution in the deep UV region is not trustable.

\subsection{Addition of flavor}

Let us now consider the addition of backreacting flavor D5-branes to the previous  ${\cal N}=(1,1)$  background. These flavor branes should  be non-compact (extended along the $\sigma$ direction) and should be extended in such a way that they  do not further  break the supersymmetry.  Actually, we will be able to find a set of suitable deformations of the unflavored background without determining previously the family of supersymmetric embeddings of the flavor branes. In principle one should use kappa symmetry to find the precise shape of these D5-branes in the ${\cal N}=(1,1)$ background. However, this analysis is quite involved and we will not attempt to perform it directly. Instead, we will try to find directly the four-form charge density distribution for the system of extended calibrated sources, $\Omega$, which is compatible with our metric ansatz (\ref{11-10d-ansatz}) and preserves all the supersymmetries of the unflavored system. We will find a general expression of $\Omega$ which we will subsequently restrict to a particular case with interesting properties.  The corresponding background metric for this $\Omega$ can be computed by numerical integration of the system of BPS equations.  In subsection \ref{microscoping} we will find a set of supersymmetric embeddings for the flavor branes and we will argue that the $\Omega$ found previously  by studying the compatibility of SUSY with our ansatz  contains the density distribution which results after averaging over these embeddings in a suitable way. 

Recall that, due to the WZ  term of the action for the smeared brane system, $T_5\int \Omega\wedge C_6$, the Bianchi identity is violated as:
\beq
dF_3\,=\,2\kappa_{10}^2\,T_5\,\Omega\,\, .
\label{Bianchi-viol}
\eeq
As in the ${\cal N}=(2,2)$ case, we will place our flavor branes at a fixed value $\rho_Q$ in the $\rho$ coordinate ($\rho_Q$ being proportional to the quark mass). Thus, the smearing form $\Omega$ should contain $\delta(\rho-\rho_Q)$ in its expression. Moreover, it is clear from (\ref{Bianchi-viol}) that $d\Omega=0$. These last two  conditions are satisfied if $\Omega$ is of the form:
\beq
2\kappa_{10}^2\,T_5\,\Omega\,=\,\delta(\rho-\rho_Q)\, d\rho\,\wedge\,
d\Lambda\,\,,
\eeq
where $\Lambda$ is a two-form depending on $\sigma$ and on the angular coordinates. Notice that we can represent $F_3$  satisfying (\ref{Bianchi-viol}) as:
\beq
F_3\,=\,dC_2\,+\,f_3\,\,,
\eeq
where $df_3=2\kappa_{10}^2\,T_5\,\Omega$. Clearly,  one can take $f_3$ to be given by:
\beq
f_3\,=\,\Theta(\rho-\rho_Q)\,d\Lambda\,\,.
\eeq
Thus, the total RR three-form field strength is just:
\beq
F_3\,=\,dC_2\,+\,\Theta(\rho-\rho_Q)\,d\Lambda\,\,.
\label{F3-Lambda}
\eeq
From the expression (\ref{F3-Lambda}) of $F_3$  the similarity between the two-form $\Lambda$ and the RR potential $C_2$ is quite evident. We will assume in this flavored case  that $C_2$ is still given by the ansatz (\ref{C2-ansatz}). It is thus natural to adopt a similar ansatz for $\Omega$, namely:
\beq
\Lambda\,=\,L_1(\sigma)\,E^1\wedge E^2\,+\,L_2(\sigma)\,
\Big[\,{\cal S}^{\xi}\wedge {\cal S}^{3}\,+\,{\cal S}^1\wedge {\cal S}^{2}\,\Big]\,\,,
\eeq
where $L_1$ and $L_2$ are functions of the coordinate $\sigma$ to be determined. Notice that the resulting value of the smearing form $\Omega$ is given by:
\bear
&&\Omega\,=\,{\delta(\rho-\rho_Q)\over 2\kappa_{10}^2\,T_5}
\,d\rho\wedge\,\Bigg[
(L_1+L_2)\,
\Big[\,E^1\wedge ({\cal S}^{\xi}\wedge {\cal S}^{2}\,-\,{\cal S}^1\wedge {\cal S}^{3}\big)\,+\,
E^2\wedge ({\cal S}^{\xi}\wedge {\cal S}^{1}\,+\,{\cal S}^2\wedge {\cal S}^{3}\big)\,\Big]\,+\,\rc\rc
\label{Omega-11}
&&
\qquad\qquad\qquad\qquad\qquad+
\dot L_1\,d\sigma\,\wedge\,E^1\wedge E^2\,+\,\dot L_2\,d\sigma\wedge
\Big[\,{\cal S}^{\xi}\wedge {\cal S}^{3}\,+\,{\cal S}^1\wedge {\cal S}^{2}\,\Big]\,\Bigg]\,\,.
\eear
Moreover, if $C_2$ is parametrized in terms of the functions $g_1$ and $g_2$ as in 
(\ref{C2-ansatz}), the total RR three-form $F_3$ for the flavored background can be written as:
\bear
&&F_3\,=\,\big[g_1'\,d\rho+\,\big(\dot g_1\,+\,\dot L_1\,\Theta(\rho-\rho_q)
\big)d\sigma\big]\,
\wedge E^1\wedge E^2\,+\,\rc\rc
&&
+\big[g_2'\,d\rho+\,\big(\dot g_2\,+\,\dot L_2\,\Theta(\rho-\rho_q)
\, \big)d\sigma\big]\wedge\big({\cal S}^{\xi}\wedge {\cal S}^{3}\,+\,{\cal S}^1\wedge {\cal S}^{2}\big)\,+\,\label{F3-11flavored} \\\rc
&&\,+\,\big(g_1+g_2+(L_1+L_2)\Theta(\rho-\rho_q)\big)\,\Big[\,
E^1\wedge ({\cal S}^{\xi}\wedge {\cal S}^{2}\,-\,{\cal S}^1\wedge {\cal S}^{3}\big)\,+\,
E^2\wedge ({\cal S}^{\xi}\wedge {\cal S}^{1}\,+\,{\cal S}^2\wedge {\cal S}^{3}\big)\,\Big]\,\,.
\nonumber
\eear
It is clear from this expression that the flavored BPS system can be obtained from the unflavored one in (\ref{BPS-10d-unflavored}) by means of the substitution:
\bear
&&\dot g_i\to \dot g_i\,+\,\dot L_i\,\Theta(\rho-\rho_Q)\,\,,\qquad (i=1,2)\,\,,\rc\rc
&&g_1+g_2\to g_1+g_2\,+\,(L_1+L_2)\,\Theta(\rho-\rho_Q)\,\,.
\eear
One gets:
\bear
&&\dot{ \Phi}\,=\,{m^2\over 2}\,{z^{{2\over 3}}\,e^{2 \Phi}\,\over \sigma^2}\,g_1'\,-\,
{\sigma e^{-2\Phi}\over z^{{7\over 3}}}\,\,,\rc\rc
&&\Phi'\,=\,-{m^2\over 2}\,{z^{2}\,e^{2 \Phi}\,\over \sigma^2}\,
\big[\,\dot g_1\,+\,\dot L_1\,\Theta(\rho-\rho_Q)\,\big]
\,-\,
{3 m^2\over \sigma  z^{{1\over 3}}}
\big[\,g_1+g_2+(L_1+L_2)\Theta(\rho-\rho_Q)\,\big]\,\,,\rc\rc
&&\dot z\,=\,{3\sigma\over z^{{4\over 3}}}\, e^{-2 \Phi}\,\,,\rc\rc
&&z'\,=\,{3m^2\over \sigma}\, z^{{2\over 3}}\,\big[\,g_1+g_2+(L_1+L_2)\Theta(\rho-\rho_Q)\,\big]\,\,,\rc\rc
&&\dot g_2\,=\,{1\over \sigma}\,\big[\,g_1+g_2+(L_1+L_2)\Theta(\rho-\rho_Q)\,\big]\,
-\,\dot L_2\,\Theta(\rho-\rho_Q)\,,\rc\rc
&&g_2'\,=\,-{\sigma\over m^2 z^{{2\over 3}}}\,e^{-2 \Phi}\,\,.
\label{BPS-flavored-singular}
\eear
By analyzing the system (\ref{BPS-flavored-singular}) one easily discovers that the $(\rho, \sigma)$ crossed derivatives of the functions $g_1$ and $g_2$ are not equal. Indeed, one can prove that:
\beq
\partial_{\rho}\,\dot g_1\,-\,\partial_{\sigma}\,g_1'\,=\,
\partial_{\sigma}\,g_2'\,-\,\partial_{\rho}\,\dot g_2\,=\,
-{1\over \sigma}\,
\big[\,L_1\,+\,L_2\,-\,\sigma \dot L_2\,\big]\,
\delta(\rho-\rho_Q)\,\,,\label{crossed}
\eeq
which, after all,  is not surprising since the first derivatives of $g_1$ and $g_2$ are potentially singular at  $\rho=\rho_Q$ according to the system (\ref{BPS-flavored-singular}). Notice that this would make $d^2C_2\not=0$ and our stating point  equation (\ref{Bianchi-viol}) would not be satisfied for the $\Omega$ written in (\ref{Omega-11}). In order to avoid dealing with this unwanted singularity, we will require the vanishing of the right-hand side of (\ref{crossed}), which is equivalent to imposing   that $L_1$ and $L_2$ satisfy the differential equation:
\beq
\dot L_2\,=\,{1\over \sigma}\,\big(\,L_1\,+\,L_2\,\big)\,\,.
\label{L12-eq}
\eeq
Notice also that, if equation (\ref{L12-eq}) holds, the equation for $\dot g_2$ becomes:
\beq
\dot g_2\,=\,{1\over \sigma}\,\big[\,g_1+g_2\,\big]\,\,,
\eeq
and the potentially singular term for $\dot g_2$ at $\rho=\rho_Q$ disappears. Furthermore, as shown in appendix \ref{eoms}, any solution of (\ref{BPS-flavored-singular}) with $L_1$ and $L_2$ satisfying (\ref{L12-eq}) also solves the equations of motion of the gravity plus (smeared) branes system.  Moreover, 
as in the unflavored system, one can write a single PDE for the function $z(\rho,\sigma)$, which now has a source term parametrized by $L_1\,+\,L_2$, namely:
\beq
z^{{4\over 3}}\,\big(\,z \ddot z\,+\,{2\over 3}\,(\dot z)^2\,\big)\,=\,{2\over 3}\,(z')^2-z\,z''\,-\,
{3m^2(L_1(\sigma)\,+\,L_2(\sigma))\over \sigma}\,z^{{5\over 3}}\,\delta(\rho-\rho_Q)\,\,.
\label{PDE-11}
\eeq
Furthermore, one can verify that the calibration conditions (\ref{calibration-conditions}) reduce to the system (\ref{BPS-flavored-singular}) when the new ansatz (\ref{F3-11flavored}) for $F_3$ is adopted.

Notice that one can use the compatibility condition (\ref{L12-eq}) to obtain the function $L_2(\sigma)$ in terms of $L_1(\sigma)$. Indeed, one can easily integrate  (\ref{L12-eq})  by applying the method of  variation of constants, with the result:
\beq
L_2(\sigma)\,=\,\Big[\,A\,+\,\int_{\sigma_0}^{\sigma}\,\,d\zeta\,\,
{L_1(\zeta)\over \zeta^2}\,\,\,\Big]\,\sigma\,\,,
\label{L2(L1)}
\eeq
where $A$ and $\sigma_0$ are constants. Eq. (\ref{L2(L1)}) is not enough to determine $L_1$ and $L_2$. For this reason one has to find some other physical conditions that could be used to constrain these functions. Recall that $\Omega$ parametrizes the volume transverse to the flavor branes, on which we smear them to produce a continuous distribution. Its Hodge dual ${}^*\Omega$ is a six-form with components parallel to the worldvolume of the flavor branes. Color and flavor branes should not overlap in the internal space.  It is thus natural to require the vanishing of  the pullback of ${}^*\Omega$ to the worldvolume of the color branes, namely:
\beq
{}^*\,\Omega{|_{color\,wv}}\,=\,0\,\,.
\label{pullback*Omega}
\eeq
Recall from the analysis of subsection \ref{probe-11} that the color branes are located at $\sigma={\rm constant}=0$ and their angular embedding is such that the pullback of the forms $E^1$ and $E^2$ is such that $\hat E^1=\xi\,\hat{\cal S}^1$ and   $\hat E^2=-\xi\,\hat{\cal S}^2$. By computing explicitly the pullback of  ${}^*\Omega$ to this submanifold, we get:
\beq
{}^*\,\Omega{|_{color\,wv}}\,=\,{m^4\over 2 \kappa_{10}^2\,T_5}\,\,\delta(\rho-\rho_Q)\,
\Big[\,e^{3\Phi}\,z^{4}\,\,{\dot L_1\over \sigma^2}\,-\,
{e^{\Phi}\over z^{{2\over 3}}}\,\xi^2\,\sigma^2\,\dot L_2\,\Big]_{\sigma=0}\,\,
dx^0\wedge dx^1\wedge {\rm Vol}(S^4)\,\,.
\eeq
Thus, to impose the condition (\ref{pullback*Omega}) one has to require:
\beq
{\dot L_1\over \sigma^2}{\Big|_{\sigma=0}}\,=\,\sigma^2\,\dot L_2\big|_{\sigma=0}\,=\,0\,\,,
\label{sigma0-conditions}
\eeq
which constrain the behavior of $L_1$ and $L_2$ near $\sigma=0$. Actually, the condition on $L_1$ just means that, if we assume that $L_1\sim \sigma^{\beta}$ near 
$\sigma=0$, then necessarily $\beta>3$. In fact, assuming this power behavior of $L_1$, it is straightforward to use (\ref{L2(L1)}) to get $L_1$ and $L_2$, namely:
\beq
L_1\,=\,B\,\sigma^{\beta}\,\,,\qquad
L_2\,=\,A\sigma \,+\,{B\over \beta-1}\,\,\sigma^{\beta-1}\,\,,
\qquad
(\beta>3)\,\,.
\label{powerlikeL12}
\eeq
Another test that the smearing form $\Omega$ must pass can be obtained by looking at the the DBI term of the action of the system of smeared branes. Taking into account that, for a SUSY configuration, the induced  volume form is just the pullback of the calibration form ${\cal K}$ and that the smearing is performed just by taking the wedge product with $\Omega$, this action takes the form:
\beq
S^{flavor}_{DBI}\,=\,-T_5\,\int_{{\cal M}_{10}}\,e^{-{\Phi}}\,\Omega\wedge {\cal K}\,\,.
\eeq
The ten-form $\Omega\wedge {\cal K}$ can be explicitly computed from (\ref{Omega-11}) and (\ref{Phi-expression}), with the result:
\beq
\Omega\wedge {\cal K}\,=\,-{m^4\over 2 \kappa_{10}^2\,T_5}\,\,
z^{-{1\over 3}}\,\,
\Big[\,6\dot L_2\,+\,z^{{7\over 3}}\,e^{2\Phi}\,\,{\dot L_1\over \sigma^2}\,\Big]\,
\delta(\rho-\rho_Q)\,\,{\rm Vol}\,({\cal M}_{10})\,\,.
\label{Omega-wedge-K}
\eeq

Notice that $\Omega\wedge {\cal K}$ can be interpreted as the mass distribution of the system of flavor branes which, being a ten-form in a ten-dimensional space,  is proportional to the volume form  ${\rm Vol}\,({\cal M}_{10})$ of the ten-dimensional manifold. The function multiplying ${\rm Vol}\,({\cal M}_{10})$ in (\ref{Omega-wedge-K}) represents the mass density of the $\rho=\rho_Q$ slice of the space in which we  are smearing the flavor branes.  Notice also that our conditions (\ref{sigma0-conditions}) ensure the regularity of $\Omega\wedge {\cal K}$ at $\sigma=0$. Moreover, from a physical point of view one should require that the mass distribution $\Omega\wedge {\cal K}$ integrated over any portion of ${\cal M}_{10}$ always  gives a positive number. This positivity condition is fulfilled if the function multiplying ${\rm Vol}\,({\cal M}_{10})$ is everywhere positive. One can check that this is the case when $L_1$ and $L_2$ are given by (\ref{powerlikeL12}) if both constants $A$ and $B$ are negative.  Actually, there is a solution of those written in (\ref{powerlikeL12}) that is particularly simple and appealing, namely:
\beq
L_1=0\,\,,\qquad\,\,L_2\,=\,A\,\,\sigma\,\,.
\label{L1L2simpler}
\eeq
In order to explore the properties of this solution, let us parametrize the constant $A$ as:
\beq
A\,=\,-2 \kappa_{10}^2\,T_5\,\,{3n_f\over 2\pi}\,\,,
\label{A}
\eeq
where we introduced the factor $2 \kappa_{10}^2\,T_5$ to absorb the one we have introduced in the definition of $L_1$ and $L_2$ in (\ref{Omega-11}) and the remaining factors  have been introduced for convenience. The constant $n_f$ characterizes the density of  flavor branes of our setup (see subsection \ref{microscoping}). Notice that, in this case, $\Omega$ can be neatly written in terms of the one-forms of the frame basis (\ref{10d-frame}) as:
\bear
&&\Omega\,=\,-{3n_f\over 2\pi}\,\,{m^4\over z^{{1\over 3}}}\,
\delta(\rho-\rho_Q)\,\,e^9\wedge\,
\Big[\,e^7\wedge (e^2\wedge e^4-e^3\wedge e^5)\,+\,
e^8\wedge (e^2\wedge e^3+e^4\wedge e^5)\,+\,
\rc\rc 
&&\qquad\qquad\qquad\qquad\qquad\qquad\qquad
+\,e^{6}\wedge (e^2\wedge e^5+e^3\wedge e^4)
\,\Big]\,\,.
\label{Omega-homogeneuos}
\eear
Moreover, the mass distribution is just given by:
\beq
\Omega\wedge {\cal K}\,=\,
{9n_f\over \pi}\,m^4\,\,\,
z^{-{1\over 3}}\,
\delta(\rho-\rho_Q)\,\,{\rm Vol}\,({\cal M}_{10})\,\,,
\label{Omega-wedge-K-special}
\eeq
and the PDE equation (\ref{PDE-11}) takes the form:
\beq
z^{{4\over 3}}\,\big(\,z \ddot z\,+\,{2\over 3}\,(\dot z)^2\,\big)\,=\,{2\over 3}\,(z')^2-z\,z''\,+\,
{9\pi n_f\over N_c}\,z^{{5\over 3}}\,\delta(\rho-\rho_Q)\,\,.
\label{PDE-11-simpler}
\eeq
Notice that the source term in (\ref{PDE-11-simpler}) is not singular at $\sigma=0$. Moreover, from the fourth equation in the flavored BPS system (\ref{BPS-flavored-singular}), we get that:
\beq
z'\,=\,3 m^2 \,z^{{3\over 3}}\,\Big[\,{g_1+g_2\over \sigma}\,-\,6\pi g_s\alpha' n_f\,
\Theta(\rho-\rho_Q)\,\Big]\,\,.
\label{zprime-11-simpler}
\eeq
In order to integrate (\ref{PDE-11-simpler}) we assume that $z(\rho,\sigma)$ is given by the unflavored solution (\ref{implicit-sol}) for $\rho\le\rho_Q$, while at $\rho=\rho_Q$ the derivative of $z$ with respect to $\rho$ jumps in the form dictated by (\ref{zprime-11-simpler}), namely:
\beq
z'(\rho_Q+\epsilon, \sigma)\,-\,z'(\rho_Q-\epsilon, \sigma)\,=\,
-{18\pi n_f\over  N_c}\,\,z^{{3\over 3}}(\rho_Q, \sigma)\,\,.
\eeq
\begin{figure}
\centering
\includegraphics[width=1\textwidth]{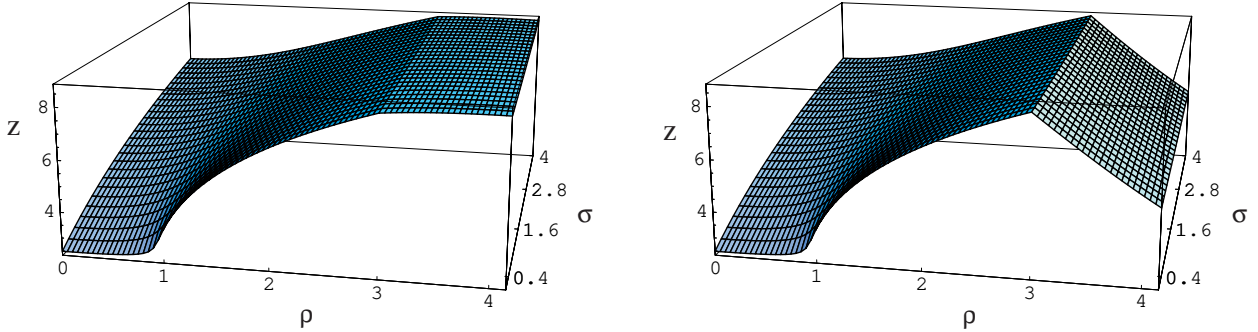} \hfill
\caption{Plot of $z(\rho,\sigma)$ obtained by the numerical integration of (\ref{PDE-11-simpler}) for $\rho_Q=3$ and $x\equiv{18\pi n_f\over  N_c}=0.2$ (left) and  $x=1$ (right). }
\label{z-11-flv}
\end{figure}
In figure \ref{z-11-flv} we have plotted the function $z(\rho,\sigma)$, resulting from the numerical integration of the PDE (\ref{PDE-11-simpler}),  for two different values of $n_f$. Notice the similarity with the results obtained for the flavored system with four supersymmetries (see figure \ref{z-phi-22-flv}).

\subsubsection{Microscopic analysis of the smearing}
\label{microscoping}
The smearing form $\Omega$ is the flavor density distribution that should be obtained after averaging over a continuous set of supersymmetric brane embeddings. For the case at hand we have not been able to find the most general embedding of a flavor brane that preserves all the supersymmetries of the unflavored background. However, we have been able to find a set of such embeddings  and we will argue that, after performing a suitable average, they give rise to some of the components of the smearing density (\ref{Omega-homogeneuos}). In order to illustrate this fact let us consider a D5-brane whose worldvolume is parametrized by the following set of coordinates:
\beq
\zeta^{a}\,=\,(x^0, x^1, \xi, \tilde\psi,\sigma,\phi)\,\,.
\eeq
Moreover, we will assume that the embedding of the brane in the ten-dimensional spacetime is given by:
\beq
\theta=\theta(\sigma)\,\,,
\qquad\qquad
\tilde\theta\,,\,\tilde\varphi={\rm constant}\,\,,\qquad
\rho=\rho_Q\,\,,
\label{ansatz-11-emb}
\eeq
where $\theta(\sigma)$ is a function to be determined and $\tilde\theta$, $\tilde\varphi$ and $\tilde\psi$ are the three angles which parametrize the $SU(2)$ left-invariant one-forms $\omega^1$, $\omega^2$ and $\omega^3$ (see (\ref{left-invariant})). By using the kappa symmetry condition $\Gamma_{\kappa}\,\epsilon\,=\,\epsilon$ one can easily prove that, in order to preserve the two supersymmetries of the background, the function $\theta(\sigma)$ must satisfy the following differential equation:
\beq
\sigma\partial_{\sigma}\theta\,=\,\cot\theta\,\,.
\eeq
The integration of this equation yields:
\beq
\cos\theta\,=\,{\sigma_*\over \sigma}\,\,,
\label{11-embedding}
\eeq
with $\sigma^*$ being a constant. Notice that in (\ref{11-embedding}) $\sigma\ge |\sigma_*|$, \ie\ $|\sigma_*|$ is the minimal value of the coordinate $\sigma$ along the brane (for which $\theta=0, \pi$). Notice that, for the ansatz (\ref{ansatz-11-emb}), the brane is embedded in the $S^4$ in such a way that the pullback of the one-forms $\omega^1$ and $\omega^2$ vanish, while $\hat \omega^3=d\tilde \psi$. Therefore,  the induced metric  on the brane worldvolume takes the form:
\beq
e^{\Phi}\,dx^2_{1,1}\,+\,{z e^{\Phi}\over m^2}\,\,
{1\over (1+\xi^2)^2}\,\,\Big[4\,d\xi^2+\xi^2\, d\tilde\psi^2\,\Big]+
{e^{-\Phi}\over m^2 z^{{4\over 3}}}\,\Big[{d\sigma^2\over \sin^2\theta}\,+\,
\sigma^2\,\sin^2\theta \big(d\phi\,-\,{\xi^2\over 1+\xi^2}\,d\tilde\psi\big)^2\,\Bigg]\,\,.
\qquad\qquad
\label{inducedmetric-fl-11}
\eeq
As a check of the supersymmetric nature of these embeddings one can verify that the induced volume form corresponding to the metric (\ref{inducedmetric-fl-11}) is equal to the pullback of the calibration form (\ref{Phi-expression}). Notice also that these embeddings determine a two-sphere inside the $S^4$.

The configuration of the flavor brane just found has a neat interpretation when one looks in detail at the structure of the normal bundle $N_3$ spanned by the coordinates $(\sigma, \theta, \phi)$. By inspecting the metric (\ref{11-10d-ansatz}) one realizes that $N_3$ is a three-dimensional flat space $\mathbb{R}^3$ fibered over the four-cycle. Accordingly, let us introduce the following set of cartesian coordinates for $N_3$:
\beq
z^1\,=\,\sigma\sin\theta\cos\phi\,\,,\qquad
z^2\,=\,\sigma\sin\theta\sin\phi\,\,,\qquad
z^3\,=\,\sigma\cos\theta\,\,.
\eeq
In terms of these coordinates, the embeddings just found correspond to the plane:
\beq
z^3\,=\,\sigma_*\,\,.
\label{z_3-constant}
\eeq
Notice that we have found a family of configurations parametrized by the constant values of $\tilde\theta$, $\tilde\varphi$ and $\sigma_*$ at $\rho=\rho_Q$. By changing $\tilde\theta$, $\tilde\varphi$ we select a different $S^2$ inside the $S^4$. Moreover, it is clear from (\ref{z_3-constant}) that changing $\sigma_*$ corresponds to choosing a plane $z^3\,=\,$constant  in $N_3$. Let us suppose that we consider a set of branes  embedded in this way that are homogeneously distributed within the $S^4$ and $N_3$. The resulting distribution form $\Gamma$ should be proportional to the  volume element of the space transverse to the branes, which is two-dimensional within the $S^4$ and one-dimensional inside the $N_3$. Accordingly,  let us write $\Gamma$ in factorized form as:
\beq
\Gamma\,=\,\delta(\rho-\rho_Q)\,d\rho\wedge\Gamma_1\wedge \Gamma_2\,\,,
\label{Gamma-tot}
\eeq
with $\Gamma_1$ and $\Gamma_2$ being the contributions of the $S^4$  and $N_3$ respectively. For the embeddings (\ref{ansatz-11-emb}) the transverse directions inside the $S^4$ are those spanned by the one-forms $\omega^1$ and $\omega^2$. By looking at the line element of $S^4$ it  is easy to get the corresponding transverse volume form, namely:
\beq
\Gamma_1\,=\,{3\over 2\pi}\,\,{\xi^2\over (1+\xi^2)^2}\,\omega^1\wedge \omega^2\,\,.
\label{Gamma1}
\eeq
The numerical coefficient included in (\ref{Gamma1}) is just the ratio between the volume occupied by the brane within the $S^4$ ($4\pi$) and the total volume of the $S^4$ ($8\pi^2/3$).

Let us now obtain $\Gamma_2$. First of all we define the function $f$ as:
\beq
f\equiv \sigma\cos\theta-\sigma_*\,\,.
\eeq
If we distribute the embeddings homogeneously in $\sigma_*$ with constant density $n_f$ the corresponding distribution is given by:
\beq
\Gamma_2\,=\,\int \delta(f)\,df\,\, n_f \,d\sigma_*\,\,.
\label{Gamma2-def}
\eeq
The integral over $\sigma_*$ can be immediately computed with the help of the $\delta$-function and the result is:
\beq
\Gamma_2\,=\,n_f\,\big[\,\cos\theta d\sigma\,-\,\sigma\,\sin\theta\,d\theta\,\big]\,\,.
\label{Gamma2-res}
\eeq
Using (\ref{Gamma1}) and (\ref{Gamma2-res}) in (\ref{Gamma-tot}) we get the following expression for $\Gamma$: 
\beq
\Gamma\,=\,-{3n_f\over 2\pi}\,\,{\xi^2\over (1+\xi^2)^2}\,\,
\delta(\rho-\rho_Q)\,d\rho\wedge \omega^1\wedge \omega^2\wedge
\big[\,\sigma\,\sin\theta\,d\theta\,-\,\cos\theta d\sigma]\,\,.
\eeq
Let us now compare this result with the one given by $\Omega$ in (\ref{Omega-11}) when the functions  $L_1$ and $L_2$ are given by (\ref{L1L2simpler}). With this purpose let us consider the terms in $\Omega$ whose components along the $S^4$ are of the form  $\omega^1\wedge \omega^2$. By inspecting  (\ref{Omega-11}) one readily concludes that there are only two such terms, namely:
\beq
\Omega\big|_{\omega^1\wedge \omega^2}\,=\,-{A\over 2\kappa^2_{10}\,T_5}\,
\delta(\rho-\rho_Q)\,d\rho\wedge \Big[\,
\sigma\,E^1\wedge {\cal S}^1\wedge {\cal S}^3\,-\,d\sigma\wedge 
 {\cal S}^1\wedge {\cal S}^2\,\Big]_{\big|_{\omega^1\wedge \omega^2}}\,\,.
 \label{Omega-w1w2}
\eeq
After using the expression of $E^1$ and of the ${\cal S}^i$'s (eqs. (\ref{E1E2}) and (\ref{rotomega})), one gets:
\bear
&&E^1\wedge {\cal S}^1\wedge {\cal S}^3\big|_{\omega^1\wedge \omega^2}\,=\,-
{\xi^2\over (1+\xi^2)^2}\,\,\sin\theta d\theta \wedge \omega^1\wedge \omega^2\,\,,\rc\rc
&&d\sigma\wedge  {\cal S}^1\wedge {\cal S}^2\big|_{\omega^1\wedge \omega^2}\,=\,-
{\xi^2\over (1+\xi^2)^2}\,\,\cos\theta\,d\sigma\wedge \omega^1\wedge \omega^2\,\,.
\eear
Plugging this result in (\ref{Omega-w1w2}) we immediately verify that, indeed, 
$\Omega\big|_{\omega^1\wedge \omega^2}$ coincides with $\Gamma$ if the constant $A$ is related to $n_f$ as in (\ref{A}). Notice that this relation between $A$ and $n_f$ depends on the prescription  we have adopted in (\ref{Gamma1}) for the global constant. If we modify this prescription nothing essential changes in our results.

The embeddings just studied can be easily generalized by changing the plane that the branes occupy  in $N_3$. Due to the fibration of $N_3$, in order to preserve supersymmetry,  the change of the plane should be accompanied by the change of the $S^2$ that the branes wrap inside the $S^4$. For example one could consider the plane 
$z^1={\rm constant}$ and embed the branes along the $S^2$ with $\omega^2=\omega^3=0$. By repeating the previous calculation one can verify that the distribution form coincides with the   components along  $\omega^2\wedge\omega^3$ of $\Omega$. Similarly, placing the brane at $z^2={\rm constant}$ and $\omega^1=\omega^3=0$ reproduces the components of  $\Omega$ along   $\omega^1\wedge\omega^3$.  Actually, one can consider a generic plane in $N_3$ and, after performing an average over all  its possible directions as in the approach of \cite{Bigazzi:2008zt}, one can see that the components of $\Omega$ along $\omega^i\wedge \omega^j$ are obtained. Presumably, all the contributions to $\Omega$ are obtained as the result of the homogeneous smearing of a more general class of embeddings, a calculation that we will not attempt to do here. Notice also that, in our calculation we are distributing the branes in a non-compact space (the transverse directions to the plane in $N_3$) which, as we argued, seems to give rise to the smearing form with the special values of $L_1$ and $L_2$ of (\ref{L1L2simpler}). It seems reasonable to think that the general $\Omega$ could be obtained   by a different (non-homogeneous) distribution. If this is the case, the density (\ref{Omega-homogeneuos}) could be regarded as describing the distribution of branes in  a region where they homogeneously fill  the $\rho=\rho_Q$ slice of the  ten-dimensional space-time.

\section{Conclusions}
\label{conclusions}
In this paper we have found solutions of type IIB supergravity which correspond to D5-branes wrapped along a four-cycle and which preserve four and two supersymmetries. After adopting an ansatz for the metric and three form, we found a system of BPS equations which are obtained by requiring the vanishing of the supersymmetric variations of the dilatino and gravitino.  In our ansatz the different functions depend on the two variables $\rho$ and $\sigma$ and, as a consequence, the resulting BPS system involves partial derivatives and it is difficult to solve. 

Quite remarkably, it turns out that an analytic solution can be found by using seven-dimensional gauged supergravity. Indeed, in seven dimensions the solution is simpler since it involves functions that depend on one radial variable. Upon uplifting to ten dimensions the expressions of the metric and RR three-form become rather complicated, with coordinates which are non-trivially fibered due to the topological twist needed to realize the  supersymmetry. However, by introducing the new coordinates $\rho$ and $\sigma$ the form of the solutions simplifies greatly and their interpretation in terms of branes wrapping a cycle in a non-trivial manifold becomes more transparent. Actually, this form of the uplifted metric inspired our ten-dimensional ans\"atze. 

We have argued that, if the integration constants are chosen appropriately, our wrapped brane solutions are dual to a slice of the Coulomb branch of SYM with ${\cal N}=(2,2)$ or ${\cal N}=(1,1)$ supersymmetry. Moreover, we have studied the deformation induced by the addition of unquenched flavor in the limit in which $N_f$ is large and $N_c/N_f$ remains finite. In this case it is justified to consider a continuous distribution of flavor branes smeared over their transverse angular directions. This brane distribution induces a violation of the Bianchi identity of the RR three-form, which can  be accounted for by modifying  the ansatz of $F_3$.  This modification of $F_3$ changes the BPS equations in such a way that they imply the equations of motion  with extended sources. 

In our study of the ${\cal N}=(1,1)$ case with flavor we have developed a formalism which does not require the precise knowledge of the family of flavor brane embeddings that make up the smeared distribution. Clearly, this formalism could be used in other backgrounds, such as the one in \cite{Gomis:2001aa, Gauntlett:2001ur} (see \cite{Gaillard:2008wt} for an attempt to add backreacting flavor to this case). Moreover, this approach could also be useful to study the unquenched background in  the Higgs branch of the setups analyzed in refs. \cite{Angel,Arean:2008az,Ramallo:2008ew}.

We have studied some implications of our gravity solutions in the dual gauge theories. However, notice that this study is limited, among other things, by the bad UV behavior of the background, a problem that is common to all gravity solutions corresponding to D5-branes. The situation would clearly improve if one considers, instead, duals of two-dimensional gauge theories constructed from D3-branes  wrapped on a two-cycle.  The corresponding background with eight supersymmetries was obtained and analyzed in ref. 
\cite{Arean:2008az}. The solution with four supersymmetries can be obtained by wrapping the D3-branes along a two-cycle of a Calabi-Yau threefold \cite{Maldacena:2000mw}.

Another interesting problem for future work would be trying to find a supergravity solution  dual to a two-dimensional gauge theory with just one supersymmetry. This background would be the two-dimensional analogue of the one of refs. \cite{CV, MN} for $d=4$ or  that of refs. \cite{Chamseddine:2001hk, Maldacena:2001pb, Schvellinger:2001ib} for $d=3$. The natural setup that would lead to a supergravity solution of this sort would be a configuration of D5-branes wrapping a Cayley four-cycle of a manifold of $Spin(7)$ holonomy.

The understanding of the deformation introduced by backreacting  flavor to the gravity solutions is a very important step in order to approach the gauge/gravity correspondence to phenomenology. Two-dimensional field theories have been always considered as a good theoretical laboratory where one can develop and test new techniques which could eventually shed light on the  study of  realistic four-dimensional theories. We hope that this will be also the case for the formalism developed here.

\section*{Acknowledgments}

We are very grateful to Paolo Merlatti and Carlos N\'u\~nez for collaborating in the initial stages of this project. We also thank Jerome Gaillard, \'Angel Paredes, Johannes Schmude, Jonathan Shock and Dimitrios Zoakos for very useful discussions. The  work of EC and AVR was funded in part by MEC and  FEDER  under grant
FPA2008-01838,  by the Spanish Consolider-Ingenio 2010 Programme CPAN (CSD2007-00042) and by Xunta de Galicia (Conseller\'\i a de Educaci\'on and grant PGIDIT06PXIB206185PR). EC is supported by a spanish FPU fellowship.

\vskip 1cm
\renewcommand{\theequation}{\rm{A}.\arabic{equation}}
\setcounter{equation}{0}
\appendix

\section{Wrapped D5-branes from gauged supergravity}
\label{gaugedsugra7d}
We shall begin by writing the bosonic part of the lagrangian of the $SO(4)$ gauged
supergravity \cite{salam,it1} where we will be looking for SUSY configurations. The field content of this theory includes the  seven-dimensional metric $g_{\mu\nu}$, a gauge field $A^{ij}$ in the adjoint representation of $SO(4)$ (the corresponding field strength will be denoted by $F^{ij}$) and ten scalar fields arranged in a symmetric $4\times 4$  matrix
$V_{ij}$ (in the following we will not distinguish between upper and lower latin indices).  From \cite{salam,it1}, one can read the  lagrangian for these bosonic fields:
\beq
{\cal L}=\sqrt{-g}\left[R-P_{\mu\;ij}\,P^\mu_{ij}-P_{\mu\,ii}\,
P^{\mu}_{jj}
-{1\over2}(V_{ik}\,V_{jl}\,F_{\mu\nu}^{kl})^2+{m^2\over2}\left(T^2
-2T_{ij}\,T_{ij}\right)\right]\,,
\label{7dso4lag}
\eeq
where the kinetic term for the scalars $P_\mu$ can be read from:
\beq
V^{-1}_{ik}\,\partial_\mu\,V_{kj}+2m\,V^{-1}_{ik}\,A^{kl}_\mu\,V_{lj}=P_{\mu\;(ij)}+
Q_{\mu\;[ij]}\,,
\eeq
while $T_{ij}$ is constructed as:
\beq
T_{ij}=V^{-1}_{ik}\,V^{-1}_{kj}\,,
\label{7dtij}
\eeq
and $T=T_{ij}\delta^{ij}$. The covariant derivative acting on the spinors takes the form:
\beq
{\cal
D}_\mu\,\psi=\left(\partial_\mu+{1\over4}Q_\mu^{ij}\,\Gamma_{ij}+{1\over4}\,
\omega^{ab}\,\gamma_{ab}\right)\psi\,.
\label{7dcovdvtve}
\eeq
The SUSY variations can be written as:
\bear
\delta\hat\psi_\mu&=&\left[{\cal
D}_\mu-{1\over4}\,\gamma_\mu\,\gamma^\nu\,V^{-1}_{ij}\,\partial_\nu\,V_{ji}+{1\over4}
\hat F^{ij}_{\mu\nu}\,\Gamma^{ij}\,\gamma^\nu\right]\,\epsilon=0\,,\rc\rc
\delta\left(\Gamma^k\,\lambda_k\right)&=&\bigg[{m\over2}\left(T_{kj}-{1\over5}T\,
\delta_{kj}\right)\Gamma^k\,\Gamma^j+{1\over2}\gamma_\mu\,P^\mu_{kj}\,\Gamma^k\,
\Gamma^j+\rc\rc
&&+{1\over16}\gamma^{\mu\nu}\left(\Gamma^k\Gamma^{ij}\Gamma^k-{1\over5}
\Gamma^{ij}\right)\hat F_{\mu\nu}^{ij}\bigg]\,\epsilon\,,\quad({\rm no\;
sum\;in\;}k)\,, 
\label{7dsusyvar}
\eear
where we have defined:
\beq
\hat F^{ij}_{\mu\nu}=V_{ik}\,V_{jl}\,F_{\mu\nu}^{kl}\,.
\eeq

In order to obtain a ten-dimensional solution corresponding to an NS5-brane  one must perform the uplift developed in \cite{uplift}\footnote{ In
order to match the formulas there we apply the following
identifications:
\beq
\hat A^{ij}=2A^{ij}\;,\qquad g=m\,,
\label{cvtnot}
\eeq
where $\hat A$ is the one-form gauge field of \cite{uplift} and $g$ is
the coupling constant there.}. Let us first write the expression for the
uplifted metric and the dilaton in string frame:
\bear
&&e^\Phi=\Delta^{-{1\over2}}\,Y^{3\over4}\,\,,\rc\rc
&&ds^2=Y^{1\over2}\left[ds^2_7+{1\over
m^2}\,\Delta^{-1}\,T^{-1}_{ij}\,D\mu^i\,D\mu^j\right]\,,
\label{upliftmetricstr}
\eear
where the $\mu^i$ ($i=1,\cdots 4$) are coordinates of the external $S^3$ that must satisfy the constraint:
\beq
\sum_{i=1}^{4}\,(\mu^i)^2\,=\,1\,\,.
\label{sum-mus}
\eeq
In eq. (\ref{upliftmetricstr}) the quantities $\Delta$ and $Y$ are given by:
\beq
\Delta=T_{ij}\,\mu^i\,\mu^j\,,\quad Y=\det(T_{ij})\,,
\label{updelta}
\eeq
and the gauge-covariant exterior derivative $D$ is defined as:
\beq
D\mu^i=d\mu^i+{2m}\,A^{ij}\,\mu^j\,,\quad
D\,T_{ij}=dT_{ij}+2m\left(A^{ik}\,T_{kj}+A^{jk}\,T_{ki}\right)\,.
\eeq
 The corresponding NSNS three-form $H_{3}$  of the ten-dimensional supergravity is given by:
\bear
&& H_{3}={1\over6}\,\epsilon_{i_1 i_2 i_3
i_4}\Big[m^{-2}\,U\,
\Delta^{-2}\,D\mu^{i_1}\wedge D\mu^{i_2}\wedge
D\mu^{i_3}\,\mu^{i_4}-\rc\rc
&&-3m^{-2}\Delta^{-2}\,D\mu^{i_1}\wedge D\mu^{i_2}\wedge
DT_{i_3j}\,T_{i_4k}\,
\mu^j\,\mu^k-3m^{-1}\Delta^{-1}F_{(2)}^{i_1 i_2}\wedge
D\mu^{i_3}\,T_{i_4j}\,\mu^j
\Big]\,\,,\qquad\qquad
\label{uplift3form}
\eear
where $U$ is defined as:
\beq
U=2\,T_{ik}\,T_{jk}\,\mu^i\,\mu^j-\Delta\,T_{ii}\,.
\label{upliftU}
\eeq
Finally, in order to get a solution corresponding to a D5-brane one has to perform an $S$-duality which, for the type of backgrounds we are considering  corresponds to just  flipping   the sign  of the dilaton, $\Phi\to-\Phi$, and relabeling  the NSNS three-form $H_3$ as the RR three-form $F_3$, while the Einstein frame metric is not changed. 

\subsection{Branes wrapping $S^2\times S^2$}
Let us consider the following  ansatz for the seven-dimensional metric:
\beq
ds_7^2=e^{2f}\left(dx_{1,1}^2+dr^2\right)+{e^{2g}\over
m^2}\left(d\theta_1^2+\sin^2\theta_1\,d\phi_1^2+d
\theta_2^2+\sin^2\theta_2\,d\phi_2^2\right)\,,
\label{7ds22metric}
\eeq
where $f$ and $g$ are functions of the radial coordinate $r$. In order to implement the topological twist in our setup, we will adopt  the following ansatz for the  one-form gauge field:
\beq
A^{12}={1\over2m}\,\left(\cos\theta_1\,d\phi_1+\cos\theta_2\,d\phi_2\right)\,,
\label{7ds22gaugef}
\eeq
while we will assume that the matrix $V_{ij}$ can be represented in terms of the scalar fields $\lambda=\lambda(r)$ and $x=x(r)$ as:
\beq
V_{ij}=e^{\lambda\over8}\,(e^{x\over2},e^{x\over2},e^{-{x\over2}},e^{-{x\over2}})\,.
\label{7ds2s2vij}
\eeq 
Then,  from the definition (\ref{7dtij}) we get that $T_{ij}$ is:
\beq
T_{ij}=V^{-1}_{ik}\,V^{-1}_{kj}=e^{-{\lambda\over4}}\,(e^{-x},e^{-x},
e^{x},e^{x})\,.
\label{7ds22tij}
\eeq
Finally, $P_{\mu\;ij}$ and $Q_{\mu\;ij}$ take the form:
\bear
&&P_{\mu\;ij}=\delta_{\mu\,r}\left[{\lambda'\over8}(1,1,1,1)+{x'\over2}(1,1,-1,-1)
\right]\,,  \rc \rc
&&Q_{\mu\;ij}=2m\,A^{ij}_\mu\,.\label{7ds22qmu}
\eear
In (\ref{7ds22qmu}) the prime denotes derivative with respect to the radial coordinate $r$.

\subsubsection{SUSY variations}
Let us now impose that our ansatz corresponds to a supersymmetric solution, which is equivalent to demanding that the right-hand side of the supersymmetry variations (\ref{7dsusyvar}) vanish for some Killing spinors satisfying certain projections. These projections are:
\beq
\gamma_{\theta_1\phi_1}\,\epsilon=\gamma_{\theta_2\phi_2}\,\epsilon\,=\,
\Gamma_{12}\,\epsilon\,,\qquad\qquad
\gamma_r\,\epsilon=\epsilon\,,
\label{7ds22projs}
\eeq
where $\gamma_{\theta_i}$,  $\gamma_{\phi_i}$ and $\gamma_r$ are constant Dirac matrices along the corresponding frame directions of the seven-dimensional metric (\ref{7ds22metric}). By analyzing the vanishing of the supersymmetric variations one readily proves that one can take:
\beq
\lambda\,=\,4f\,\,.
\eeq
Moreover, if we define $h$ as:
\beq
h\,=\,g-f\,\,,
\label{h-def}
\eeq
one arrives, after some algebra, at the following system of first-order differential equations:
\bear
&&h'+{m\over2}\,e^{-2h+x}=0\,,\rc \rc
&&x'-2m\sinh{x}-m\,e^{-2h+x}=0\,,
\rc \rc
&&f'+{2m\over5}\cosh{x}-{m\over5}\,e^{-2h+x}=0\,,
\label{7ds22bps}
\eear
where, as before,  the prime denotes derivative with respect to $r$. Let us now define a new radial variable $z$ as:
\beq
z\,=\,e^{2h}\,\,.
\eeq
From the first equation in (\ref{7ds22bps}), one gets that the two radial variables $r$ and $z$ are related as:
\beq
dz=-m\,e^x\,dr\,.
\label{7ds22rad}
\eeq
Moreover, if the dot denotes derivatives with respect to $z$, one arrives at the following BPS system:
\bear
&&\dot x-e^{-2x}+1+{1\over z}=0\,,\rc \rc
&&\dot f-{1\over5}e^{-2x}+{1\over5z}-{1\over5}=0\,.
\label{7ds22bps-z}
\eear
The first equation in (\ref{7ds22bps-z}) can easily be integrated yielding:
\beq
e^{2x}={1-2z+2z^2+c\,e^{-2z}\over2z^2}\,,
\label{7ds22xsol}
\eeq
with $c$  being an integration constant. Notice that $e^{2x}$ is just the function that was defined in section \ref{22sol} (see eq. (\ref{e2x})).   Moreover, 
from the second equation in  (\ref{7ds22bps-z}) one gets:
\beq
e^{5f}=e^{2z}\,e^{x(z)}=e^{2z}\left({1-2z+2z^2+c\,e^{-2z}\over2z^2}
\right)^{1\over2}\,.
\label{7ds22fsol}
\eeq
\subsubsection{Uplift to ten dimensions}
Let us now write the metric in ten dimensions by using the uplifting formula (\ref{upliftmetricstr}). With this purpose let us first parametrize the coordinates $\mu^i$ $(i=1,\dots,4)$ that span the external $S^3$. They must satisfy the constraint written  in (\ref{sum-mus}), which can be solved as:
\bear
&&\mu^1=\cos\tilde\theta\cos\tilde\phi_1\,,\quad
\mu^2=\cos\tilde\theta\sin\tilde\phi_1\,,\rc\rc
&&\mu^3=\sin\tilde\theta\cos\tilde\phi_2\,,\quad
\mu^4=\sin\tilde\theta\sin\tilde\phi_2\,,
\label{7dmuipar1}
\eear
where $0\leq\tilde\theta\leq\pi/2\,$, $0\leq\tilde\phi_{1,2}\leq2\pi\,$\,. Let us write the ten-dimensional metric in these angular coordinates for the D5-brane in the string frame. In order to find this result, we will have to rewrite the metric (\ref{upliftmetricstr}) in the Einstein frame and then we must apply an S-duality transformation. After this process, the final result for the string frame metric is:
\bear
ds^2&=&e^{\Phi}\,\Big[\,dx_{1,1}^2+{e^{-2x}\over m^2}\,dz^2
+{z\over m^2}\left(d\theta_1^2+\sin^2\theta_1
\,d\phi_1^2+d\theta_2^2+\sin^2\theta_2\,d\phi_2^2\right)+{1
\over m^2}\,d\tilde\theta^2+\rc\rc&&+
{e^x\over m^2\,\Sigma}\,\cos^2\tilde\theta\left(d\tilde\phi_1-
\cos\theta_1 d\phi_1-\cos\theta_2 d\phi_2\right)^2+
{e^{-x}\over m^2\,\Sigma}\,\sin^2\tilde\theta\,d\tilde\phi_2^2\,\Big]
\,,
\label{uplmetrics2s2str}
\eear
with the quantity $\Sigma$ being given by:
\beq
\Sigma=e^{-x}\,\cos^2\tilde\theta+e^{x}\,\sin^2\tilde\theta\,.
\label{omegas22}
\eeq
In (\ref{uplmetrics2s2str}) $\Phi$ is the dilaton of the D5-brane solution, whose explicit solution is:
\beq
e^{2\Phi}\,=\,e^{2z+x}\,\,\Sigma\,=\,e^{2z}\,\,\big(
\cos^2\tilde\theta\,+\,e^{2x}\,\sin^2\,\tilde\theta\,\big)\,\,.
\label{22-gaugedsugra-dilaton}
\eeq
Similarly, by using the uplifting formula (\ref{uplift3form}) for the three-form, we can write the corresponding RR field strength, namely:
\bear
F_{3}=&&-{2\sin\tilde\theta\cos\tilde\theta\over m^2\,\Sigma^2}\left(
d\tilde\theta+\sin\tilde\theta\cos\tilde\theta\,\dot x\,dz\right)\wedge
\left(d\tilde\phi_1-\cos\theta_1\,d\phi_1-\cos\theta_2\,d\phi_2\right)
\wedge d\tilde\phi_2-\rc\rc &&-{e^x\sin^2\tilde\theta\over m^2\,\Sigma}
\left(\sin\theta_1\,d\theta_1\wedge d\phi_1+\sin\theta_2\,d\theta_2
\wedge d\phi_2\right)\wedge d\tilde\phi_2\,.
\label{10d3forms22}
\eear
In order to match the ten-dimensional ansatz (\ref{D5-newmetric}), let us introduce two new radial variables $\rho$ and $\sigma$, defined as:
\beq
\rho\,=\,\sin\tilde\theta\, e^{z}\,\,,\qquad\qquad
\sigma\,=\,z\,\cos\tilde\theta\,e^{z+x}\,\,.
\label{change-variables}
\eeq
It is now straightforward to verify that the metric (\ref{uplmetrics2s2str}) can be written in the form (\ref{D5-newmetric}) if we identify the angles $(\tilde\phi_1, \tilde\phi_2)$ of the gauged supergravity approach with the $(\psi, \chi)$ variables used in the ten-dimensional analysis of section \ref{22section} by means of the relation:
\beq
\tilde\phi_1\,=\,2\pi-\psi\,\,,\qquad\qquad
\tilde\phi_2\,=\,2\pi-\chi\,\,.
\eeq
Notice also that the implicit solution (\ref{implict-22}) is obtained from the definition  (\ref{change-variables}) just by using that $\sin^2\tilde\theta+\cos^2\tilde\theta\,=\,1$.  Moreover, the dilaton $\Phi$ written in (\ref{22-gaugedsugra-dilaton}) reduces to the one in (\ref{g-gaugedsugra}) once  it is expressed in the new coordinates. One can also check that $F_3$, as given by (\ref{10d3forms22}), can be written as in (\ref{F3ansatz}), with the  function $g$ being identified with:
\beq
m^2\,g\,=\,{e^{x}\sin^2\tilde\theta\over \Sigma}\,=\,
{e^{2x}\sin^2\tilde\theta\over \cos^2\tilde\theta+e^{2x}\sin^2\tilde\theta}\,\,.
\label{m2g}
\eeq
After using the change of variables (\ref{change-variables}),  the expression of $g$ written in (\ref{m2g}) becomes the one displayed in (\ref{g-gaugedsugra}). 

\begin{figure}
\centering
\includegraphics[width=0.4\textwidth]{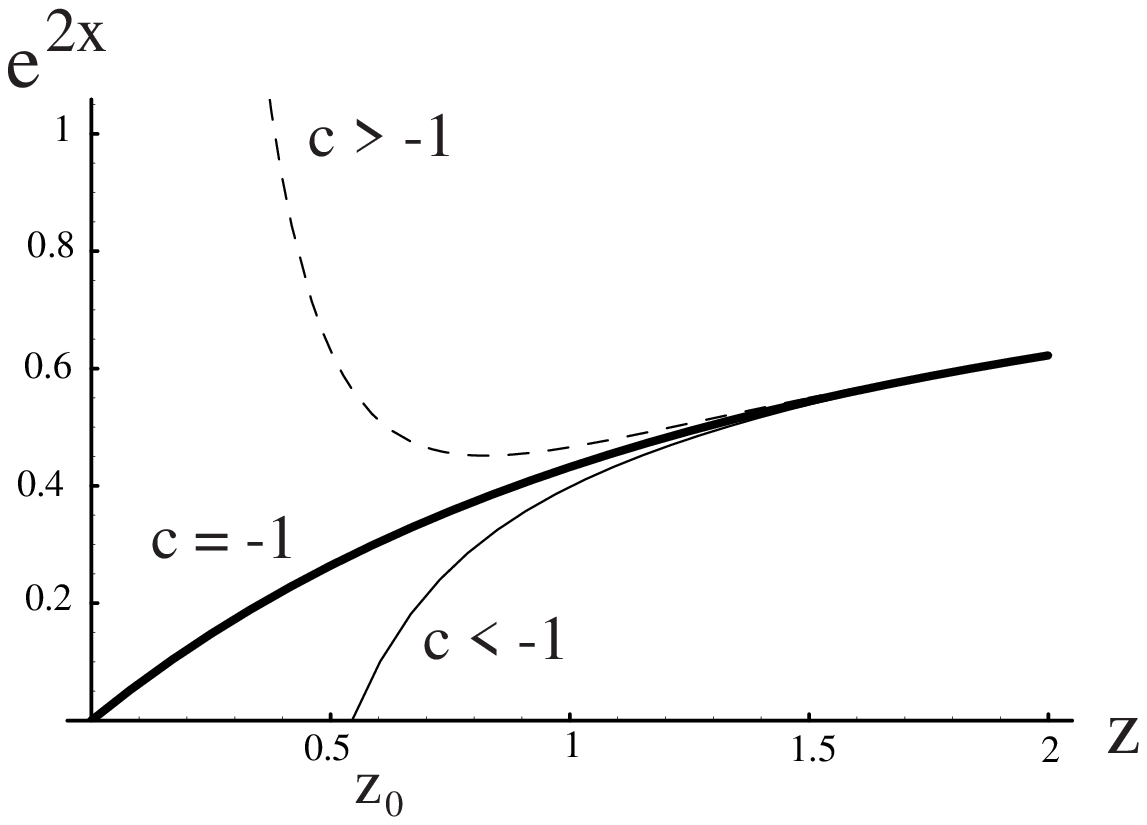} \hfill
\includegraphics[width=0.4\textwidth]{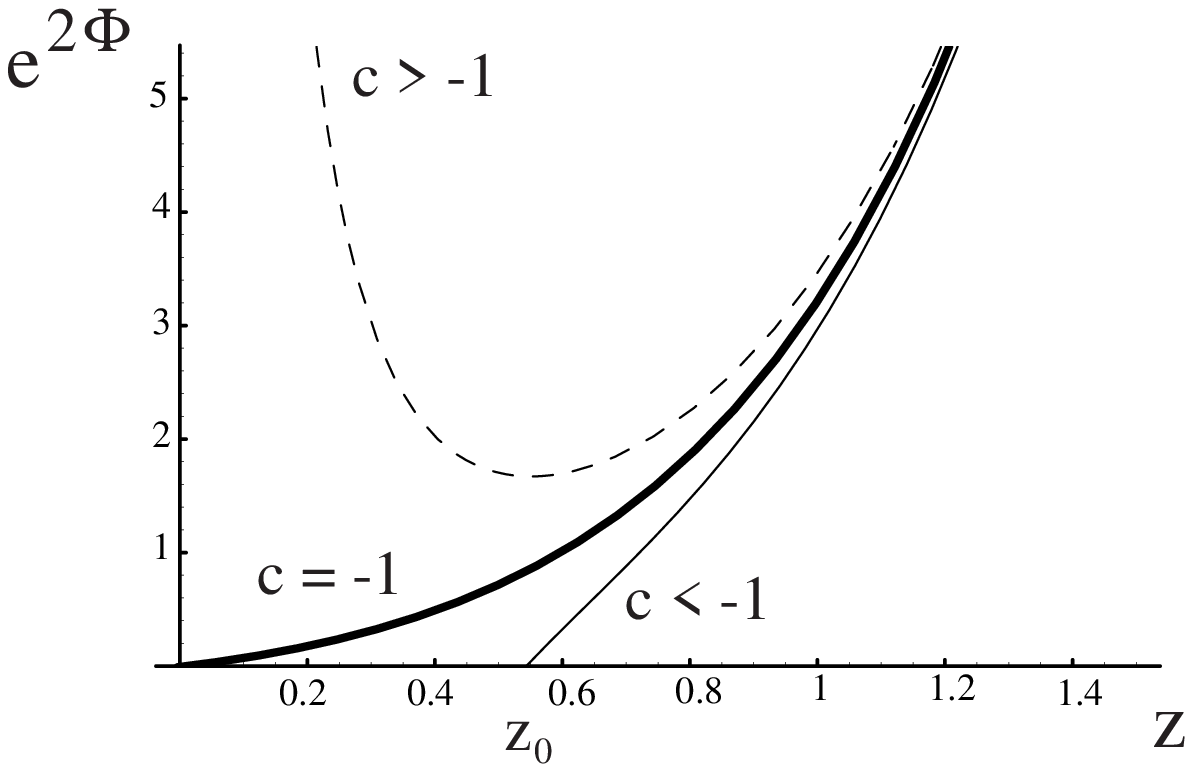}
\caption{On the left we plot $e^{2x}$, as given by (\ref{7ds22xsol}),  for the three ranges of $c$. On the right we represent the corresponding values of $e^{2\Phi}$ for $\tilde\theta=\pi/2$ and different values of $c$.   }
\label{e2x-2phi-22}
\end{figure}

The behavior and interpretation of the solution changes significantly with the value of the integration constant $c$ in (\ref{7ds22xsol}) (see figure \ref{e2x-2phi-22}). 
On the one hand, $e^{2x}$ tends to one at $z\to
\infty$ for any $c$, while for $z\to0$, $e^{2x}$  diverges for $c>-1$ and vanishes for $c=-1$. In contrast, when $c<-1$, $e^{2x}$ vanishes at some finite value $z_0>0$ and becomes negative for $z<z_0$. We would like to argue, following closely a similar analysis in ref. \cite{Gauntlett:2001ps}, that this last $c<-1$ case is the one that has a cleaner physical interpretation.  With this purpose, let us look at the form of the dilaton (\ref{22-gaugedsugra-dilaton}) for $\tilde\theta=\pi/2$.  As shown in figure \ref{e2x-2phi-22}, the dilaton either diverges or vanishes at some some value of $z$, making the ten-dimensional  metric (\ref{uplmetrics2s2str}) singular at that point (see also (\ref{D5-newmetric})). According to the criteria of ref. \cite{Maldacena:2000mw} a singularity is ``good" if the norm of the time-like Killing vector field in the Einstein frame is decreasing as one approaches the singularity. This norm is given by the $G_{tt}$ component of the Einstein frame metric, which in our case is simply  $e^{\Phi/2}$. Looking at the behavior of the dilaton in figure \ref{e2x-2phi-22} we immediately conclude that the singularity is good only for $c<-1$, as claimed in the main text.

\subsubsection{UV and IR limits}
Let us now discuss in detail the behavior of the solution in the UV and IR regions. The UV limit of the metric and dilaton is obtained by taking $z\to\infty$. In this limit one readily verifies that $e^{2x}, \Sigma\to 1$ and, therefore, the metric (\ref{uplmetrics2s2str}) and the dilaton (\ref{22-gaugedsugra-dilaton}) become:
\bear
&&ds^2\,\approx\,e^{\Phi}\,\Bigg[
dx_{1,1}^2
+{z\over m^2}\left(d\theta_1^2+\sin^2\theta_1
\,d\phi_1^2+d\theta_2^2+\sin^2\theta_2\,d\phi_2^2\right)+{dz^2\over m^2}\,+
\qquad\qquad\qquad\qquad\rc\rc
&&\qquad\qquad
+{1\over m^2}\,\Big[\,\,d\tilde\theta^2\,+
\,\sin^2\tilde\theta\,d\tilde\phi_2^2\,+\,
\cos^2\tilde\theta\left(d\tilde\phi_1-
\cos\theta_1 d\phi_1-\cos\theta_2 d\phi_2\right)^2
\,\Big]\Bigg]\,\,,\rc\rc
&&e^{2\Phi}\,\approx\,e^{2z}\,\,.
\eear
This form of the metric and dilaton do indeed correspond to a stack of  D5-branes with worldvolume $\mathbb{R}^{1,1}\times S^2\times S^2$, with the appropriate twisting. 

Let us next consider the IR limit. As explained in the previous subsection, the IR behavior of the solution strongly depends on the value of the integration constant $c$. From now on we will concentrate on the case $c<-1$. In this case the space ends at  $z=z_0$, where $z_0$ is such that $e^{2x(z)}$ vanishes and is given by the solution of the equation:
\beq
e^{-2z_0}\,=\,-{1\over c}\,\big[\,1\,-\,2z_0\,+2z_0^2\,\big]\,\,.
\eeq
Near $z=z_0$ the function $e^{2x(z)}$ can be expanded as:
\beq
e^{2z}\,\approx 2(z-z_0)\,\,.
\eeq
Notice that, in the plane of the ten-dimensional coordinates $(\rho,\sigma)$, $z=z_0$ is not a point but rather the line $(\sigma=0, 0\le \rho\le e^{z_0})$, parametrized by the angle $\tilde\theta$ (see  (\ref{change-variables})).  The dilaton becomes singular for $z=z_0$ and $\tilde\theta={\pi\over 2}$. Let us expand the solution near this point. For this purpose we introduce new coordinates $(\zeta, \psi)$, defined as:
\beq
\sqrt{2(z-z_0)}\,=\,\sqrt{\zeta} \,\sin \Big({\psi\over 2}\Big)\,\,,\qquad\qquad
{\pi\over 2}\,-\,\tilde\theta\,=\,\sqrt{\zeta}\,\cos \Big({\psi\over 2}\Big)\,\,.
\eeq
Notice that in these coordinates the singularity is just located at $\zeta=0$. One can check that, at first order in $\zeta$:
\beq
\Sigma\,\approx\,{\sqrt{\zeta}\over \sin \big({\psi\over 2}\big)}\,\,.
\eeq
Using this result one can readily verify that, near the singularity, the metric and dilaton can be written as:
\bear
&&ds^2\,\approx\,e^{\Phi}\,\Bigg[
dx_{1,1}^2
+{z_0\over m^2}\left(d\theta_1^2+\sin^2\theta_1
\,d\phi_1^2+d\theta_2^2+\sin^2\theta_2\,d\phi_2^2\right)+
\qquad\qquad\qquad\qquad\rc\rc
&&+{1\over 4 m^2 \zeta}\,\Big[d\zeta^2\,+\,\zeta^2\,d\psi^2\,+\,
\zeta^2\,\sin^2\psi\,
\left(d\tilde\phi_1-
\cos\theta_1 d\phi_1-\cos\theta_2 d\phi_2\right)^2+\,d\tilde\phi_2^2\,\Big]\Bigg]\,\,,\rc\rc
&&e^{2\Phi}\,\approx e^{2z_0}\,\zeta\,\,,
\eear
which is precisely the form of the metric and dilaton for a stack of D5-branes located at 
$\zeta=0$ (or $z=z_0$, $\tilde\theta=\pi/2$) and smeared on a circle along the $\tilde\phi_2$ direction (notice that the space transverse to the branes is (fibered) $\mathbb{R}^3\times S^1$ and $1/\zeta$ is a harmonic function in this space). This result implies that our solutions can be interpreted as the gravity duals of a particular slice of the Coulomb branch of the ${\cal N}=(2,2)$ SYM theory, namely the one that is generated when the wrapped D5-branes are moved along an angular direction which is transverse both to the $CY_3$ and the D5-brane worldvolume.

\subsection{Fivebranes wrapping a co-associative four-cycle}
Let us now  try to find a solution of $SO(4)$ seven-dimensional gauged supergravity
corresponding to a fivebrane wrapping a co-associative four-cycle as described in
\cite{gaun}. The ansatz for the 7d metric is:
\beq
ds^2=e^{2f}\left(dx_{1,1}^2+dr^2\right)+{e^{2g}\over
m^2}\,d\Omega_4^2
\,,\label{7dS4metric}
\eeq
where $d\Omega_4^2$ is the metric of a four-sphere,  which was defined in  (\ref{dOmega4}) in terms of the $SU(2)$ left-invariant one-forms $\omega^i$ and the coordinate $\xi$.  We shall
then consider the following one-form frame:
\bear
&&e^{x^\mu}=e^f\,dx^\mu\;,\qquad e^r=e^f\,dr\;,\qquad
e^4={e^g\over m}\,{2\over1+\xi^2}\,d\xi\,,\rc\rc &&
e^i={e^g\over m}\,{\xi\over1+\xi^2}\,\omega^i\;,\quad(i=1,2,3)
\,.
\label{7ds4vielbein}
\eear
Following \cite{gaun}, in order to implement the topological twist, we shall take the one-form gauge field to be proportional to the anti-self dual part of the spin connection along the four-sphere, namely:
\beq
A^{12}={1\over 2m}\,{\xi^2\over 1+\xi^2}\,\,\omega^1\,,\qquad
A^{13}=-{1\over 2m}\,{\xi^2\over 1+\xi^2}\,\,\omega^2\,,\qquad
A^{23}=-{1\over 2m}\,{\xi^2\over 1+\xi^2}\,\,\omega^3\,.
\label{7ds4a1}
\eeq
Moreover, we will take the following ansatz for the scalars:
\beq
V_{ij}=e^{\lambda\over8}\,(e^{x\over2},e^{x\over2},e^{x\over2},e^{-{3\over2}x})\,,
\label{7ds4vij}
\eeq 
where $\lambda$ and $x$ are fields that depend only on the radial coordinate $r$. 
Hence, $T_{ij}$ is given by:
\beq
T_{ij}=V^{-1}_{ik}\,V^{-1}_{kj}=e^{-{\lambda\over4}}\,(e^{-x},e^{-x},
e^{-x},e^{3x})\,,
\label{7ds4tij}
\eeq
whereas $P_{\mu\;ij}$ and $Q_{\mu\;ij}$ take the form:
\beq
P_{\mu\;ij}=\delta_{\mu\,r}\left[{\lambda'\over8}(1,1,1,1)+{x'\over2}(1,1,1,-3)
\right]\,,\qquad\qquad
Q_{\mu\;ij}=2m\,A^{ij}_\mu\,,
\eeq
where the prime denotes derivative with respect to $r$. 

\subsubsection{BPS equations}

The SUSY variations will result in a system of first-order differential equations
upon imposing the following set of four independent projections:
\beq
\gamma_{1234}\,\epsilon=\epsilon\,,\quad
\gamma_{12}\,\epsilon=\Gamma_{23}\,\epsilon\,,\quad
\gamma_{13}\,\epsilon=\Gamma_{31}\,\epsilon\,,\quad
\gamma_r\,\epsilon=\epsilon\,,
\label{7ds4indprojs}
\eeq
where the indices of the $\gamma$'s refer to the frame (\ref{7ds4vielbein}). From the analysis of the supersymmetry variations of the gravitino and dilatino one readily concludes that, also in this case, one can take $\lambda=4f$. The system of first-order differential equations one arrives at is:
\bear
&&h'+{3m\over2}\,e^{-2h+x}=0\,,\rc \rc
&&x'+{m\over2}\left(e^{-x}-e^{3x}\right)-m\,e^{-2h+x}=0\,,
\rc \rc
&&f'+{m\over10}\left(3e^{-x}+e^{3x}\right)-{3m\over5}\,e^{-2h+x}=0\,,
\label{7ds4bps}
\eear
where, as in (\ref{h-def}),  we have defined $h=g-f$. After defining a new radial
variable $z=e^{2h}$, one gets:
\bear
&&\dot x+{1\over3}\sinh{2x}+{1\over3z}=0\,, \rc\rc
&&\dot f-{1\over30}\left(3e^{-2x}+e^{2x}\right)+{1\over5z}=0\,,
\label{7ds4bps-xf}
\eear
with the dot denoting derivative with respect to $z$. Moreover, it follows from the first equation in (\ref{7ds4bps}) that
the two radial variables $z$ and $r$ are related as:
\beq
dz=-3m\,e^x\,dr\,.
\label{7ds4rad}
\eeq

It turns out that the system  (\ref{7ds4bps-xf})  can be solved analytically in terms of Bessel functions. First, for $x(z)$ one gets the following family of
solutions:
\beq
e^{2x}={I_{-{5\over6}}\,({z\over3})+c\,I_{{5\over6}}\,({z\over3})
\over I_{1\over6}\,({z\over3})+c\,I_{-{1\over6}}\,({z\over3})}\,,
\label{s4xsol}
\eeq
where $c$ is an integration constant. Note that $x(z)$ is just the function defined in (\ref{exz}).  Then, plugging this solution into the second equation in  (\ref{7ds4bps-xf}) it yields:
\beq
e^f=\left[\left(I_{-{5\over6}}\left({z\over3}\right)+c\,
I_{{5\over6}}\left({z\over3}\right)\right)^3\left(
I_{1\over6}\,({z\over3})+c\,I_{-{1\over6}}\,({z\over3})
\right)z^{2\over3}\right]^{1\over10}\,.
\label{s4fsol}\,
\eeq
where $c$ is the same real constant  as in (\ref{s4xsol}). 

\subsubsection{Ten-dimensional solution}

Let us now uplift the solution just found to ten-dimensions. In this case it is more useful to use the following parametrization of the $\mu^i$'s: 
\bear
&&\mu^1=\sin{\psi}\,\sin{\theta}\,\cos{\phi}\,,\quad
\mu^2=\sin{\psi}\,\sin{\theta}\,\sin{\phi}\,,\rc\rc
&&\mu^3=\sin{\psi}\,\cos{\theta}\,,\quad
\mu^4=\cos{\psi}\,,
\label{7dmuipar2}
\eear
with $0\leq\theta,\psi\leq\pi\,$, and $0\leq\phi\leq2\pi\,$. Using these coordinates, the quantities $Y$ and $\Delta$ defined in (\ref{updelta}) take the form:
\beq
Y=e^{-4f}\,\,,\qquad\qquad
\Delta\,=\,e^{x-f}\,\,\Theta\,\,,
\eeq
where $\Theta$ is defined as:
\beq
\Theta=e^{-2x}\,\sin^2\psi+e^{2x}\,\cos^2\psi\,. 
\label{Theta}
\eeq
Let us now write the complete ten-dimensional solution for a D5-brane. After applying an S-duality transformation to the result obtained from (\ref{upliftmetricstr}), we get that the dilaton is given by:
\beq
e^{2\Phi}\,=\,e^{5f+x}\,\,\Theta\,\,,
\label{dilaton-11-gauged}
\eeq
while the string frame metric becomes:
\beq
ds^2=e^{\Phi}\,\Bigg[\,
dx_{1,1}^2+{e^{-2x}\over9\,m^2}\,dz^2+{z\over
m^2}\,d\Omega^2_4\,
+\,{e^{-2x}\over m^2}\,d\psi^2\,
+\,{\sin^2\psi\over m^2 \Theta}\left[(E^1)^2+(E^2)^2\right]
\,\Bigg]\,,
\label{uplmetrics4nice}
\eeq
where $E^1$ and $E^2$ are the one-forms defined in (\ref{E1E2}). 
The calculation of the RR three-form $F_3$ from (\ref{uplift3form}) is straightforward but rather tedious. The final result can be compactly written in terms of the two-form potential $C_2$ (see (\ref{F3-C2})), which can be recast in terms of two functions $g_1$ and $g_2$ as in (\ref{C2-ansatz}). The value of these functions is rather simple, namely:
\beq
g_1\,=\,{1\over m^2}\,\,\Big(\,{\sin\psi\cos\psi\over \Theta}\,e^{2x}\,-\,\psi\,\Big)\,\,,
\qquad\qquad
g_2\,=\,{\psi\over m^2}\,\,.
\label{g1-g2-gauged}
\eeq

In order to make contact with the ten-dimensional approach of section \ref{11-10d}, let us now define the new set of variables:
\beq
\sigma\,=\,z^{{2\over 3}}\,e^{{5\over 2}f\,+\,{x\over 2}}\,\sin\psi
\,\,,\qquad\qquad
\rho\,=\,e^{{5\over 2}f\,-\,{3x\over 2}}\,\cos\psi\,\,.
\label{sigma-rho-def}
\eeq
In terms of the dilaton $ \Phi$, the above relations can be written as:
\beq
\sigma\,=\,{z^{{2\over 3}}\,e^{ \Phi}\over \sqrt{\Theta}}\,\sin\psi
\,\,,\qquad\qquad
\rho\,=\,\,{e^{ \Phi-2x}\over  \sqrt{ \Theta}}
\,\cos\psi\,\,,
\eeq
where we have taken into account the relation (\ref{dilaton-11-gauged}). By using (see (\ref{7ds4bps-xf})):
\beq
{5\over 2}\dot f\,+\,{\dot x\over 2}\,=\,{e^{-2x}\over 3}\,-\,{2\over 3z}\,\,,\qquad\qquad
{5\over 2}\dot f\,-\,{3\dot x\over 2}\,=\,{1\over 3} e^{2x}\,\,,
\eeq
one can prove straightforwardly that:
\bear
&&d\sigma\,=\,{e^{\Phi}\over \sqrt{\Theta}}
\,\,z^{{2\over 3}}\,\Big[\,{e^{-2x}\over 3}\sin\psi\,dz\,+\,\cos\psi\,d\psi\,\Big]\,\,,\rc\rc
&&d\rho\,=\,{e^{\Phi}\over \sqrt{\Theta}}
\Big[\,{\cos\psi\over 3}\,dz\,-\,\sin\psi\,e^{-2x}\,d\psi\,\Big]\,\,.
\label{jacobian}
\eear
The inverse of this relation is:
\bear
&&dz\,=\,{e^{-\Phi}\over \sqrt{\Theta}}\,\,\Big(\,3 z^{-{2\over 3}}\,\sin\psi\,d\sigma\,+\,
3 e^{2x}\,\cos\psi\,d\rho\,\Big)\,\,,\rc\rc
&&d\psi\,=\,{e^{-\Phi}\over \sqrt{\Theta}}\,\,
\Big(\, z^{-{2\over 3}}\,e^{2x}\,\cos\psi\,d\sigma\,-\,
 \sin\psi\,d\rho\,\Big)\,\,.
 \label{inverse-jacobian}
\eear
It is now immediate to verify that the metric (\ref{uplmetrics4nice}) can be written as in our ten-dimensional  ansatz (\ref{11-10d-ansatz}). Moreover, the gauged supergravity analysis provides a particular highly non-trivial solution of the BPS system (\ref{BPS-10d-unflavored}). In order to write this solution entirely in terms of the ten-dimensional variables $\rho$ and $\sigma$, let us notice that, after using (\ref{s4xsol}) and (\ref{s4fsol}),  we get:
\beq
e^{{5\over 2}f\,+\,{x\over 2}}\,=\,z^{{1\over 6}}\,\Big|\,
I_{-{5\over6}}\,({z\over3})+c\,I_{{5\over6}}\,({z\over3})\,\Big|\,\,\,,
\qquad\qquad
e^{{5\over 2}f\,-\,{3x\over 2}}\,=\,z^{{1\over 6}}\,\Big|\,
I_{{1\over6}}\,({z\over3})+c\,I_{-{1\over6}}\,({z\over3})\,\Big|\,\,\,.
\eeq
Thus, we can rewrite (\ref{sigma-rho-def}) as:
\beq
\sigma\,=\,z^{{5\over 6}}\,\Big|\,
I_{-{5\over6}}\,({z\over3})+c\,I_{{5\over6}}\,({z\over3})\,\Big|\,\sin\psi\,\,,
\qquad\qquad
\rho\,=\,z^{{1\over 6}}\,\Big|\,
I_{{1\over6}}\,({z\over3})+c\,I_{-{1\over6}}\,({z\over3})\,\Big|\,\cos\psi\,\,.
\label{sigma-rho-z-psi}
\eeq
After eliminating the angle $\psi$ in (\ref{sigma-rho-z-psi}), we get the implicit solution written in (\ref{implicit-sol}) of the PDE (\ref{PDE-unflavored}). Moreover, one can check that, when expressed in terms of  the variables $\rho$ and $\sigma$, the dilaton $\Phi$ and the functions $g_1$ and $g_2$ written in eqs. (\ref{dilaton-11-gauged}) and (\ref{g1-g2-gauged}) coincide with the ones of (\ref{dilaton-11-10d}) and (\ref{g12-11-10d}). 

\begin{figure}
\centering
\includegraphics[width=0.4\textwidth]{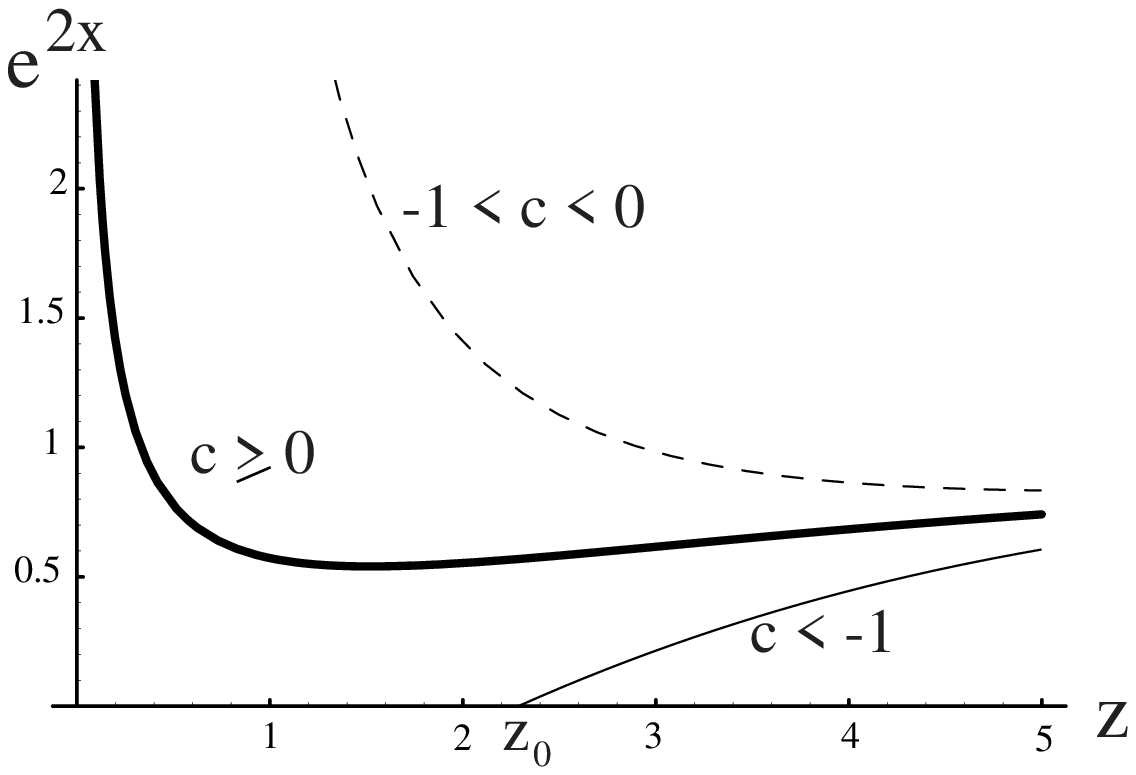} \hfill
\includegraphics[width=0.4\textwidth]{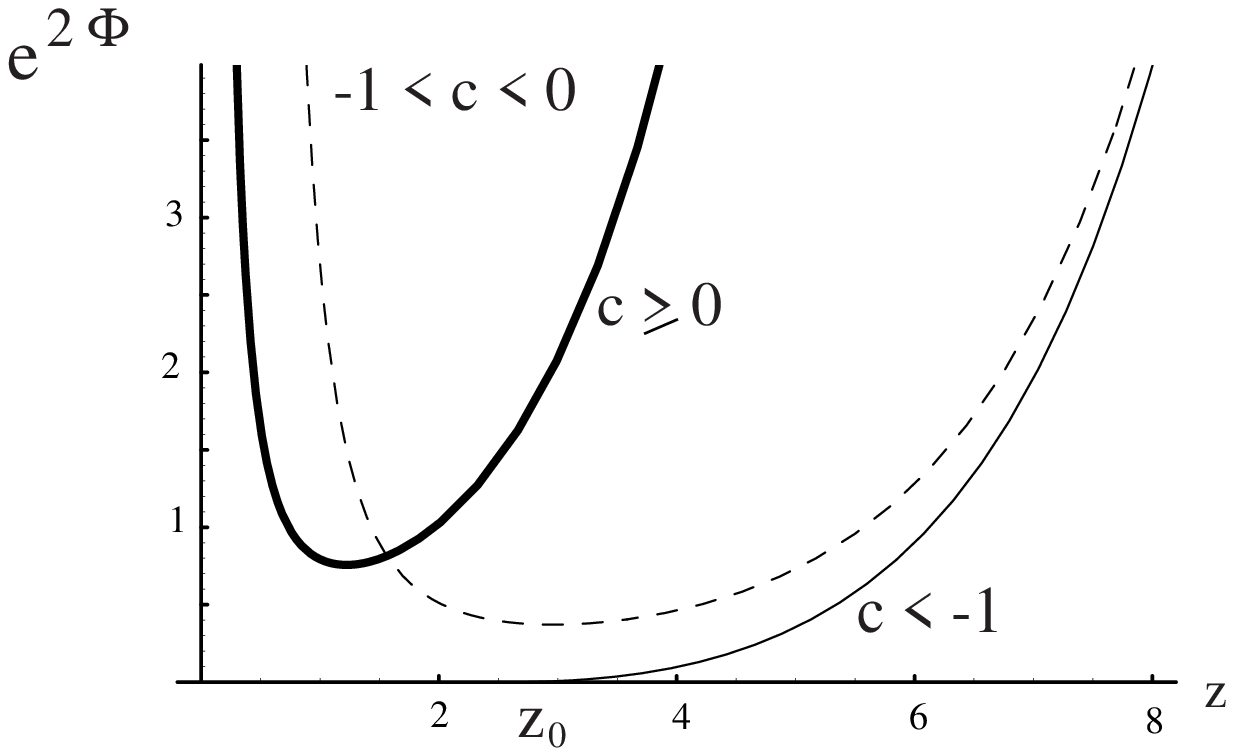}
\caption{On the left we plot $e^{2x}$ for three ranges of the constant $c$ in (\ref{s4xsol}). The plots on the right represent $e^{2\Phi}$, taken from (\ref{dilaton-11-gauged}), for $\psi=0$.}
\label{e2x-2phi-11}
\end{figure}

Let us now study the behavior of the solution for the different values  of the integration constant $c$  of eqs.  (\ref{s4xsol}) and (\ref{s4fsol}). The different behaviors of the function $e^{2x}$ are plotted in figure \ref{e2x-2phi-11} for three ranges of $c$. When $c\ge 0$ $e^{2x}$ diverges at $z=0$. On the contrary if $-1<c<0$ the function $e^{2x}$ diverges at some point $\tilde z_0$ such that 
$I_{1\over6}({\tilde z_0\over3})+c\,I_{-{1\over6}}({\tilde z_0\over3})=0$, which marks the end of the space. The solution is not well-defined for $c=-1$ since in this case 
$e^{2x}$ is negative for $z>0$. Finally, when $c<-1$ $e^{2x}$ vanishes at $z=z_0$, where $z_0$ has been defined in (\ref{z0}). The behavior of the dilaton for the three ranges of $c$ is also displayed in figure \ref{e2x-2phi-11}. From this figure we conclude that only when $c<-1$ the $tt$ component of the Einstein frame metric, namely $e^{\Phi/2}$, decreases when we approach the singularity and, thus, only in this $c<-1$ case is the singularity good. Actually, $z=z_0$ defines a line of singularities. This fact can be checked by analyzing the form of the  dilaton around $z=z_0$. Indeed, by using the properties of the modified Bessel functions one can verify that, for $z$ close to $z_0$, one has:
\bear
&&e^{2x}\,\approx {1\over 3}\,\,(z-z_0)\,,\rc\rc
&&I_{-{5\over6}}\Big({ z\over3}\Big)+c\,I_{{5\over6}}\Big({
z\over3}\Big)\,\approx\,{\gamma\over 3}\,(z-z_0)\,\,,\rc\rc
&&e^{5f}\,\approx\,{z_0^{{1\over 3}}\gamma^2\over 3\sqrt{3}}\,\,(z-z_0)^{{3\over 2}}\,\,,
\label{IRexpansion-11}
\eear
where $\gamma$ is a constant (depending on $z_0$ and $c$) defined as:
\beq
\gamma\equiv 
I_{{1\over6}}\Big({ z_0\over 3}\Big)+c\,I_{-{1\over6}}\Big({
z_0\over 3}\Big)\,\,.
\eeq
Using this result in (\ref{dilaton-11-gauged}) one arrives at the following expression of $e^{2\Phi}$ for 
$z\to z_0$:
\beq
e^{2\Phi}\,\approx\,{z_0^{{1\over 3}}\gamma^2\over 3}\,(z-z_0)\,\Big[\,
\sin^2\psi\,+\,{(z-z_0)^2\over 9}\,\,\cos^2\psi\,\Big]\,\,,
\eeq
which shows that $e^{2\Phi}\to 0$ for any value of $\psi$ in the interval $[0,\pi]$. Notice that, in terms of the variables $\sigma$ and  $\rho$ of (\ref{sigma-rho-z-psi}), the singularities lie in the segment $\sigma=0, |\rho|\le \rho_c$, where $\rho_c$ has been defined in (\ref{rho_c}), as claimed at the end of subsection \ref{11-solution}.

\subsubsection{UV and IR limits}
We will now discuss the UV and IR limits of the solution found, following the analysis of ref. \cite{Gauntlett:2001ur}.  First of all, we consider the UV limit of the metric, by taking $z\to\infty$. In order to take this limit in our solution it is quite useful to recall the asymptotic behavior of the modified Bessel functions, namely:
\beq
I_{\nu}\Big({z\over 3}\Big)\,\approx \sqrt{{3\over 2\pi}}\,\,{e^{{z\over 3}}\over 
\sqrt{z}}\,\,,\qquad\qquad z\to\infty\,\,.
\eeq
It follows from this behavior that
$e^{2x}\to 1$ and $\Omega\to 1$ in the large $z$ limit. Therefore, in the UV  the metric and dilaton take the  form:
\bear
&&ds^2\approx\,e^{\Phi}\,\Bigg[\,
 dx^2_{1,1}\,+\,{z\over m^2}\,d\Omega_4^2\,+\,{1\over m^2}\,\,
\Big[\,{1\over 9}\,dz^2\,+\,d\psi^2\,+\,\sin^2\psi\,
\big[\,\big(E^1\big)^2\,+\,\big(E^2\big)^2\,\big]\Big]\,\Bigg]
\,\,,\rc\rc
&&e^{2\Phi}\,\approx\,{3\over 2\pi}\,\,{e^{{2z\over 3}}\over z^{{2\over 3}}}\,\,,
\eear
which, indeed, corresponds to a  stack of D5-branes with worldvolume $\mathbb{R}^{1,1}\times S^4$. 

Let us next explore the IR  behavior of our background. We will concentrate on the case in which $c<-1$, which is singular at a point $z=z_0$, with $z_0$ being defined in (\ref{z0}). In order to study the limiting form of the metric and dilaton near the singularity we have to expand all functions around $z=z_0$ as in (\ref{IRexpansion-11}). 
If we define a new radial variable $y$ as:
\beq
y={1\over 3}\,\,(z-z_0)\,\,,
\eeq
then, the metric and dilaton take the form (for $\psi\not=0,\pi$):
\bear
&&ds^2\,\approx\,
e^{\Phi}\,\Bigg[\,
dx^2_{1,1}\,+\,{z_0\over m^2}\,d\Omega_4^2\,+\,
{1\over m^2\,y}\,\Big[\,dy^2\,+\,y^2\,
\big[\,\big(E^1\big)^2\,+\,\big(E^2\big)^2\,\big]\,+\,d\psi^2\,\Big]
\,\Bigg]\,\,,\rc\rc
&&e^{-2\Phi}\,\approx\,\gamma^2\,z_0^{{1\over 3}}\,y\,\sin^2\psi\,\,.
\eear
This form of the metric indicates that the singularity is generated by a distribution of branes parametrized by the the coordinate $\psi$. As $0\le\psi\le\pi$ this distribution has the topology of a segment and can be regarded as a linear distribution of branes which is dual to a slice of the Coulomb branch of the  ${\cal N}=(1,1)$ SYM.   In order to confirm this interpretation,  let us analyze, following ref. \cite{Gauntlett:2001ur},  the solution near $\psi=0$. With this purpose we introduce new coordinates $r$ and $\alpha$, defined as:
\bear
&&{1\over 3}\,\,(z-z_0)\,=\,\sqrt{r}\,\,\sin\Big({ \alpha\over 2}\Big)\,\,,\rc\rc
&&\psi\,=\,\sqrt{r}\,\,\cos\Big({ \alpha\over 2}\Big)\,\,.
\eear
In these variables, for small $\psi$,  one can check that $\Theta$, as defined in (\ref{Theta}) is given by:
\beq
\Theta\,\approx\,{\sqrt{r}\over \sin\big({ \alpha\over 2}\big)}\,\,.
\eeq
Then, one can check that metric and dilaton are:
\bear
&&ds^2\,\approx\,e^{\Phi}\,\Bigg[\,
dx^2_{1,1}\,+\,{z_0\over m^2}\,d\Omega_4^2\,+\,
{1\over 4 m^2 r^{{3\over 2}}\,\sin\big({ \alpha\over 2}\big)}\,\,
\Big[\,dr^2\,+\,r^2\,\Big(\,d\alpha^2\,+\,\sin^2\alpha\,
\big[\,\big(E^1\big)^2\,+\,\big(E^2\big)^2\,\big]\,\Big)\,
\Big]\,\Bigg]\,\,,\rc\rc
&&e^{-2\Phi}\,\approx\,\gamma^2\,z_0^{{1\over 3}}\,r^{{3\over 2}}\,
\sin\big({ \alpha\over 2}\big)\,\,,
\eear
which, as argued in \cite{Gauntlett:2001ur}, is consistent with the interpretation of the solution given above since $1/r^{{3\over 2}}\,\sin\big({ \alpha\over 2}\big)$ is a harmonic function in $\mathbb{R}^4$.

\vskip 1cm
\renewcommand{\theequation}{\rm{B}.\arabic{equation}}
\setcounter{equation}{0}

\section{Additional fivebrane solutions}
\label{additional}
In this appendix we will show that there exist solutions of the unflavored BPS systems (\ref{BPS}) and (\ref{BPS-10d-unflavored}) that are different from the ones obtained in gauged supergravity. We will find these solutions directly by adopting  an ansatz in which the two ten-dimensional variables $\rho$ and $\sigma$ are separated.  We will consider first the BPS system for the background with four supersymmetries and, subsequently, we will attack the system of BPS equations corresponding to the preservation of two SUSYs. 

\subsection{Backgrounds with four supersymmetries}

Let us try to solve the BPS system (\ref{BPS}) by means of the ansatz:
\beq
g\,=\,g(\sigma)\,\,,
\qquad\qquad
\Phi=\Phi(\rho)\,\,,
\eeq
Notice that the third equation in  (\ref{BPS}) is automatically solved by the ansatz. As  the last equation  is a consequence of the other three equations, we only need to solve the first two equations in (\ref{BPS}). By integrating  the second equation in (\ref{BPS}) over $\sigma$, we get:
\beq
z^3\,=\,{3\sigma^2\over 2}\,\,e^{-2\Phi}\,+\,f(\rho)\,\,,
\eeq
where $f(\rho)$ is an unknown function to be determined.  Moreover, by integrating the first equation in  (\ref{BPS}) we get:
\beq
z\,=\,B(\sigma)\,+\,m^2\,g(\sigma)\,\log(\rho)\,\,,
\eeq
where $B(\sigma)$ is a function to be determined. By combining these last two equations, we get:
\beq
{3\sigma^2\over 2}\,\,e^{-2\Phi}\,+\,f(\rho)\,=\,
\Big(\,B(\sigma)\,+\,m^2\,g(\sigma)\,\log(\rho)\,\Big)^3\,\,.
\label{z-compatibility}
\eeq
By comparing the $\sigma$ dependence of  both sides of (\ref{z-compatibility}) one concludes straightforwardly that  the four functions $B^{3}(\sigma)$, $B^{2}(\sigma)\,g(\sigma)$,  $B(\sigma)\,g^{2}(\sigma)$ and $g^{3}(\sigma)$ must be polynomials in $\sigma$ of second degree. This, in turn, is only possible if $B(\sigma)$ and $g(\sigma)$ are proportional to each other. Let us write:
\beq
m^2\,g(\sigma)\,=\,\kappa B(\sigma)\,\,,
\qquad\qquad
 B^{3}(\sigma)\,=\,b_2\,\sigma^2\,+\,b_1\,\sigma\,+\,b_0\,\,,
 \eeq
where $\kappa$, $b_0$, $b_1$ and $b_2$ are constants. Actually, the absence of linear terms in $\sigma$ on the left-hand-side of (\ref{z-compatibility}) implies that $b_1=0$. Thus, we write:
\beq
B(\sigma)\,=\,(\,b_2\,\sigma^2\,+\,b_0\,)^{{1\over 3}}\,\,.
\eeq
By using this expression of $B(\sigma)$ in  (\ref{z-compatibility}) one easily gets the form of the dilaton $\Phi$ and the integration function $f$, namely:
\beq
e^{2\Phi(\rho)}\,=\,{3\over 2 b_2}\,\,{1\over 
\big(\,1\,+\,\kappa\,\log(\rho)\,\big)^3}\,\,,
\qquad\qquad
f(\rho)\,=\,b_0\,\big(\,1\,+\,\kappa\,\log(\rho)\,\big)^3\,\,.
\eeq
The corresponding expressions for $z$ and $g$ are:
\beq
z\,=\,\big(\,b_2\,\sigma^2\,+\,b_0\,\big)^{{1\over 3}}\,
\big(\,1\,+\,\kappa\,\log(\rho)\,\big)\,\,,
\qquad\qquad
g\,=\,{\kappa\over m^2}\,\,\big(\,b_2\,\sigma^2\,+\,b_0\,\big)^{{1\over 3}}\,\,.
\eeq
Let us now change from the radial variable $\sigma$ to a new one $r$, defined as:
\beq
r^2\,=\,{6\over m^2}\,\,\sqrt{{3\over 2 b_2}}\,
\big(\,b_2\,\sigma^2\,+\,b_0\,\big)^{{1\over 3}}\,\,,
\eeq
and let us introduce a new mass parameter $\mu$, which has the form:
\beq
\mu^2\,=\,\sqrt{{3\over 2 b_2}}\,\,m^2\,\,.
\eeq
Then, after rescaling the Minkowski coordinates in the appropriate way and by redefining the $\rho$ coordinate as $\rho/\mu\to\rho$,  the metric takes the form:
\bear
&&ds^2\,=\,{dx^2_{1,1}\over 
\big(\,1\,+\,\kappa\,\log(\mu\rho)\,\big)^{{3\over 2}}}\,+\,
\big(\,1\,+\,\kappa\,\log(\mu\rho)\,\big)^{{3\over 2}}\,
\big(\,d\rho^2\,+\,\rho^2\,d\chi^2\,\big)\,+\,\rc\rc
&&\qquad\qquad\qquad\qquad\qquad\qquad
+\,\big(\,1\,+\,\kappa\,\log(\mu\rho)\,\big)^{-{1\over 2}}\,ds^2_6\,\,,
\label{reg-con-10d}
\eear
where $ds^2_6$ is the metric of the conifold with a blown up four-cycle (the regularized conifold):
\bear
&&ds^2_6\,=\,{r^2\over 6}\,\,\Big(\,d\theta_1^2\,+\,\sin^2\theta_1\,d\phi_1^2\,+\,
d\theta_2^2\,+\,\sin^2\theta_2\,d\phi_2^2\,\Big)\,+\,
{dr^2\over 1\,-\,{a^6\over r^6}}\,+\,\rc\rc
&&\qquad\qquad\qquad\qquad
+{r^2\over 9}\,\,\Big(1\,-\,{a^6\over r^6}\Big)\,
\Big(\,d\psi\,+\,\cos\theta_1 d\phi_1\,+\,\cos\theta_2 d\phi_2\,\Big)^2\,\,,
\label{regul-conifold}
\eear
with the constant $a$ being:
\beq
a\,=\,{(b_0)^{{1\over 6}}\over m}\,\,\Big({54\over b_2}\Big)^{{1\over 4}}\,\,.
\eeq
Notice that the variables $r$ and $\rho$ in  (\ref{reg-con-10d}) and (\ref{regul-conifold}) are dimensionful. Moreover, the dilaton and $g$ are given by:
\beq
e^{\Phi}\,=\,{\mu^2\over m^2}\,\,{1\over 
\big(\,1\,+\,\kappa\,\log(\mu\rho)\,\big)^{{3\over 2}}}\,\,,\qquad\qquad
g\,=\,\kappa\,{\mu^2\over m^2}\,{r^2\over 6}\,\,.
\eeq
The RR field strength $F_3$ that corresponds to the function $g$ written above is:
\bear
&&F_3\,=\,\kappa\,{\mu^2\over m^2}\,
\Big[\,{r^2\over 6}\,\Big(\sin\theta_1 d\theta_1\wedge d\phi_1\,+\,
\sin\theta_2 d\theta_2\wedge d\phi_2\Big)\,+\,\rc\rc
&&\qquad\qquad\qquad
+{r\over 3}\,\Big(d\psi\,+\,\cos\theta_1 d\phi_1\,+\,\cos\theta_2 d\phi_2\,\Big)\wedge dr
\,\Big]\,\wedge d\chi\,\,.
\eear
Let us rewrite this expression for $F_3$ in a more compact form. With this purpose, let us introduce the following vielbein basis for the conifold metric:
\bear
&&E^1\,=\,{r\over \sqrt{6}}\,d\theta_1\,\,,\qquad\qquad\qquad
E^2\,=\,{r\over \sqrt{6}}\,\sin\theta_1\,d\phi_1\,\,,\rc\rc
&&E^3\,=\,{r\over \sqrt{6}}\,d\theta_2\,\,,\qquad\qquad\qquad
E^4\,=\,{r\over \sqrt{6}}\,\sin\theta_2\,d\phi_2\,\,,\rc\rc
&&E^5\,=\,{r\over 3}\,\Big(1\,-\,{a^6\over r^6}\Big)^{{1\over 2}}\,
\Big(d\psi\,+\,\cos\theta_1 d\phi_1\,+\,\cos\theta_2 d\phi_2\,\Big)\,\,,\rc\rc
&&E^6\,=\,{dr\over \Big(1\,-\,{a^6\over r^6}\Big)^{{1\over 2}}}\,\,.
\eear
In terms of this basis the K\"ahler form of the conifold takes the form:
\beq
J\,=\,E^1\wedge E^2\,+\,E^3\wedge E^4\,+\,E^5\wedge E^6\,\,,
\eeq
and the RR three form $F_3$ is just given by:
\beq
F_3\,=\,\kappa\,{\mu^2\over m^2}\,J\wedge d\chi\,\,.
\eeq

\subsection{Backgrounds with two supersymmetries}

In this subsection we want to obtain new solutions of the first-order BPS system 
 (\ref{BPS-10d-unflavored}), different from the one we got from gauged supergravity.
 We have been able to find two different ans\"atze to solve this system that we detail in the next two subsections.

\subsubsection{First ansatz}
As a first possibility for getting new solutions of the BPS system (\ref{BPS-10d-unflavored}), let us adopt the following ansatz for $z$, $g_{1,2}$ and $\Phi$:
\beq
z\,=\,z(\sigma)\,\,,\qquad
g_1(\rho)\,=\,-g_2(\rho)\,\,,\qquad
\Phi\,=\,\Phi(\sigma)\,\,.
\eeq
Notice that ${\rm 2}^{\rm nd}$, ${\rm 4}^{\rm th}$ and ${\rm 5}^{\rm th}$ equations in  (\ref{BPS-10d-unflavored}) are automatically solved by the ansatz.  Moreover, the ${\rm 6}^{\rm th}$ equation in  (\ref{BPS-10d-unflavored}), namely:
\beq
g_2'\,=\,-{\sigma \over m^2 z^{{2\over 3}}}\,e^{-2\Phi}\,\,,
\eeq
is such that the left-hand side only depends on $\rho$, whereas the right-hand side depends of $\sigma$.  Separating variables, we get:
\beq
g_2'\,=\,c_1\,\,,\qquad\qquad
\sigma\,e^{-2\Phi}\,=\,-m^2 c_1\,z^{{2\over 3}}\,\,,
\eeq
where $c_1$ is a constant. The equation for $g_2$ can be immediately integrated, namely:
\beq
g_2\,=\,c_1+c_2\rho\,\,,
\eeq
where $c_2$ is a new constant. Furthermore,  using the third equation in the BPS system  (\ref{BPS-10d-unflavored}) we can obtain the following equation for $z(\sigma)$:
\beq
z^{{2\over 3}}\,\dot z\,=\,-3m^2\,c_1\,\,,
\eeq
which can also be readily integrated. It is also easy to get the expression of the dilaton. Let us write this solution in terms of new constants
$\lambda_{1,2,3}$ defined as $\lambda_{1,2}=-5m^2 c_{1,2}$ and $\lambda_3=5c_3$. Then, one has:
\bear
&&g_1=-g_2\,=\,{1\over 5m^2}\,\big(\,\lambda_1\,\rho+\lambda_2\,\big)\,\,,\rc\rc
&&z=\big[\,\lambda_3\,+\,\lambda_1\sigma\,\big]^{{3\over 5}}\,\,,
\qquad\qquad
e^{2\Phi}\,=\,{5\sigma\over \lambda_1}\,
\big[\,\lambda_3\,+\,\lambda_1\sigma\,\big]^{-{2\over 5}}\,\,.
\eear
It is  now convenient to define the function $H(\sigma)$ as:
\beq
H(\sigma)\,=\,\sqrt{\lambda_1\over 5\sigma}\,\,\big[\,\lambda_3\,+\,\lambda_1\sigma\,\big]^{{1\over 5}}\,\,.
\eeq
Then, $z$ and $\Phi$ can be rewritten as:
\beq
z\,=\,\Big({5\sigma\over \lambda_1}\Big)^{{3\over 2}}\,\Big[\,H(\sigma)\,\Big]^{3}\,\,,
\qquad\qquad
e^{\Phi}\,=\,\Big[\,H(\sigma)\,\Big]^{-1}\,\,,
\eeq
and the metric takes the form:
\bear
&&ds^2_{10}\,=\,H^{-1}\,dx^2_{1,1}\,+\,
{1\over m^2}\,\Big({5\sigma\over \lambda_1}\Big)^{{3\over 2}}\,H^2\,d\Omega^2_{4}\,+\,
\rc\rc
&&\qquad\qquad+\,
{1\over m^2}\,\Big({\lambda_1 \over 5\sigma}\Big)^{2}\,\,H^{-3}\,
\Big[\,d\sigma^2\,+\,\sigma^2\,
\big(\,(E^1)^2+(E^2)^2\,\big)\,\Big]\,+\,{H\over m^2}\,(d\rho)^2\,\,.
\eear
Notice that, by plugging the solution  for $g_1$ and $g_2$ into the expression (\ref{F3-10d-unflavored}) of $F_3$, we get:
\beq
F_3\,=\,{\lambda_1\over 5m^2}\,\Big[\,E^1\wedge E^2\,-\,
{\cal S}^{\xi}\wedge {\cal S}^{3}\,-\,{\cal S}^1\wedge {\cal S}^{2}\,\Big]\wedge d\rho\,\,.
\eeq
Let us now rewrite this solution by means of a change of the radial variable: we will change from $\sigma$ to a new variable $r$, related to the former as:
\beq
r\,=\,{2\sqrt{5}\over 3}\,\lambda_1^{{3\over 10}}\,\sigma^{{3\over 10}}\,\,
\Big(\,1\,+\,{\lambda_3\over \lambda_1\,\sigma}\,\Big)^{{3\over 10}}\,\,.
\eeq
Moreover, instead of the constant $\lambda_3$, we will use the constant $l$, defined as:
\beq
l\,=\,{2\sqrt{5}\over 3}\,\lambda_3^{{3\over 10}}\,\,.
\eeq
From these definitions one can check that:
\beq
1\,-\,{l^{{10\over 3}}\over r^{{10\over 3}}}\,=\,
\Big(\,1\,+\,{\lambda_3\over \lambda_1\,\sigma}\,\Big)^{-1}\,\,,
\qquad\qquad
{dr\over r}\,=\,{3\over 10}\,\,\Big(\,1\,+\,{\lambda_3\over \lambda_1\,\sigma}\,\Big)^{-1}\,
{d\sigma\over\sigma}\,\,.
\eeq
Moreover, after a convenient  rescaling of the Minkowski coordinates,  the metric and dilaton take the form:
\bear
&&ds^2_{10}\,=\,e^{2k}\,\Bigg[\,dx^2_{1,1}\,+\,
\Big(\,1\,-\,{ l^{{10\over 3}}\over  r^{{10\over 3}}}\,\Big)^{-1}\,\,(dr)^2\,+\,
{9\over 20}\,r^2\,d\Omega_4^2\,+\,\rc\rc
&&\qquad\qquad\qquad\qquad+\,
{9\over 100}\,r^2\,\,\Big(\,1\,-\,{ l^{{10\over 3}}\over  r^{{10\over 3}}}\,\Big)\,
\Big[\,(E^1)^2\,+\,(E^2)^2\Big]\,\,\Bigg]\,+\,{e^{-2k}\over m^4}\,\,\big(d\rho\big)^2\,\,,\rc\rc
&&e^{2\Phi}\,=\,m^4\,e^{4k}\,\,,
\label{1st-add-11}
\eear
where $k$ is defined as:
\beq
e^{2k}\,=\,{3\over 2\lambda_1 m^2}\,\,r\,\,
\Big(\,1\,-\,{l^{{10\over 3}}\over r^{{10\over 3}}}\,\Big)^{{1\over 2}}\,\,.
\eeq
In order to identify this background let us perform a T-duality transformation along the coordinate $\rho$ (shifting $\rho$ is an isometry of the background (\ref{1st-add-11})). 
According to the T-duality rules in \cite{Bergshoeff:1995as}, one should perform the changes:
\beq
g_{\rho\rho}\,\rightarrow {1\over g_{\rho\rho}}\,\,,\qquad\qquad
e^{2\Phi}\rightarrow {e^{2\Phi}\over g_{\rho\rho}}\,\,.
\eeq
Also the RR three-form $F_3$ is converted into an RR two-form $F_2$. The resulting type IIA background is:
\bear
&&ds^2_{10}\,=\,e^{2k}\,\Bigg[\,dx^2_{1,2}\,+\,
\Big(\,1\,-\,{ l^{{10\over 3}}\over  r^{{10\over 3}}}\,\Big)^{-1}\,\,(dr)^2\,+\,
{9\over 20}\,r^2\,d\Omega_4^2\,+\,\rc\rc
&&\qquad\qquad\qquad\qquad\qquad\qquad+\,
{9\over 100}\,r^2\,\,\Big(\,1\,-\,{ l^{{10\over 3}}\over  r^{{10\over 3}}}\,\Big)\,
\Big[\,(E^1)^2\,+\,(E^2)^2\Big]\,\,\Bigg]\,\,,\rc\rc
&&e^{2\Phi}\,=\,m^8\,e^{6k}\,=\,\Bigg({3 m^{{2\over 3}}\over 2\lambda_1}\Bigg)^3\,
r^3\,\,\Big(\,1\,-\,{ l^{{10\over 3}}\over  r^{{10\over 3}}}\,\Big)^{{3\over 2}}\,\,,\rc\rc
&&F_2\,=\,{\lambda_1\over 5 m^2}\,\Big[\,E^1\wedge E^2\,-\,
{\cal S}^{\xi}\wedge {\cal S}^{3}\,-\,{\cal S}^1\wedge {\cal S}^{2}\,\Big]\,\,.
\eear
This solution corresponds to  a D6-brane wrapped along a four-cycle of  a $G_2$-cone, which is an analogue in the type IIA theory of the solution found in \cite{Chamseddine:2001hk}-\cite{Schvellinger:2001ib}.

\subsubsection{Second ansatz}

As a second ansatz, let us now try to find solutions of (\ref{BPS-10d-unflavored}) in which the functions $g_1$ and $g_2$ are proportional to each other. Let us write $g_1=\alpha g_2$, with $\alpha$ being a constant and let us separate variables as:
\beq
g_1=\alpha g_2\,=\,\alpha R(\rho)\,S(\sigma)\,\,,
\qquad\qquad
z\,=\,z_1(\rho)\,z_2(\sigma)\,\,.
\eeq
From the ${\rm 5}^{\rm th}$ equation in the BPS system, we get that $S(\sigma)$ satisfies the equation:
\beq
{\dot S\over S}\,=\,{1+\alpha\over \sigma}\,\,,
\eeq
which can be immediately integrated as:
\beq
S(\sigma)\,=\,c_1\,\sigma^{1+\alpha}\,\,.
\label{S-sigma}
\eeq
By substituting the ansatz in the ${\rm 4}^{\rm th}$ equation of (\ref{BPS-10d-unflavored}), using the result just found for $S(\sigma)$ and separating variables, we arrive at:
\beq
{1\over R}\,{z_1'\over z_1^{{2\over 3}}}\,=\,3(1+\alpha)m^2\,c_1\,
\sigma^{\alpha}\, z_2^{-{1\over 3}}\,=\,c_2\,\,,
\eeq
where $c_2$ is a new constant. From this equation we can immediately get 
$z_2(\sigma)$, namely:
\beq
z_2\,=\,\Bigg[{3(1+\alpha)m^2\,c_1\over c_2}\Bigg]^3\,\,\sigma^{3\alpha}\,\,,
\label{z2-sigma}
\eeq
as well as an equation for $z_1(\rho)$ and $R(\rho)$, namely:
\beq
z_1^{-{2\over 3}}\,z_1'\,=\,c_2\,R\,\,.
\label{z1-R-1}
\eeq
Let us now combine the ${\rm 3}^{\rm rd}$ and ${\rm 6}^{\rm th}$ equations in (\ref{BPS-10d-unflavored}) to give:
\beq
{\dot z\over g_2'}\,=\,-{3 m^2\over z^{{2\over 3}}}\,\,.
\eeq
Using the results just found for $z_2(\sigma)$ and $S(\sigma)$ (eqs. (\ref{z2-sigma}) and (\ref{S-sigma})) and separating variables we get:
\beq
{z_1^{{5\over 3}}\over R'}\,=\,-{m^2 c_1\over \alpha}\,\,
{\sigma^{2+\alpha}\over z_2^{{5\over 3}}}\,=\,c_3\,\,,
\label{c3-eq}
\eeq
where $c_3$ is a new constant.  The consistency of the $\sigma$-dependent part of equation (\ref{c3-eq}) with (\ref{z2-sigma}) fixes the value of $\alpha$ and relates the constant $c_3$ with the other constants $c_1$ and $c_2$, namely:
\beq
\alpha={1\over 2}\,\,,\qquad\qquad
c_3\,=\,-2m^2 c_1\,\,\Big({2c_2\over 9 m^2 c_1}\Big)^5\,\,.
\eeq
On the other hand, the $\rho$-dependent part of (\ref{c3-eq}), gives rise to the following relation between $z_1$ and $R$:
\beq
z_1^{{5\over 3}}\,=\,c_3\,R'\,\,.
\label{z1-R-2}
\eeq
By combining (\ref{z1-R-1}) and (\ref{z1-R-2}) one can get $z_1(\rho)$ in terms of $R(\rho)$, namely:
\beq
z_1(\rho)\,=\,\sqrt{c_2 \,c_3\,R(\rho)^2\,+\,c_4}\,\,,
\eeq
with $c_4$ being a new constant of integration. Moreover, $R(\rho)$ can be obtained as the solution of the equation:
\beq
R'(\rho)\,=\,{1\over c_3}\,\Big[\,c_2\,c_3R(\rho)^2\,+\,c_4\,\Big]^{{5\over 6}}\,\,.
\eeq
In order to rewrite this solution, let us  define new constants $\lambda$ and $\kappa$ as:
\beq
\lambda^2\,=\,-c_2 \,c_3\,\,,\qquad\qquad
\kappa\,=\,{9 m^2 c_1\over 2 c_2}\,\,.
\eeq
The constants $c_{1,2,3}$ can be written in terms of $\lambda$ and $\kappa$ as:
\beq
c_1\,=\,{\lambda\over 3 m^2}\,\kappa^3\,\,,\qquad\qquad
c_2\,=\,{3\over 2}\, \lambda\kappa^2\,\,,\qquad\qquad
c_3\,=\,-{2\lambda\over 3\kappa^2}\,\,.
\eeq
In terms of these new constants $\kappa$ and $\lambda$, the different functions of the ansatz take the form:
\bear
&&z\,=\,\kappa^3\,\Big[\,c_4-\lambda^2\,R^2(\rho)\,\Big]^{{1\over 2}}
\,\sigma^{{3\over 2}}\,\,,\rc\rc
&&g_1\,=\,{1\over 2}\,g_2\,=\,{\kappa^3\,\lambda\over 6 m^2}\,R(\rho)\,\sigma^{{3\over 2}}\,\,,\rc\rc
&&e^{-2\Phi}\,=\,{\kappa^7\over 2}\,\,\Big[\,c_4\,-\,\lambda^2\,R(\rho)^2\,\Big]^{{7\over 6}}\,\,\sigma^{{3\over 2}}\,\,,
\eear
with $R(\rho)$ satisfying:
\beq
\lambda\, R'(\rho)\,=\,-{3\kappa^2\over 2}\,\,
\Big[\,c_4-\lambda^2\,R^2(\rho)\,\Big]^{{5\over 6}}\,\,.
\label{Rprime-rho}
\eeq
In order to write the background obtained in a more convenient way, let us change from the $\rho$ coordinate to a new coordinate $r$, defined as:
\beq
r\,=\,{R(\rho)\over \sqrt{c_4}}\,\,.
\eeq
Moreover, we will redefine again the integration constants  and we will introduce new constants $a$ and $b$, which are related to $\lambda$, $\kappa$ and $c_4$ as follows:
\beq
a\,=\,{1\over \lambda}\,\,,\qquad\qquad
b\,=\,\kappa^{{1\over 2}}\,\,c_4^{{1\over 12}}\,\,.
\eeq
One can check that $z$, $g_1$, $g_2$ and the dilaton $\Phi$ can be written in terms of the new  variables and constants as:
\bear
&&z\,=\,b^6\,\Big[\,1\,-\,{r^2\over a^2}\,\Big]^{{1\over 2}}
\,\sigma^{{3\over 2}}\,\,,\rc\rc
&&g_1\,=\,{1\over 2}\,g_2\,=\,{b^6\,\over 6   m^2 a}\,\,r\,\sigma^{{3\over 2}}\,\,,\rc\rc
&&e^{\Phi}\,=\,{\sqrt{2}\over b^7}\,\,
\Big[\,1\,-\,{r^2\over a^2}\,\Big]^{-{7\over 12}}\,\,\sigma^{-{3\over 4}}\,\,,
\eear
while, from (\ref{Rprime-rho}), we find that the jacobian of the change $\rho\to r$ is given by:
\beq
{ d\rho\over d r}\,=\,-{2\over 3 a b^4}\,\,\Big[\,1\,-\,{r^2\over a^2}\,\Big]^{-{5\over 6}}\,\,.
\eeq
Using these results, we can readily get the form of the metric in the new variables, namely:
\bear
&&ds^2_{10}\,=\,{\sqrt{2}\over b^7}\,
\Big[\,1\,-\,{r^2\over a^2}\,\Big]^{-{7\over 12}}\,\,\sigma^{-{3\over 4}}\,dx^2_{1,1}\,+\,
{1\over \sqrt{2} m^2\,b}\,
\Big[\,1\,-\,{r^2\over a^2}\,\Big]^{-{1\over 12}}\,\,\sigma^{-{5\over 4}}\,
\Bigg[\,d\sigma^2\,+\,\qquad\qquad\rc\rc
&&\qquad\qquad+\sigma^2\Big[{4\over 9 a^2}\,
\Big(\,1\,-\,{r^2\over a^2}\,\Big)^{-1}\,dr^2\,+\,2d\Omega^2_4\,+\,
(E^1)^2\,+\,(E^2)^2\,\Big]\Bigg]\,\,.
\eear
In order to reabsorb some of the constants of integration, 
let us redefine further the coordinates $r$ and  $\sigma$ as $r\to a r$ and $\sigma \to 2^{2/3}\,b^{4/3}\, m^{8/3}\,\sigma$,  and let us introduce a new constant $N$, defined as $N=m^2\,b^8$. Then, the metric, dilaton and the functions $g_1$ and $g_2$ parametrizing   the RR three-form are:
\bear
&&ds^2_{10}\,=\,{1\over N}\,
\big[\,1\,-\,r^2\,\big]^{-{7\over 12}}\,\,\sigma^{-{3\over 4}}\,dx^2_{1,1}\,+\,
\big[\,1\,-\,r^2\,\big]^{-{1\over 12}}\,\,\sigma^{-{5\over 4}}\,
\Big[\,d\sigma^2\,+\,\qquad\qquad\rc\rc
&&\qquad\qquad\qquad\qquad\qquad\qquad
+\sigma^2\Big({4dr^2\over 9(\,1\,-\,r^2\,\big)}\,
+\,2d\Omega^2_4\,+\,
(E^1)^2\,+\,(E^2)^2\,\Big)\Big]\,\,,\rc\rc
&&e^{-\Phi}\,=\,N\big[\,1\,-\,r^2\,\big]^{{7\over 12}}\,\,\sigma^{{3\over 4}}\,\,,\rc\rc
&&g_1\,=\,{1\over 2}\,g_2\,=\,{N\over 3}\, r\,\sigma^{{3\over 2}}\,\,.
\eear
Notice that the coordinate $r$ lies in the range $0\le r\le 1$. Actually, it can be regarded as an angular coordinate. To make this fact more explicit we could change  to a new coordinate $\alpha$ such that, for example,  $\sin\alpha=r$. This metric is singular. Its scalar curvature is:
\beq
R\,=\,{3\over 32}\,\,{5+16 r^2\over \big[\,1\,-\,r^2\,\big]^{{29\over 24}}\sigma^{9\over 8}}\,\,,\eeq
which diverges at $\sigma=0$ and $r=1$.

\vskip 1cm
\renewcommand{\theequation}{\rm{C}.\arabic{equation}}
\setcounter{equation}{0}

\section{Equations of motion}
\label{eoms}

In this appendix we verify that the any solution of the  first-order BPS  equations (\ref{BPS-flavored}) and (\ref{BPS-flavored-singular}) also satisfies the equations of motion for the gravity plus smeared branes systems with four and two supersymmetries. In order to carry out this calculation it is more convenient to work in the  Einstein frame. Recall that the metric in Einstein  and string frame are related as:
\beq
G_{MN}^{Einstein}\,=\,e^{-{\Phi\over 2}}\,G_{MN}^{string}\,\,.
\eeq
From now on we will suppress the extra label in $G$ and we will assume that all metric elements are written in the Einstein frame. The total  action of the system is of the form:
\beq
S\,=\,S_{IIB}\,+\,S_{flavor}\,\,,
\label{S-total}
\eeq
where $S_{IIB}$ is the action of type IIB supergravity for the metric, dilaton and RR three-form, namely:
\beq
S_{IIB}\,=\,{1\over 2\kappa_{10}^2}\,\,\int\,d^{10} x\,
\sqrt{-G}\,\,\Big(\,R\,-\,{1\over 2}\,\partial_M\,\Phi\,\partial^{M}\,\Phi\,-\,
{1\over 12}\,e^{\Phi}\,F_3^2\,\Big)\,\,.
\eeq
In (\ref{S-total})  $S_{flavor}$ is the smeared version  of the DBI+WZ action for the flavor branes which, in Einstein frame, is given by:
\beq
S_{flavor}\,=\,-T_5\,\int_{{\cal M}_{10}}\,e^{{\Phi\over 2}}\,\Omega\wedge {\cal K}\,+\,
T_5\,\int_{{\cal M}_{10}}\,\Omega\wedge C_6 \,\,,
\label{flavor-action}
\eeq
where ${\cal K}$ is the calibration form and $\Omega$ the smearing form. To write the DBI term in (\ref{flavor-action}) (the first term) we have assumed that there is no worldvolume gauge field and we  have taken into account that for a SUSY configuration the induced  volume form is just the pullback of the calibration form ${\cal K}$. After the smearing we can substitute this pullback by the wedge product with the smearing form $\Omega$. Notice that, in the Einstein frame, the calibration form ${\cal K}$ is still given by eqs. (\ref{calK-explicit}) and (\ref{Phi-expression}), with the $e^a$'s being the forms of the basis (\ref{frame}) and (\ref{10d-frame}) multiplied by $e^{-{\Phi\over 4}}$.

Recall that the WZ term is the one responsible for the violation of the Bianchi identity for $F_3$, namely $dF_3\,=\,2 \kappa^2_{10}\,T_5\,\Omega$. This equation is automatically incorporated in our ansatz. 
Moreover, the Maxwell  equation for $F_3$, namely $d\big( e^{\Phi}\, {}^*\,F_3)=0$ is also automatic as a consequence of the first calibration condition in (\ref{calibration-conditions}) which, in the Einstein frame, reads 
$-e^{\Phi}\, {}^*\,F_3\,=\,d\big( e^{{\Phi\over 2}}\,{\cal K}\big)$ (actually,  $e^{{\Phi\over 2}}{\cal K}$ can be taken as the six-form potential  $C_6$).

The equation of motion of the dilaton derived from (\ref{S-total})  is:
\beq
d\big({}^*\,d\,\Phi\big)\,\,=\,
{1\over 2}\,e^{\Phi}\,F_3\wedge {}^*\,F_3\,+\,
{1\over 2}\,e^{{\Phi\over 2}}\,{\cal K}\wedge dF_3\,\,.
\label{dilaton-eom}
\eeq
Notice that we have  written the equation of $\Phi$ in terms of differential forms. One can show that (\ref{dilaton-eom}) is satisfied as a consequence of the BPS equations 
(\ref{BPS-flavored}) or (\ref{BPS-flavored-singular}). In fact, in order to evaluate the left-hand side of (\ref{dilaton-eom}) one has to compute the derivatives of the BPS equations 
(\ref{BPS-flavored}) and (\ref{BPS-flavored-singular}). In this process one generates some terms in which the Heaviside function $\Theta(\rho-\rho_Q)$ is differentiated  and, therefore, the Dirac delta function $\delta(\rho-\rho_Q)$ is produced. These terms match precisely the one containing $dF_3$ in (\ref{dilaton-eom}), while the remaining ones correspond to the $F_3^2$ term.  It also worth pointing out that, in the ${\cal N}=(1,1)$ case, the consistency equation (\ref{L12-eq}) must also be used to verify the fulfillment of (\ref{dilaton-eom}).

The Einstein equations, in Einstein frame,  are:
\bear
&&R_{MN}\,-\,{1\over 2} G_{MN}\,R\,=\,{1\over 2}\,\Big(\,
\partial_M\Phi\,\partial_N\Phi\,-{G_{MN}\over 2}\,
\partial_P\Phi\,\partial^P\Phi\,\Big)\,+\,\rc\rc
&&\qquad\qquad\qquad\qquad+
{e^{\Phi}\over 12}\,\Big(\,3F^{(3)}_{MPQ}\,F^{(3)\,\,PQ}_{N}\,-\,{G_{MN}\over 2}\,
F_3^2\,\Big)\,+\,T_{MN}\,\,,
\label{Einstein}
\eear
where $T_{MN}$ is the energy-momentum tensor of the smeared flavor branes, defined as:
\beq
T_{MN}\,=\,-{2\kappa_{10}^2\over \sqrt{-G}}\,{\delta S_{DBI}\over \delta G^{MN}}\,\,.
\eeq
It is interesting to write the different components $T_{ab}$  of $T$ in flat components with respect to our frame basis $\{ e^{a}\}$.  This can be done by using the Palatini formalism and computing directly the derivatives of $S_{DBI}$ with respect to the vielbeins.  In order to rewrite $T_{ab}$ in a suggestive form, let us introduce the following (standard) definition. 
Let $A^{(p)}$ be an arbitrary $p$-form which, in the basis of the frame one-forms $e^{a}$, can be written as
\beq
A^{(p)}\,=\,{1\over p!}\,A^{(p)}_{a_1\cdots a_p}\,e^{a_1}\wedge\cdots e^{a_p}\,\,.
\eeq
Then, we define $\iota_{e^{a}}\,\big[A^{(p)}]$ as the following $(p-1)$-form
\beq
\iota_{e^{a}}\,\big[A^{(p)}]\,=\,{1\over (p-1)!}\,A^{(p)}_{a a_2\cdots a_p}\,
e^{a_2}\wedge\cdots e^{a_p}\,\,.
\label{iA}
\eeq
It can then be verified that  $T_{ab}$ can be written as:
\beq
T_{ab}\,{\rm Vol}({\cal M}_{10})\,=\,-\kappa_{10}^2\,T_5\,e^{{\Phi}\over 2}\,
\Omega\wedge e^{a}\wedge \iota_{e^{a}}\,\big[{\cal K}]\,\,\eta_{ab}\,\,.
\label{Tab}
\eeq
Taking into account that the calibration form ${\cal K}$ is diagonal in our frame basis (see (\ref{calK-explicit}) and (\ref{Phi-expression})), 
this equation can be recast in a simpler form as follows.  Let us denote by $ {\cal K}_a$ the part of ${\cal K}$ that contains $e^a$ in its expression. Then, one can verify that the $T_{ab}$ written in (\ref{Tab}) can be compactly expressed as:
\beq
T_{ab}\,{\rm Vol}({\cal M}_{10})\,=\,-\kappa_{10}^2\,T_5\,e^{{\Phi}\over 2}\,
\Omega\wedge {\cal K}_a\,\eta_{ab}\,\,.
\label{T-vierbein}
\eeq
Let us now particularize the general expression (\ref{T-vierbein}) to our two cases. For the system with four supersymmetries, after using (\ref{calK-explicit}) and (\ref{Omega-22}), one gets:
\bear
&&T_{x^{\mu}x^{\nu}}\,=\,-{g_s\alpha'N_f\over 2}\,{m^4 e^{{3\Phi\over 2}}\over \rho z}\,\,
\delta(\rho-\rho_Q)\,\eta_{\mu\nu}\,\,,\rc\rc
&&T_{22}=T_{33}=T_{44}=T_{55}=
-{g_s\alpha'N_f\over 4}\,{m^4 e^{{3\Phi\over 2}}\over \rho z}\,\,
\delta(\rho-\rho_Q)\,\,,\rc\rc
&&T_{66}=T_{77}=-{g_s\alpha'N_f\over 2}\,{m^4 e^{{3\Phi\over 2}}\over \rho z}\,\,
\delta(\rho-\rho_Q)\,\,,\rc\rc
&&T_{88}=T_{99}=0\,\,.
\label{T-components-22}
\eear
Similarly, for the system with two supersymmetries one gets:
\bear
&&-T_{00}\,=\,T_{11}\,=\,{m^4\,e^{\frac{3\Phi}{2}}\over 2\,\sigma^2\, z^{{1\over 3}}}\,
\delta (\rho-\rho_Q)\,\big(\,4\sigma\,\big(L_1+L_2\big)+\,2\,\sigma^2\,\dot L_2\,+\,z^{{7\over 3}}\,e^{2\Phi}\,\dot L_1\,\big)\,,\rc\rc
&&T_{22}\,=\,T_{33}\,=\,T_{44}\,=\,T_{55}\,=\,
{m^4\,e^{\frac{3\Phi}{2}}\over 2\,\sigma^2\, z^{{1\over 3}}}\,
\delta (\rho-\rho_Q)\,\big(\,2\sigma\,\big(L_1+L_2\big)+\,\sigma^2\,\dot L_2\,+\,z^{{7\over 3}}\,e^{2\Phi}\,\dot L_1\,\big)\,,\rc\rc
&&T_{66}\,=\,{2m^4\,e^{\frac{3\Phi}{2}}\over \sigma\, z^{{1\over 3}}}\,
\delta (\rho-\rho_Q)\,\big(\,L_1+L_2\,\big)\,,\rc\rc
&&T_{77}\,=\,T_{88}\,=\,{m^4\,e^{\frac{3\Phi}{2}}\over \sigma\, z^{{1\over 3}}}\,
\delta (\rho-\rho_Q)\,\big(\,L_1+L_2\,+\,\sigma\,\dot L_2\,\big)\,,\rc\rc
&&T_{99}\,=\,0\,\,,
\label{T-components-11}
\eear
where (\ref{Phi-expression}) and (\ref{Omega-11}) have been used. It is now straightforward, although tedious, to verify that the  Einstein equations (\ref{Einstein}) are satisfied as a consequence of the BPS equations (\ref{BPS-flavored}) and (\ref{BPS-flavored-singular}). As before, the consistency condition (\ref{L12-eq}) is also needed to satisfy (\ref{Einstein}) in the ${\cal N}=(1,1)$ case.

\medskip

%%%%%%%%%%%%%%%%%%%%%%%%%%%%%%%%%%%%%%%%%%%%%%%%%%%%%%%%%%%%%%%%%%%%%%%%%%

\end{document}